\documentclass[floatfix,twocolumn,prl,amsmath]{revtex4}
\usepackage{graphicx}
\usepackage{bm}
\usepackage{amssymb}
\usepackage{dcolumn}
\usepackage{subfigure}
\usepackage{threeparttable}
\usepackage{multirow}
\usepackage{mathrsfs}
\DeclareMathAlphabet{\mathscrbf}{OMS}{mdugm}{b}{n}
\usepackage{color}

\begin{document}
\newcommand{\vn}[1]{{\boldsymbol{#1}}}
\newcommand{\vht}[1]{{\boldsymbol{#1}}}
\newcommand{\matn}[1]{{\bf{#1}}}
\newcommand{\matnht}[1]{{\boldsymbol{#1}}}
\newcommand{\bege}{\begin{equation}}
\newcommand{\ee}{\end{equation}}
\newcommand{\bal}{\begin{aligned}}
\newcommand{\defbar}{\overline}
\newcommand{\SM}{\scriptstyle}
\newcommand{\gretke}{G_{\vn{k} }^{\rm R}(\mathcal{E})}
\newcommand{\gret}{G^{\rm R}}
\newcommand{\gadv}{G^{\rm A}}
\newcommand{\gmat}{G^{\rm M}}
\newcommand{\gles}{G^{<}}
\newcommand{\ghat}{\hat{G}}
\newcommand{\sigmahat}{\hat{\Sigma}}
\newcommand{\glesone}{G^{<,{\rm I}}}
\newcommand{\glestwo}{G^{<,{\rm II}}}
\newcommand{\sigmaret}{\Sigma^{\rm R}}
\newcommand{\sigmales}{\Sigma^{<}}
\newcommand{\sigmalesone}{\Sigma^{<,{\rm I}}}
\newcommand{\sigmalestwo}{\Sigma^{<,{\rm II}}}
\newcommand{\sigmalesthree}{\Sigma^{<,{\rm III}}}
\newcommand{\sigmaadv}{\Sigma^{A}}
\newcommand{\Bxc}{\Omega}
\newcommand{\mubo}{\mu_{\rm B}^{\phantom{B}}}
\newcommand{\rmd}{{\rm d}}
\newcommand{\rme}{{\rm e}}
\newcommand{\crea}[1]{{c_{#1}^{\dagger}}}
\newcommand{\annihi}[1]{{c_{#1}^{\phantom{\dagger}}}}
\newcommand{\intkspa}{\int\!\!\frac{\rmd^d k}{(2\pi)^d}}
\newcommand{\eal}{\end{aligned}}
\newcommand{\udot}{\overset{.}{u}}
\newcommand{\exponential}[1]{{\exp(#1)}}
\newcommand{\phandot}[1]{\overset{\phantom{.}}{#1}}
\newcommand{\phandag}{\phantom{\dagger}}
\newcommand{\Trace}{\text{Tr}}
\setcounter{secnumdepth}{2}
\title{
Dynamical and current-induced Dzyaloshinskii-Moriya interaction:
Role for damping, gyromagnetism, and current-induced torques in noncollinear magnets
}
\author{Frank Freimuth$^{1,2}$}
\email[Corresp.~author:~]{f.freimuth@fz-juelich.de}
\author{Stefan Bl\"ugel$^{1}$}
\author{Yuriy Mokrousov$^{1,2}$}
\affiliation{$^1$Peter Gr\"unberg Institut and Institute for Advanced Simulation,
Forschungszentrum J\"ulich and JARA, 52425 J\"ulich, Germany}
\affiliation{$^2$ Institute of Physics, Johannes Gutenberg University Mainz, 55099 Mainz, Germany
}
\begin{abstract}
Both applied electric
currents and magnetization dynamics
modify the Dzyaloshinskii-Moriya interaction (DMI), 
which we call 
current-induced DMI (CIDMI) and dynamical DMI (DDMI), respectively.
We report a theory of CIDMI and DDMI.
The inverse of CIDMI consists in
charge pumping by a time-dependent
gradient of magnetization $\partial^2 \vn{M}(\vn{r},t)/\partial
\vn{r}\partial t$,
while the inverse of DDMI 
describes the torque generated 
by $\partial^2 \vn{M}(\vn{r},t)/\partial\vn{r}\partial t$.
In noncollinear magnets CIDMI and DDMI depend on the
local magnetization direction. The resulting spatial gradients
correspond to torques that need to be included into the theories of 
Gilbert damping, gyromagnetism, and
current-induced torques (CITs)
in order to satisfy the Onsager reciprocity relations.
CIDMI is related to the modification of orbital magnetism
induced by magnetization dynamics, which we call dynamical orbital magnetism (DOM),
and spatial gradients of DOM contribute to charge pumping.
We present applications of this formalism to the CITs
and to the torque-torque correlation in textured Rashba ferromagnets.
\end{abstract}

\maketitle
\section{Introduction}
Since the Dzyaloshinskii-Moriya interaction (DMI) 
controls the magnetic texture of domain walls and
skyrmions, methods to tune this chiral interaction
by external means have exciting prospects.
Application of gate voltage~\cite{voltage_induction_dmi_AuFeMgO,
controlling_DMI_layer_stacking_capping_electric_field,
large_voltage_tuning_dmi_dynamic_control_skyrmion_chirality}
or laser pulses~\cite{ultrafast_modification_exchange_interaction} 
are promising ways to modify DMI.
Additionally, theory predicts that in 
magnetic trilayer structures the DMI in the top magnetic
layer can be controlled by the magnetization 
direction in the bottom magnetic
layer~\cite{sotcocucotrila}.
Moreover, methods to generate spin currents may be used to  
induce DMI, which is predicted
by the relations between the two~\cite{spicudmi,dmi_doppler_shift}.
Recent 
experiments show that also electric currents
modify DMI in metallic magnets,
which leads to large changes in the domain-wall velocity~\cite{cidmi_karnad,cidmi_hayashi}.
However, a rigorous theoretical formalism for the investigation
of current-induced DMI (CIDMI) in metallic
magnets has been lacking so far, and the development of
such a formalism is one goal of this paper.

Recently, a Berry phase theory of 
DMI~\cite{mothedmisot,spicudmi,phase_space_berry} 
has been developed,
which formally resembles the modern theory of orbital
magnetization~\cite{shi_quantum_theory_orbital_mag,om_insulators_mlwfs,om_crystals_mlwfs}. 
Orbital magnetism is modified
by the application of an electric field, which is known as 
the orbital magnetoelectric response~\cite{theory_OM_response}. 
In the case
of insulators it is straightforward to
derive the expressions for the magnetoelectric response
directly. 
However, in metals it is much easier to derive expressions instead for
the inverse of the magnetoelectric response, i.e., for the generation of 
electric currents by time-dependent magnetic 
fields~\cite{gyrotropic_magnetic_effect_magnetic_moment_fermi_surface}.
The inverse current-induced
DMI (ICIDMI) consists in charge pumping by time-dependent gradients
of magnetization. Due to the analogies between orbital magnetism
and the Berry phase theory of DMI one may expect that in metals
it is convenient to obtain expressions for ICIDMI, which can
then be used to describe the CIDMI by exploiting the reciprocity
between CIDMI and ICIDMI. We will show in this paper that this
is indeed the case.

In noncentrosymmetric ferromagnets spin-orbit interaction (SOI)
generates torques on the magnetization -- the so-called 
spin-orbit torques (SOTs) -- 
when an electric current
is applied~\cite{sot_review}.
The Berry phase theory of DMI~\cite{mothedmisot,spicudmi,phase_space_berry} 
establishes a relation to SOTs.
The formal analogies between orbital
magnetism and DMI have been shown to be a very
useful guiding principle in the development of the 
theory of SOTs driven by 
heat currents~\cite{itsot}.
In particular, it is fruitful to consider the DMI
coefficients as a spiralization, which is formally
analogous to magnetization.
In the theory of thermoelectric effects in magnetic systems the curl of magnetization describes
a bound current, which cannot be measured in transport experiments
and needs to be subtracted from the Kubo linear response in order
to obtain the measurable current~\cite{energy_magnetization_and_thermal_hall_effect,berry_anomalous_thermoelectric_xiao,thermoelectric_response_magnetic_field}. 
Similarly, in the theory of the thermal spin-orbit torque
spatial gradients of the DMI spiralization, which result from
the temperature gradient together with the temperature dependence
of DMI, need to be subtracted in order to obtain
the measurable torque and to satisfy a Mott-like relation~\cite{mothedmisot,itsot}.
In noncollinear magnets 
the question arises whether gradients of the spiralization that are due to
the magnetic texture correspond to torques like those from
thermal gradients.
We will show that indeed the spatial gradients of CIDMI
need to be included into the
theory of current-induced torques (CITs) in noncollinear magnets in order to satisfy the
Onsager reciprocity relations~\cite{spin_motive_forces_current_induced_torques}. 

When the system is driven out of equilibrium by 
magnetization dynamics rather than electric current 
one may expect DMI to be modified as well.
The inverse effect of this dynamical DMI (DDMI)
consists in the generation of torques by
time-dependent magnetization gradients.
In noncollinear magnets the DDMI spiralization varies
in space. We will show that the resulting gradient
corresponds to a torque that needs to be considered in the
theory of Gilbert damping and gyromagnetism in noncollinear magnets.

This paper is structured as follows.
In section~\ref{sec_formalism_notation_cit} we 
give an overview of CIT in noncollinear magnets
and introduce the notation.
In section~\ref{sec_formalism_response_current_tidegra}
we describe the formalism used to calculate the
response of electric current to time-dependent magnetization
gradients. In section~\ref{sec_formalism_cidmi} we show
that current-induced DMI (CIDMI) and electric current driven
by time-dependent magnetization gradients are reciprocal effects.
This allows us to obtain an expression for CIDMI based on the
formalism of section~\ref{sec_formalism_response_current_tidegra}.
In section~\ref{sec_direct_inverse_DDMI} we discuss that time-dependent
magnetization gradients generate additionally torques on the magnetization
and show that the inverse effect consists in the modification of DMI
by magnetization dynamics, which we call dynamical DMI (DDMI).
In section~\ref{sec_dom} we demonstrate that magnetization dynamics
induces orbital magnetism, which we call dynamical orbital
magnetism (DOM) and show that DOM is related to CIDMI.
In section~\ref{sec_cidmi_dom_in_cit} we explain how the
spatial gradients of CIDMI and DOM contribute to the direct
and to the inverse CIT,
respectively.
In section~\ref{sec_ddmi_torque_torque} we discuss how the
spatial gradients of DDMI contribute to the torque-torque
correlation.
In section~\ref{sec_formalism_cit} we complete the formalism used to
calculate the CIT
in noncollinear magnets by adding the chiral contribution of
the torque-velocity correlation.
In section~\ref{sec_formalism_icit} we finalize the theory of the inverse CIT
by adding the chiral contribution of the velocity-torque correlation.
In section~\ref{sec_formalism_damping_gyromag} we finish
the computational formalism of gyromagnetism and damping by
adding the chiral contribution of the torque-torque correlation
and the response of the torque to the time-dependent magnetization
gradients. 
In section~\ref{sec_symmetry} we discuss the symmetry
properties of the response to time-dependent magnetization gradients. 
In section~\ref{sec_results_cit}
we present the results for the chiral contributions to 
the direct and the inverse CIT in the Rashba model
and show that both the
perturbation by the time-dependent magnetization gradient 
and the spatial gradients of CIDMI and DOM need
to be included to ensure that they are reciprocal.
In section~\ref{sec_results_gyrodamp}
we present the results for the 
chiral contribution to the torque-torque correlation in
the Rashba model and show that both the
perturbation by the time-dependent magnetization gradient 
and the spatial gradients of DDMI need
to be included to ensure that it
satisfies the Onsager symmetry relations.
This paper ends with a summary in section~\ref{sec_summary}.
\section{Formalism}
\subsection{Direct and inverse current-induced torques
in noncollinear magnets}
\label{sec_formalism_notation_cit}
Even in collinear magnets the application
of an electric field $\vn{E}$ generates 
a torque $\vn{T}^{\rm CIT1}$ on the magnetization when inversion 
symmetry is broken~\cite{sot_review,ibcsoit}:
\bege\label{eq_t1}
T_{i}^{\rm CIT1}=\sum_{j}
t_{ij}(\hat{\vn{M}})
E_{j},
\ee
where $t_{ij}(\hat{\vn{M}})$
is the torkance tensor, which depends on the
magnetization direction $\hat{\vn{M}}$.
This torque is called spin-orbit torque (SOT),
but we denote it here CIT1, because it is one
contribution to the current-induced torques (CITs)
in noncollinear magnets.
Inversely, magnetization dynamics pumps a
charge current $\vn{J}^{\rm ICIT1}$ according to~\cite{invsot}
\bege\label{eq_j1}
J_{i}^{\rm ICIT1}=\sum_{j}
t_{ji}(-\hat{\vn{M}})
\hat{\vn{e}}_{j}\cdot
\left[
\hat{\vn{M}}\times
\frac{\partial \hat{\vn{M}}}{\partial t}
\right],
\ee
where
$\hat{\vn{e}}_{j}$ is a unit vector that points into the
$j$-th spatial direction.
Generally, $\vn{J}^{\rm ICIT1}$ can be explained by the
inverse spin-orbit torque~\cite{invsot}
or the magnonic charge pumping~\cite{charge_pumping_Ciccarelli}.
We denote it here by ICIT1, because it is one
contribution to the inverse CIT in noncollinear magnets.
In the special case of magnetic bilayers 
one important mechanism responsible for $\vn{J}^{\rm ICIT1}$
arises from the combination of spin pumping
and the inverse spin Hall effect~\cite{prl_mosendz_spin_pumping,prl_czeschka_spin_pumping}.

In noncollinear magnets there is a second contribution to the
CIT, which is proportional to the spatial derivatives of 
magnetization~\cite{RALPH20081190}:
\bege\label{eq_t2}
T_{i}^{\rm CIT2}=\sum_{jkl}
\chi^{\rm CIT2}_{ijkl}
E_{j}
\hat{\vn{e}}_{k}\cdot
\left[
\hat{\vn{M}}\times
\frac{\partial \hat{\vn{M}}}{\partial r_{l}}
\right].
\ee
The description of noncollinearity by the derivatives $\partial
\hat{\vn{M}}/\partial r_{l}$ is only 
applicable when the
magnetization
direction changes slowly in space like in magnetic skyrmions with large radius
and in wide magnetic domain walls.
In order to treat noncollinear
magnets such as Mn$_{3}$Sn~\cite{ahe_Mn3Sn},
where the magnetization direction varies strongly on the scale of
one unit cell, Eq.~\eqref{eq_t2} needs to be modified, which is beyond
the scope of the present paper.
The adiabatic and the non-adiabatic~\cite{nonadiabatic_stt_garate_macdonald} 
spin transfer torques
are two important contributions to $\chi^{\rm CIT2}_{ijkl}$, but
the interplay between broken inversion symmetry, SOI,
and noncollinearity can lead to a large number of 
additional mechanisms~\cite{current_induced_torques_rashba_ferromagnets,spin_motive_forces_current_induced_torques}.
Similarly, the current pumped by magnetization dynamics
contains a contribution that is proportional to the 
spatial derivatives of magnetization~\cite{spin_motive_forces_current_induced_torques,universal_electromotive_force_induced_by_dwm,generalized_faraday_barnes_maekawa}:
\bege\label{eq_j2}
J_{i}^{\rm ICIT2}=\sum_{jkl}\chi_{ijkl}^{\rm ICIT2}
\hat{\vn{e}}_{j}\cdot
\left[
\hat{\vn{M}}\times\frac{\partial \hat{\vn{M}}}{\partial t}
\right]
\hat{\vn{e}}_{k}\cdot
\left[
\hat{\vn{M}}\times
\frac{\partial \hat{\vn{M}}}{\partial r_{l}}
\right].
\ee
$T_{i}^{\rm CIT2}$ and $J_{i}^{\rm ICIT2}$
can be considered as \textit{chiral contributions} to the
CIT and to the ICIT, respectively, because they distinguish
between left- and right-handed spin spirals. 
Due to the reciprocity between direct and inverse 
CIT~\cite{invsot,spin_motive_forces_current_induced_torques}  
the coefficients $\chi^{\rm ICIT2}_{ijkl}$ 
and $\chi^{\rm CIT2}_{jikl}$ are related
according to
\bege
\label{eq_cit_icit_recipro}
\chi^{\rm ICIT2}_{ijkl}(\hat{\vn{M}})
=
\chi^{\rm CIT2}_{jikl}(-\hat{\vn{M}}).
\ee
\subsection{Response of electric current to time-dependent
magnetization gradients}
\label{sec_formalism_response_current_tidegra}
In order to compute $\vn{J}^{\rm ICIT2}$
based on the Kubo linear response formalism it is necessary to 
split it 
into two contributions,  $\vn{J}^{\rm ICIT2a}$ 
and $\vn{J}^{\rm ICIT2b}$.
While $\vn{J}^{\rm ICIT2a}$ is obtained as linear response 
to the perturbation by a \textit{time-dependent magnetization gradient}
in a collinear ferromagnet,
$\vn{J}^{\rm ICIT2b}$ is obtained as linear response 
to the perturbation by magnetization dynamics in a
noncollinear ferromagnet. 
Therefore, as will become clear below, $\vn{J}^{\rm ICIT2a}$
can be expressed by a correlation function of two operators,
because it describes the response of the current to a time-dependent magnetization
gradient: A time-dependent magnetization gradient is
a \textit{single} perturbation, which is described by a single
perturbing operator.
In contrast, $\vn{J}^{\rm ICIT2b}$ involves the correlation of three 
operators, because it describes the response of the current to magnetization dynamics
in the presence of perturbation by noncollinearity. These are
\textit{two} perturbations:
One perturbation by the magnetization dynamics, and a second
perturbation to describe the noncollinearity. In the
Kubo formalism the expressions for the response one the one hand to a
time-dependent magnetization gradient, which is described by a single
perturbing operator, and the response on the other hand to a time-dependent
magnetization
in the presence of a magnetization gradient, which is described
by two perturbing operators, are different. Therefore, we split
$\vn{J}^{\rm ICIT2}$ into these two contributions, which we call
$\vn{J}^{\rm ICIT2a}$ and $\vn{J}^{\rm ICIT2b}$.
In the remainder of this section we discuss the calculation of the 
contribution $\vn{J}^{\rm ICIT2a}$. 
The contribution $\vn{J}^{\rm ICIT2b}$ is discussed in 
section~\ref{sec_formalism_icit} below.

$\vn{J}^{\rm ICIT2a}$ is 
determined by the second derivative
of magnetization with respect to time and space variables
and can be written as
\bege\label{eq_icit2a}
J_{i}^{\rm ICIT2a}=\sum_{jk}\chi_{ijk}^{\rm ICIT2a}
\frac{\partial^2 \hat{M}_{j}}{\partial r_{k}\partial t}.
\ee
A nonzero second derivative $\frac{\partial^2 \hat{M}_{j}}{\partial r_{k}\partial t}$
is what we refer to as a 
\textit{time-dependent magnetization gradient}.
We will show below 
that in special cases $\frac{\partial^2 \hat{M}_{j}}{\partial r_{k}\partial t}$
can be expressed in terms of the 
products $\frac{\partial \hat{M}_{l}}{\partial r_{k}}  \frac{\partial\hat{M}_{l}}{\partial t}$,
which will allow us to rewrite $J_{i}^{\rm ICIT2a}$ in the
form of Eq.~\eqref{eq_j2} in the cases relevant for the chiral ICIT.
However, as will become clear below, 
Eq.~\eqref{eq_icit2a} is the most general expression
for the response to time-dependent magnetization gradients, and
it cannot generally be rewritten in the form of Eq.~\eqref{eq_j2}:
This
is only possible when it describes a contribution to the chiral ICIT.

$\vn{J}^{\rm ICIT2a}$ occurs 
in two different situations, which
need to be distinguished. 
In one case the magnetization gradient varies
in time like $\sin(\omega t)$ everywhere in
space. An example is
\bege\label{eq_example1}
\hat{\vn{M}}(\vn{r},t)=
\begin{pmatrix}
\eta \sin(\vn{q}\cdot \vn{r})\sin(\omega t)\\
0\\
1
\end{pmatrix},
\ee
where $\eta$ is the amplitude and the 
derivatives at $t=0$ and $\vn{r}=0$ are
\bege
\left.
\frac{\partial \hat{\vn{M}}(\vn{r},t)}
{\partial r_{i} }
\right |_{\vn{r}=t=0}=
\left.
\frac{\partial \hat{\vn{M}}(\vn{r},t)}
{ \partial t}
\right |_{\vn{r}=t=0}=
0
\ee
and
\bege\label{eq_second_deriv_example1}
\left.
\frac{\partial^2 \hat{\vn{M}}(\vn{r},t)}
{\partial r_{i} \partial t}
\right |_{\vn{r}=t=0}=
\begin{pmatrix}
\eta q_i \omega\\
0\\
0
\end{pmatrix}.
\ee

In the other case the magnetic texture varies like
a propagating wave, i.e., proportional 
to $\sin(\vn{q}\cdot\vn{r}-\omega t)$. An example
is given by
\bege\label{eq_example2}
\hat{\vn{M}}(\vn{r},t)=
\begin{pmatrix}
\eta \sin(\vn{q}\cdot \vn{r}-\omega t)\\
0\\
1-\frac{\eta^2}{2}\sin^2(\vn{q}\cdot \vn{r}-\omega t)
\end{pmatrix},
\ee
where the derivatives at $t=0$ and $\vn{r}=0$ are
\bege\label{eq_first_deriv_r}
\left.
\frac{\partial \hat{\vn{M}}(\vn{r},t)}
{\partial r_{i} }
\right |_{\vn{r}=t=0}=
\begin{pmatrix}
\eta q_i \\
0\\
0
\end{pmatrix},
\ee
\bege\label{eq_first_deriv_t}
\left.
\frac{\partial \hat{\vn{M}}(\vn{r},t)}
{ \partial t}
\right |_{\vn{r}=t=0}=
\begin{pmatrix}
-\eta  \omega\\
0\\
0
\end{pmatrix}
\ee
and
\bege\label{eq_second_deriv_example2}
\left.
\frac{\partial^2 \hat{\vn{M}}(\vn{r},t)}
{\partial r_{i} \partial t}
\right |_{\vn{r}=t=0}=
\begin{pmatrix}
0\\
0\\
\eta^2 q_i \omega\\
\end{pmatrix}.
\ee
In the latter example, Eq.~\eqref{eq_example2}, 
the second derivative, 
Eq.~\eqref{eq_second_deriv_example2}, is 
along the magnetization $\hat{\vn{M}}(\vn{r}=0,t=0)$,
while in the former example, Eq.~\eqref{eq_example1}, 
the second derivative, Eq.~\eqref{eq_second_deriv_example1},
is perpendicular to the magnetization 
when $\vn{r}=0$ and $t=0$.

We assume that the Hamiltonian is given by
\bege
\begin{aligned}
H(\vn{r},t)=
&-\frac{\hbar^2}{2m_e}\Delta+V(\vn{r})+
\mubo
\hat{\vn{M}}(\vn{r},t)\cdot \vn{\sigma}
\Bxc^{\rm xc}(\vn{r})+\\
&+
\frac{1}{2 e c^2}\mubo
\vn{\sigma}\cdot
\left[
\vn{\nabla}V(\vn{r})\times\vn{v}
\right],
\end{aligned}
\ee
where the first term describes the kinetic energy,
the second term is a scalar potential,
$\Bxc^{\rm xc}(\vn{r})$ in the third term is the exchange field,
and the last term describes the spin-orbit interaction.
Around $t=0$ and $\vn{r}=0$ we can
decompose the Hamiltonian as $H(\vn{r},t)=H_{0}+\delta H(\vn{r},t)$,
where $H_{0}$ is obtained from $H(\vn{r},t)$ 
by replacing $\hat{\vn{M}}(\vn{r},t)$ by $\hat{\vn{M}}(\vn{r}=0,t=0)$
and
\bege\label{eq_pert_type1}
\begin{aligned}
\delta H(\vn{r},t)&=
\frac{\partial H_0}{\partial \hat{M}_{x}}
\eta
\sin(\vn{q}\cdot\vn{r})
\sin(\omega t)
\\
&=
\mubo
\Bxc^{\rm xc}(\vn{r})
\sigma_{x}\eta
\sin(\vn{q}\cdot\vn{r})
\sin(\omega t)
\end{aligned}
\ee
in the case of the first example, Eq.~\eqref{eq_example1}.
In the case of the second example, Eq.~\eqref{eq_example2},
\bege\label{eq_pert_type2}
\begin{aligned}
\delta H(\vn{r},t)
\simeq
&\frac{\partial H}{\partial \hat{M}_{x} }\eta
\sin(\vn{q}\cdot\vn{r}-\omega t)\\
+&\frac{\partial H}{\partial \hat{M}_{z} }
\eta^2
\sin(\vn{q}\cdot\vn{r})
\sin(\omega t),
\end{aligned}
\ee
where for small $\vn{r}$ and $t$ only the second term
on the right-hand side contributes 
to $\frac{\partial^2 H(\vn{r},t)}{\partial r_k \partial t}$.
We consider here only the time-dependence of the exchange field direction
and
ignore the time-dependence of the exchange field magnitude $\Bxc^{\rm
  xc}(\vn{r})$
that is induced by the time-dependence of the exchange field direction.
While the variation of the exchange field magnitude drives currents
and torques as well, as shown in Ref.~\cite{longitex},  
the variation of the exchange field magnitude is a small response
and therefore
these secondary responses are suppressed in magnitude when
compared to the direct primary responses of the current and torque
to the variation in the exchange field direction. We will use the perturbations
Eq.~\eqref{eq_pert_type1} 
and Eq.~\eqref{eq_pert_type2} in order to compute the response of current and
torque
within the Kubo response formalism. An alternative approach for the 
calculation of the response to 
time-dependent fields is 
variational linear-response, which has been applied
to the spin susceptibility by Savrasov~\cite{PhysRevLett.81.2570}. 

The perturbation by the time-dependent gradient
can be written as
\bege\label{eq_pert_general_tidegra}
\delta H=
\frac{\partial H}{\partial  \hat {\vn{M}}}
\cdot
\frac{\partial^2 \hat {\vn{M}}}{\partial r_i \partial t}
\frac{\sin(q_{i}r_{i})}{q_i}
\frac{\sin(\omega t)}{\omega},
\ee
which turns into Eq.~\eqref{eq_pert_type1}
when Eq.~\eqref{eq_second_deriv_example1} is inserted.
When Eq.~\eqref{eq_second_deriv_example2} is inserted
it turns into the second term in Eq.~\eqref{eq_pert_type2}.

In Appendix~\ref{app_time_dependent_gradients}
we derive the linear response to perturbations of
the type of Eq.~\eqref{eq_pert_general_tidegra}
and show that the corresponding
coefficient $\chi_{ijk}^{\rm ICIT2a}$ 
in Eq.~\eqref{eq_icit2a}
can be expressed
as 
\bege\label{eq_chi_icit2a}
\begin{aligned}
\chi_{ijk}^{\rm ICIT2a}=
&
\frac{ie}{4\pi\hbar^2}
\intkspa\int
d \mathcal{E}
f(\mathcal{E})
{\rm Tr}\Bigl[\\
&
v_{i}
Rv_{k}RR 
\mathcal{O}_{j}
R
+
v_{i}
RRv_{k}R 
\mathcal{O}_{j}
R
+
\\
-&
v_{i}
RR 
\mathcal{O}_{j}
Rv_{k}R
-
v_{i}
Rv_{k}R 
\mathcal{O}_{j}
AA
\\
+&
v_{i}
R 
\mathcal{O}_{j}
Av_{k}AA
+
v_{i}
R
\mathcal{O}_{j}
AAv_{k}A\\
-&
v_{i}
Rv_{k}RR
\mathcal{O}_{j}
A
-
v_{i}
RRv_{k}R 
\mathcal{O}_{j}
A
\\
+&
v_{i}
RR 
\mathcal{O}_{j}
Av_{k}A
+
v_{i}
Av_{k}A 
\mathcal{O}_{j}
AA\\
-&
v_{i}
A 
\mathcal{O}_{j}
Av_{k}AA
-
v_{i}
A 
\mathcal{O}_{j}
AAv_{k}A
\Bigr],
\\
\end{aligned}
\ee
where $R=\gret_{\vn{k}}(\mathcal{E})$ 
and $A=\gadv_{\vn{k}}(\mathcal{E})$ 
are shorthands for the retarded 
and advanced Green's functions, respectively,
and $\mathcal{O}_{j}=\partial H/\partial \hat{M}_{j}$. 
$e>0$ is the positive elementary charge.

In the case of the
perturbation of the type Eq.~\eqref{eq_example1} 
the second derivative 
$\frac{\partial^2 \hat {\vn{M}}}{\partial r_i \partial t}$
is perpendicular to $\vn{M}$.
In this case it is convenient to 
rewrite Eq.~\eqref{eq_icit2a}
as
\bege\label{eq_chi_icidmi_ijk}
J_{i}^{\rm ICIT2a}=\sum_{jk}\chi_{ijk}^{\rm ICIDMI}
\hat{\vn{e}}_{j}\cdot
\left[
\hat{\vn{M}}\times
\frac{\partial^2 \hat{\vn{M}}}{\partial r_{k}\partial t}
\right]
,
\ee
where the coefficients $\chi_{ijk}^{\rm ICIDMI}$
are given by 
\bege\label{eq_icidmi}
\begin{aligned}
\chi_{ijk}^{\rm ICIDMI}=
&
\frac{ie}{4\pi\hbar^2}
\intkspa\int
d \mathcal{E}
f(\mathcal{E})
{\rm Tr}\Bigl[\\
&
v_{i}
Rv_{k}RR 
\mathcal{T}_{j}
R
+
v_{i}
RRv_{k}R 
\mathcal{T}_{j}
R
+
\\
-&
v_{i}
RR 
\mathcal{T}_{j}
Rv_{k}R
-
v_{i}
Rv_{k}R 
\mathcal{T}_{j}
AA
\\
+&
v_{i}
R 
\mathcal{T}_{j}
Av_{k}AA
+
v_{i}
R
\mathcal{T}_{j}
AAv_{k}A\\
-&
v_{i}
Rv_{k}RR
\mathcal{T}_{j}
A
-
v_{i}
RRv_{k}R 
\mathcal{T}_{j}
A
\\
+&
v_{i}
RR 
\mathcal{T}_{j}
Av_{k}A
+
v_{i}
Av_{k}A 
\mathcal{T}_{j}
AA\\
-&
v_{i}
A 
\mathcal{T}_{j}
Av_{k}AA
-
v_{i}
A 
\mathcal{T}_{j}
AAv_{k}A
\Bigr],
\\
\end{aligned}
\ee
and
\bege
\vn{\mathcal{T}}=\hat{\vn{M}}\times
\frac{\partial H}{\partial \hat{\vn{M}}}
\ee
is the torque operator.
In Sec.~\ref{sec_formalism_cidmi} we will explain that $\chi_{ijk}^{\rm ICIDMI}$
describes the inverse of current-induced DMI (ICIDMI).

In the case of the perturbation of the type 
of Eq.~\eqref{eq_example2} the second 
derivative $\frac{\partial^2 \hat{M}_{j}}{\partial r_{k}\partial t}$
may be rewritten as product of the first 
derivatives $\frac{\partial \hat{M}_{l}}{\partial t}$
and $\frac{\partial \hat{M}_{l}}{\partial r_{k}}$.
This may be seen as follows:
\bege
\begin{aligned}
&\frac{\partial H}
{
\partial \hat{\vn{M}}
}
\cdot\frac{\partial^2 \hat{\vn{M}}}{\partial r_{i}\partial t}
=\frac{\partial^2 H}{\partial t \partial r_{i}}=\\
&=
\frac{\partial}{\partial t}
\left[
\left(
\hat{\vn{M}}
\times
\frac{\partial H}
{
\partial \hat{\vn{M}}
}
\right)
\cdot
\left(\hat{\vn{M}}\times
\frac{
\partial \hat{\vn{M}}
}{
\partial r_{i}
}
\right)
\right]=\\
&=
\left[
\left(
\frac{\partial
\hat{\vn{M}}
}{\partial t}
\times
\frac{\partial H}
{
\partial \hat{\vn{M}}
}
\right)
\cdot
\left(\hat{\vn{M}}\times
\frac{
\partial \hat{\vn{M}}
}{
\partial r_{i}
}
\right)
\right]=\\
&=
\left[
\left(
\left(
\hat{\vn{M}}\times
\frac{\partial
\hat{\vn{M}}
}{\partial t}
\right)
\times
\hat{\vn{M}}
\right)
\times
\frac{\partial H}
{
\partial \hat{\vn{M}}
}
\right]
\cdot
\left[\hat{\vn{M}}\times
\frac{
\partial \hat{\vn{M}}
}{
\partial r_{i}
}
\right]=\\
&=-
\left[
\hat{\vn{M}}\times
\frac{
\partial \hat{\vn{M}}
}{
\partial t
}
\right]
\cdot
\left[
\hat{\vn{M}}\times
\frac{
\partial \hat{\vn{M}}
}{
\partial r_{i}
}
\right]
\left[
\hat{\vn{M}}\cdot
\frac{
\partial H
}{
\partial  \hat{\vn{M}}
}
\right]=\\
&=-
\frac{
\partial \hat{\vn{M}}
}{
\partial t
}
\cdot
\frac{
\partial \hat{\vn{M}}
}{
\partial r_{i}
}
\left[
\hat{\vn{M}}\cdot
\frac{
\partial H
}{
\partial  \hat{\vn{M}}
}
\right].\\
\end{aligned}
\ee
This expression is indeed satisfied by
Eq.~\eqref{eq_first_deriv_r}, 
Eq.~\eqref{eq_first_deriv_t}
and
Eq.~\eqref{eq_second_deriv_example2}:
\bege\label{eq_product_first_derivatives}
\frac{\partial \hat{\vn{M}}}
{\partial r_{i} }
\cdot
\frac{\partial \hat{\vn{M}}}
{ \partial t}
=-
\frac{\partial^2 \hat{\vn{M}}}
{\partial r_{i} \partial t}
\cdot
\hat{\vn{M}}
\ee
at $\vn{r}=0$, $t=0$.
Consequently,
Eq.~\eqref{eq_icit2a} can be 
rewritten as
\bege\label{eq_icit2a_convert_to_standard}
\begin{aligned}
J_{i}^{\rm ICIT2a}&=\sum_{jk}\chi_{ijk}^{\rm ICIT2a}
\frac{\partial^2 \hat{M}_{j}}{\partial r_{k}\partial t}=\\
&=-\sum_{jkl}\chi_{ijk}^{\rm ICIT2a}
\frac{\partial \hat{M}_{l}}{\partial r_{k}}
\frac{\partial \hat{M}_{l}}{\partial t}
[1-\delta_{jl}]\\
&=\sum_{jkl}\chi_{ijkl}^{\rm ICIT2a}
\hat{\vn{e}}_{j}\cdot
\left[
\hat{\vn{M}}\times\frac{\partial \hat{\vn{M}}}{\partial t}
\right]
\hat{\vn{e}}_{k}\cdot
\left[
\hat{\vn{M}}\times
\frac{\partial \hat{\vn{M}}}{\partial r_{l}}
\right],
\end{aligned}
\ee
where
\bege\label{eq_convert_icit2a}
\chi_{ijkl}^{\rm ICIT2a}=-\sum_{m}\chi_{iml}^{\rm ICIT2a}[1-\delta_{jm}]\delta_{jk}.
\ee
Thus, Eq.~\eqref{eq_icit2a_convert_to_standard}
and
Eq.~\eqref{eq_convert_icit2a}
can be used to express $J_{i}^{\rm ICIT2a}$
in the form of Eq.~\eqref{eq_j2}.

\subsection{Direct and inverse CIDMI}
\label{sec_formalism_cidmi}
Eq.~\eqref{eq_icidmi} describes the response 
of the electric current to time-dependent
magnetization gradients of the type
Eq.~\eqref{eq_pert_type1}. 
The reciprocal process
consists in the current-induced
modification of DMI. This can be shown
by expressing the DMI coefficients as~\cite{mothedmisot}
\bege
\begin{aligned}
D_{ij}
&=\frac{1}{V}
\sum_{n}f(\mathcal{E}_{\vn{k}n})
\int d^3 r
(\psi_{\vn{k}n}(\vn{r}))^{*}
\mathcal{D}_{ij}
\psi_{\vn{k}n}(\vn{r})\\
&=\frac{1}{V}
\sum_{n}f(\mathcal{E}_{\vn{k}n})
\int d^3 r
(\psi_{\vn{k}n}(\vn{r}))^{*}
\mathcal{T}_{i}(\vn{r})
r_{j}
\psi_{\vn{k}n}(\vn{r})
,
\end{aligned}
\ee
where we defined the DMI-operator $\mathcal{D}_{ij}=\mathcal{T}_{i}r_{j}$.
Using the Kubo formalism the current-induced modification of
DMI may be written as
\bege
D_{ij}^{\rm CIDMI}=\sum_{k}\chi^{\rm CIDMI}_{kij} E_{k}
\ee
with
\bege\label{eq_cidmi_dmiop}
\chi^{\rm CIDMI}_{kij}=
\frac{1}{V}\lim_{\omega\rightarrow 0}
\left[
\frac{e}{\hbar\omega}
{\rm Im}
\langle\langle \mathcal{D}_{ij} ; v_k \rangle\rangle^{\rm R}
(\hbar\omega)
\right],
\ee
where 
\bege
\langle\langle \mathcal{D}_{ij} ; v_k \rangle\rangle^{\rm R}
(\hbar\omega)=
-i\int\limits_{0}^{\infty}dte^{i \omega t}
\langle
[\mathcal{D}_{ij}(t),v_{k}(0)]_{-}
\rangle
\ee
is the Fourier transform of a retarded function
and $V$ is the volume of the unit cell.

Since the position operator $\vn{r}$ in the
DMI operator $\mathcal{D}_{ij}=\mathcal{T}_{i}r_{j}$
is not compatible with Bloch periodic boundary conditions, we do
not use 
Eq.~\eqref{eq_cidmi_dmiop} for numerical calculations of CIDMI.
However, it is convenient to use Eq.~\eqref{eq_cidmi_dmiop}
in order to demonstrate the reciprocity between direct
and inverse CIDMI.

Inverse CIDMI (ICIDMI) 
describes the electric current that responds to the perturbation
by a time-dependent magnetization gradient
according to
\bege
J_{k}^{\rm ICIDMI}=\sum_{ij}
\chi^{\rm ICIDMI}_{kij}
\hat{\vn{e}}_{i}\cdot
\left[
\hat{\vn{M}}\times
\frac{\partial^2 \hat{\vn{M}}}{\partial t \partial r_j}
\right]
.
\ee
The perturbation
by a time-dependent magnetization gradient
may be written as
\bege\label{eq_timedep_maggrad_dmiop}
\begin{aligned}
\delta H&=-\sum_{j}
\vn{m}\cdot
\frac{\partial^2 \hat{\vn{M}}}{\partial t \partial r_j}
r_{j}
\Omega^{\rm xc}(\vn{r})
\frac{\sin(\omega  t)}{\omega}=\\
&=\sum_{j}
\vn{\mathcal{T}}\cdot
\left[
\hat{\vn{M}}\times
\frac{\partial^2 \hat{\vn{M}}}{\partial t \partial r_j}
\right]r_{j}
\frac{\sin(\omega  t)}{\omega}\\
&=\sum_{ij}
\mathcal{D}_{ij}
\hat{\vn{e}}_{i}
\cdot
\left[
\hat{\vn{M}}\times
\frac{\partial^2 \hat{\vn{M}}}{\partial t \partial r_j}
\right]
\frac{\sin(\omega  t)}{\omega}.\\
\end{aligned}
\ee
Consequently, 
the coefficient $\chi^{\rm ICIDMI}_{kij}$
is given by
\bege
\chi^{\rm ICIDMI}_{kij}=
\frac{1}{V}
\lim_{\omega\rightarrow 0}
\left[
\frac{e}{\hbar\omega}
{\rm Im}
\langle\langle v_k ; \mathcal{D}_{ij} \rangle\rangle^{\rm R}(\hbar\omega)
\right].
\ee
Using
\bege
\langle\langle \mathcal{D}_{ij}; v_k \rangle\rangle^{\rm R}(\hbar\omega,\hat{\vn{M}})=-
\langle\langle v_k ; \mathcal{D}_{ij} \rangle\rangle^{\rm R}(\hbar\omega,-\hat{\vn{M}})
\ee
we find that
CIDMI and ICIDMI are related through the
equations
\bege\label{eq_cidmi_icdidmi_recipro}
\chi^{\rm CIDMI}_{kij}(\hat{\vn{M}})=-
\chi^{\rm ICIDMI}_{kij}(-\hat{\vn{M}}).
\ee
In order to calculate CIDMI we use Eq.~\eqref{eq_icidmi}
for ICIDMI and then use Eq.~\eqref{eq_cidmi_icdidmi_recipro}
to obtain CIDMI.

The perturbation Eq.~\eqref{eq_pert_type2}
describes a different kind of time-dependent 
magnetization gradient, for which the reciprocal
effect consists in the modification of the expectation value 
$\langle\vn{\sigma}\cdot\hat{\vn{M}} r_{j}\rangle$.
However, while the modification of $\langle\mathcal{T}_{i}r_{j}\rangle$ by an
applied current can be measured~\cite{cidmi_karnad,cidmi_hayashi} 
from the change of the
DMI constant $D_{ij}$, the 
quantity  $\langle\vn{\sigma}\cdot\hat{\vn{M}} r_{j}\rangle$
has not been considered so far in ferromagnets. In noncollinear
magnets the quantity $\langle\vn{\sigma} r_{j}\rangle$ can be used to define
spin toroidization~\cite{spin_toroidization}. 
Therefore, while the perturbation of 
the type of Eq.~\eqref{eq_pert_type1} is related to CIDMI and ICIDMI, which
are both accessible experimentally~\cite{cidmi_karnad,cidmi_hayashi}, 
in the case of the 
perturbation of 
the type of Eq.~\eqref{eq_pert_type2} we expect that only the
effect of driving current by the time-dependent 
magnetization gradient
is easily accessible experimentally, while its inverse
effect is difficult to measure.

\subsection{Direct and inverse dynamical DMI}
\label{sec_direct_inverse_DDMI}
Not only applied electric currents modify DMI, but also
magnetization dynamics, which we call
dynamical DMI (DDMI).
DDMI can be expressed as
\bege
D_{ij}^{\rm DDMI}=
\sum_{k}\chi_{kij}^{\rm DDMI}
\hat{\vn{e}}_{k}\cdot
\left[
\hat{\vn{M}}\times
\frac{\partial \hat{\vn{M}}}{\partial t}
\right].
\ee
In Sec.~\ref{sec_ddmi_torque_torque} we will show that the 
spatial gradient of DDMI contributes to damping
and gyromagnetism in noncollinear magnets.
The perturbation used to describe magnetization dynamics is
given by~\cite{invsot}
\bege\label{eq_pert_magn_dyn}
\delta H=
\frac{\sin(\omega t)}{\omega}
\left(
\hat{\vn{M}}
\times
\frac{\partial\hat{\vn{M}}}{\partial t}
\right)
\cdot
\vht{\mathcal{T}}.
\ee
Consequently, the coefficients $\chi_{kij}^{\rm DDMI}$
may be written as
\bege\label{eq_ddmi_dmiop}
\chi^{\rm DDMI}_{kij}=-
\frac{1}{V}
\lim_{\omega\rightarrow 0}
\left[
\frac{1}{\hbar\omega}
{\rm Im}
\langle\langle \mathcal{D}_{ij} ; \mathcal{T}_k \rangle\rangle^{\rm R}(\hbar\omega)
\right].
\ee

Since the position operator in $\mathcal{D}_{ij}$ is not
compatible with Bloch periodic boundary conditions, we
do not use Eq.~\eqref{eq_ddmi_dmiop}
for numerical calculations of DDMI,
but instead we obtain it from its inverse effect, which
consists in the generation of torques on the
magnetization due to time-dependent magnetization gradients.
These torques can be written as
\bege\label{eq_torque_iddmi}
T_{k}^{\rm IDDMI}=\sum_{ij}
\chi_{kij}^{\rm IDDMI}
\hat{\vn{e}}_{i}\cdot
\left[
\hat{\vn{M}}\times
\frac{\partial^2 \hat{\vn{M}}}{\partial t \partial r_j}
\right],
\ee
where the coefficients $\chi_{kij}^{\rm IDDMI}$
are
\bege\label{eq_iddmi_dmiop}
\chi^{\rm IDDMI}_{kij}=
\frac{1}{V}\lim_{\omega\rightarrow 0}
\left[\frac{1}{\hbar\omega}
{\rm Im}
\langle\langle \mathcal{T}_{k};\mathcal{D}_{ij} \rangle\rangle^{\rm R}
(\hbar\omega)
\right],
\ee
because the perturbation by the time-dependent gradient
can be expressed in terms of $\mathcal{D}_{ij}$
according to Eq.~\eqref{eq_timedep_maggrad_dmiop}
and because the torque on the magnetization is
described by $-\vn{\mathcal{T}}$~\cite{ibcsoit}.
Consequently,
DDMI and IDDMI are related by
\bege\label{eq_ddmi_iddmi_recipro}
\chi^{\rm DDMI}_{kij}(\hat{\vn{M}})=-
\chi^{\rm IDDMI}_{kij}(-\hat{\vn{M}}).
\ee

For numerical calculations of IDDMI we use
\bege\label{eq_iddmi}
\begin{aligned}
\chi_{ijk}^{\rm IDDMI}=
&
\frac{i}{4\pi\hbar^2}
\intkspa\int
d \mathcal{E}
f(\mathcal{E})
{\rm Tr}\Bigl[\\
&
\mathcal{T}_{i}
Rv_{k}RR
\mathcal{T}_{j}
R
+
\mathcal{T}_{i}
RRv_{k}R
\mathcal{T}_{j}
R
+
\\
-&
\mathcal{T}_{i}
RR
\mathcal{T}_{j}
Rv_{k}R
-
\mathcal{T}_{i}
Rv_{k}R
\mathcal{T}_{j}
AA
\\
+&
\mathcal{T}_{i}
R
\mathcal{T}_{j}
Av_{k}AA
+
\mathcal{T}_{i}
R
\mathcal{T}_{j}
AAv_{k}A\\
-&
\mathcal{T}_{i}
Rv_{k}RR
\mathcal{T}_{j}
A
-
\mathcal{T}_{i}
RRv_{k}R
\mathcal{T}_{j}
A
\\
+&
\mathcal{T}_{i}
RR
\mathcal{T}_{j}
Av_{k}A
+
\mathcal{T}_{i}
Av_{k}A
\mathcal{T}_{j}
AA\\
-&
\mathcal{T}_{i}
A
\mathcal{T}_{j}
Av_{k}AA
-
\mathcal{T}_{i}
A
\mathcal{T}_{j}
AAv_{k}A
\Bigr],
\\
\end{aligned}
\ee
which is derived in Appendix~\ref{app_time_dependent_gradients}.
In order to obtain DDMI we calculate IDDMI
from Eq.~\eqref{eq_iddmi} 
and use
the reciprocity relation Eq.~\eqref{eq_ddmi_iddmi_recipro}.

Eq.~\eqref{eq_torque_iddmi} is valid for
time-dependent magnetization gradients that
lead to perturbations of the type of Eq.~\eqref{eq_pert_type1}.
Perturbations of the second type,  Eq.~\eqref{eq_pert_type2},
will induce torques on the magnetization as well.
However, the inverse effect is difficult to measure
in that case, because it corresponds to the modification
of the expectation value $\langle\vn{\sigma}\cdot\hat{\vn{M}}r_j \rangle$
by magnetization dynamics. Therefore, while in the case
of Eq.~\eqref{eq_pert_type1} both direct and inverse
response are expected to be measurable and correspond
to IDDMI and DDMI, respectively, we expect that in the
case of Eq.~\eqref{eq_pert_type2} only the direct 
effect, i.e., the response of the torque to the perturbation,
is easy to observe.

\subsection{Dynamical orbital magnetism (DOM)}
\label{sec_dom}
Magnetization dynamics does not only induce DMI, but also 
orbital magnetism, which we call dynamical orbital magnetism (DOM).
It can be written as
\bege
M^{\rm DOM}_{ij}=\sum_{k}\chi_{kij}^{\rm DOM}
\hat{\vn{e}}_{k}\cdot
\left[
\hat{\vn{M}}
\times
\frac{\partial \hat{\vn{M}}}{\partial t}
\right],
\ee
where
we introduced the notation
\bege\label{eq_generalized_morb}
M_{ij}^{\rm DOM}=\frac{e}{V}
\langle v_{i} r_{j}\rangle^{\rm DOM},
\ee 
which defines a generalized orbital magnetization, such that 
\bege\label{eq_morb_from_mij}
M_{i}^{\rm DOM}=\frac{1}{2}\sum_{jk}\epsilon_{ijk}M_{jk}^{\rm DOM}
\ee
corresponds to the usual definition of orbital magnetization.
The coefficients $\chi_{kij}^{\rm DOM}$
are given by
\bege
\chi^{\rm DOM}_{kij}=-\frac{1}{V}
\lim_{\omega\rightarrow 0}
\left[
\frac{e}{\hbar\omega}
{\rm Im}
\langle\langle v_{i}r_{j} ; \mathcal{T}_{k} \rangle\rangle^{\rm R}(\hbar\omega)
\right],
\ee
because the perturbation by magnetization dynamics
is described by Eq.~\eqref{eq_pert_magn_dyn}.
We will discuss in Sec.~\ref{sec_cidmi_dom_in_cit} that the spatial
gradient of DOM contributes
to the inverse CIT.
Additionally, we will show below that DOM and CIDMI are related to each other.

In order to obtain an expression for DOM it is convenient to consider
the inverse effect, i.e., the generation of a torque by the application of
a time-dependent magnetic field $\vn{B}(t)$ that acts only on the orbital degrees
of freedom of the electrons and not on their spins. 
This torque can be written as
\bege
\mathcal{T}^{\rm IDOM}_{k}=\frac{1}{2}
\sum_{ijl}
\chi^{\rm IDOM}_{kij}
\epsilon_{ijl}\frac{\partial B_{l}}{\partial t},
\ee
where
\bege
\chi^{\rm IDOM}_{kij}=-\frac{1}{V}
\lim_{\omega\rightarrow 0}
\left[
\frac{e}{\hbar\omega}
{\rm Im}
\langle\langle \mathcal{T}_{k} ; v_{i}r_{j} \rangle\rangle^{\rm R}(\hbar\omega)
\right],
\ee
because the perturbation by the time-dependent magnetic field
is given by
\bege\label{eq_pert_tidegra_bfi}
\delta H=-\frac{e}{2}
\sum_{ijk}\epsilon_{ijk}
v_{i}r_{j}
\frac{\partial B_{k}}{\partial t}
\frac{\sin(\omega t)}{\omega}.
\ee
Therefore, the coefficients of DOM
and IDOM are related by
\bege
\chi_{kij}^{\rm DOM}(\hat{\vn{M}})
=-\chi_{kij}^{\rm IDOM}(-\hat{\vn{M}}).
\ee

In Appendix~\ref{app_time_dependent_gradients}
we show that the 
coefficient $\chi_{ijk}^{\rm IDOM}$ can be expressed
as
\bege\label{eq_chi_idom}
\begin{aligned}
\chi^{\rm IDOM}_{ijk}=
&
\frac{-ie}{4\pi\hbar^2}
\intkspa\int
d \mathcal{E}
f(\mathcal{E})
{\rm Tr}\Bigl[\\
&
\mathcal{T}_{i}
Rv_{k}RR
v_{j}
R
+
\mathcal{T}_{i}
RRv_{k}R
v_{j}
R
+
\\
-&
\mathcal{T}_{i}
RR
v_{j}
Rv_{k}R
-
\mathcal{T}_{i}
Rv_{k}R
v_{j}
AA
\\
+&
\mathcal{T}_{i}
R
v_{j}
Av_{k}AA
+
\mathcal{T}_{i}
R
v_{j}
AAv_{k}A\\
-&
\mathcal{T}_{i}
Rv_{k}RR
v_{j}
A
-
\mathcal{T}_{i}
RRv_{k}R
v_{j}
A
\\
+&
\mathcal{T}_{i}
RR
v_{j}
Av_{k}A
+
\mathcal{T}_{i}
Av_{k}A
v_{j}
AA\\
-&
\mathcal{T}_{i}
A
v_{j}
Av_{k}AA
-
\mathcal{T}_{i}
A
v_{j}
AAv_{k}A
\Bigr].
\\
\end{aligned}
\ee

Eq.~\eqref{eq_chi_idom} and 
Eq.~\eqref{eq_icidmi} differ only in 
the positions
of the two velocity operators and
the torque operator between the
Green functions. As a consequence,
IDOM are ICIDMI are related.
In Table~\ref{table_icidmi_idom_inplanex}
and Table~\ref{table_icidmi_idom_inplaney} 
we list the relations between IDOM and ICIDMI
for the Rashba model Eq.~\eqref{eq_rashba_model}.
We will explain in Sec.~\ref{sec_symmetry}
that IDOM and ICIDMI are zero in the Rashba model 
when the magnetization is along the $z$ direction.
Therefore, we discuss in Table~\ref{table_icidmi_idom_inplanex}
the case where the magnetization lies in the $xz$ plane,
and in Table~\ref{table_icidmi_idom_inplaney} we discuss 
the case where the magnetization lies in the $yz$ plane.
According to Table~\ref{table_icidmi_idom_inplanex}
and Table~\ref{table_icidmi_idom_inplaney}
the relation between IDOM and ICIDMI is of the form
$\chi^{\rm IDOM}_{ijk}=\pm\chi^{\rm ICIDMI}_{jik}$.
This is expected, because the index $i$ in $\chi^{\rm IDOM}_{ijk}$ is connected to
the torque operator, while the index $j$ in $\chi^{\rm ICIDMI}_{ijk}$
is connected to the torque operator.

\begin{threeparttable}
\caption{Relations between the inverse 
of the magnetization-dynamics induced orbital magnetism (IDOM)
and inverse current-induced DMI (ICIDMI) in the 2d Rashba
model
when $\hat{\vn{M}}$ lies in the $zx$ plane.
The components of $\chi_{ijk}^{\rm IDOM}$
(Eq.~\eqref{eq_chi_idom})
and $\chi_{ijk}^{\rm ICIDMI}$ 
(Eq.~\eqref{eq_icidmi})
are denoted by
the three indices $(ijk)$.
}
\label{table_icidmi_idom_inplanex}
\begin{ruledtabular}
\begin{tabular}{c|c}
ICIDMI
&IDOM\\
\hline
(211)
&(121)\\
\hline
(121)
&(211)\\
\hline
-(221)
&(221)\\
\hline
(112)
&(112)\\
\hline
-(212)
&(122)\\
\hline
-(122)
&(212)\\
\hline
(222)
&(222)\\
\hline
(231) & (321)\\
\hline
(132) & (312) \\
\hline
-(232)& (322) \\
\end{tabular}
\end{ruledtabular}
\end{threeparttable}

\begin{threeparttable}
\caption{Relations between IDOM
and ICIDMI in the 2d Rashba model when $\hat{\vn{M}}$ lies in the
$yz$ plane.}
\label{table_icidmi_idom_inplaney}
\begin{ruledtabular}
\begin{tabular}{c|c}
ICIDMI
&IDOM\\
\hline
(111)
&(111)\\
\hline
-(211)
&(121)\\
\hline
-(121)
&(211)\\
\hline
(221)
&(221)\\
\hline
-(112)
&(112)\\
\hline
(212)
&(122)\\
\hline
(122)
&(212)\\
\hline
-(131)
&(311)\\
\hline
(231)
&(321)\\
\hline
(132)
&(312)\\
\end{tabular}
\end{ruledtabular}
\end{threeparttable}

\subsection{Contributions from CIDMI
and DOM to direct and inverse CIT}
\label{sec_cidmi_dom_in_cit}
In electronic transport theory the continuity equation determines the
current only up to a curl field~\cite{PhysRevB.101.235430}. 
The curl of magnetization corresponds
to a bound current that cannot be measured in 
electron transport experiments such that
\bege\label{eq_transport_current_standard}
\vn{J}=\vn{J}^{\rm Kubo}-\vn{\nabla}\times\vn{M}
\ee
has to be used to extract the transport current $\vn{J}$ from the
current $\vn{J}^{\rm Kubo}$ obtained from the Kubo linear response.
The subtraction of $\vn{\nabla}\times\vn{M}$ has been
shown to be important when calculating
the thermoelectric response~\cite{PhysRevB.101.235430} and
the anomalous Nernst effect~\cite{berry_anomalous_thermoelectric_xiao}.
Similarly, in the theory of the thermal spin-orbit torque~\cite{mothedmisot,itsot}
the gradients of the DMI spiralization have to be subtracted in
order to obtain the measurable torque:
\bege\label{eq_measurable_torque_tsot}
T_{i}=T_{i}^{\rm Kubo}-\sum_{j}\frac{\partial}{\partial r_j}D_{ij},
\ee
where the spatial derivative of the spiralization arises from
its temperature dependence and
the temperature gradient.

Since CIDMI and DOM depend on the magnetization direction,
they vary spatially in noncollinear
magnets. Similar to Eq.~\eqref{eq_measurable_torque_tsot}
the spatial derivatives of the current-induced spiralization
need to be included into the theory of CIT.
Additionally, the
gradients of DOM correspond to currents 
that need to be considered in the theory of the inverse CIT, similar
to Eq.~\eqref{eq_transport_current_standard}.
In section IV we explicitly show that Onsager
reciprocity
is violated if spatial gradients of DOM and CIDMI are not subtracted
from the Kubo response expressions. By trial-and-error we find that the following
subtractions are necessary to obtain response currents and torques
that satisfy this fundamental symmetry:
\bege\label{eq_extract_icit}
\begin{aligned}
J^{\rm ICIT}_{i}
&=J_{i}^{\rm Kubo}-\frac{1}{2}\sum_{j}\frac{\partial \hat{\vn{M}}}{\partial r_j}\cdot
\frac{\partial M_{ij}^{\rm DOM}}{\partial \hat{\vn{M}}}
\end{aligned}
\ee
and
\bege\label{eq_extract_cit}
\begin{aligned}
T_{i}^{\rm CIT}
&=T_{i}^{\rm Kubo}-\frac{1}{2}\sum_{j}\frac{\partial  \hat{\vn{M}} }{\partial r_j}
\cdot
\frac{\partial D^{\rm CIDMI}_{ij}}{\partial \hat{\vn{M}}},
\end{aligned}
\ee
where $J_{i}^{\rm ICIT}$ is the current driven by magnetization
dynamics, and $T_{i}^{\rm CIT}$ is the current-induced 
torque. 

Interestingly, we find that
also the diagonal elements $M^{\rm DOM}_{ii}$ are nonzero.
This shows that the generalized definition Eq.~\eqref{eq_generalized_morb}
is necessary, because the diagonal elements $M^{\rm DOM}_{ii}$
do not contribute in the usual definition of $M_{i}$ according 
to Eq.~\eqref{eq_morb_from_mij}.
These differences in the symmetry properties between equilibrium and nonequilibrium
orbital magnetism can be traced back to symmetry breaking by the
perturbations.
Also in the case of the spiralization tensor $D_{ij}$ the
nonequilibrium correction $\delta D_{ij}$ has different
symmetry properties than the equilibrium part (see Sec.~\ref{sec_symmetry}). 

The contribution of DOM to $\chi^{\rm ICIT2}_{ijkl}$ 
can be written as
\bege\label{eq_icit2c}
\chi^{\rm ICIT2c}_{ijkl}=-\frac{1}{2}\hat{\vn{e}}_{k}\cdot
\left[
\hat{\vn{M}}\times
\frac{
\partial \chi^{\rm DOM}_{jil}
}
{
\partial \hat{\vn{M}}
}
\right]
\ee
and the contribution of CIDMI to $\chi^{\rm CIT2}_{ijkl}$ 
is given by
\bege\label{eq_cit2b}
\chi^{\rm CIT2b}_{ijkl}=-\frac{1}{2}
\hat{\vn{e}}_{k}\cdot
\left[
\hat{\vn{M}}\times
\frac{
\partial \chi^{\rm CIDMI}_{jil}
}
{
\partial \hat{\vn{M}}
}
\right].
\ee

\subsection{Contributions from DDMI to
gyromagnetism and damping}
\label{sec_ddmi_torque_torque}
The response to magnetization dynamics 
that is described by the torque-torque
correlation function
consists of torques that are related to damping 
and gyromagnetism~\cite{invsot}.
The chiral contribution to these torques
can be written
as
\bege\label{eq_t_tt2}
T_{i}^{\rm TT2}=
\sum_{jkl}\chi_{ijkl}^{\rm TT2}
\hat{\vn{e}}_{j}\cdot
\left[
\hat{\vn{M}}\times\frac{\partial \hat{\vn{M}}}{\partial t}
\right]
\hat{\vn{e}}_{k}\cdot
\left[
\hat{\vn{M}}\times
\frac{\partial \hat{\vn{M}}}{\partial r_{l}}
\right],
\ee
where the coefficients $\chi_{ijkl}^{\rm TT2}$
satisfy the Onsager relations
\bege
\label{eq_torquetorque_onsa}
\chi^{\rm TT2}_{ijkl}(\hat{\vn{M}})
=
\chi^{\rm TT2}_{jikl}(-\hat{\vn{M}}).
\ee

Since DDMI depends on the magnetization direction,
it varies spatially in noncollinear magnets
and the resulting gradients of DDMI contribute 
to the damping and to the gyromagnetic
ratio:
\bege\label{eq_extract_tt}
\begin{aligned}
T_{i}^{\rm TT}
&=T_{i}^{\rm Kubo}-\frac{1}{2}\sum_{j}\frac{\partial  \hat{\vn{M}} }{\partial r_j}
\cdot
\frac{\partial D^{\rm DDMI}_{ij}}{\partial \hat{\vn{M}}}
\end{aligned}.
\ee
The resulting contribution of the spatial derivatives of DDMI 
to the coefficient $\chi_{ijkl}^{\rm TT2}$ is
\bege\label{eq_chi_tt2c}
\chi^{\rm TT2c}_{ijkl}=-\frac{1}{2}
\hat{\vn{e}}_{k}\cdot
\left[
\hat{\vn{M}}\times
\frac{\partial\chi_{jil}^{\rm DDMI}(\hat{\vn{M}})}
{\partial \hat{\vn{M}}}
\right].
\ee

\subsection{Current-induced torque (CIT) in noncollinear magnets}
\label{sec_formalism_cit}
The chiral contribution to CIT
consists of the spatial gradient of CIDMI, $\chi^{\rm CIT2b}_{ijkl}$
in Eq.~\eqref{eq_cit2b},
and the Kubo linear response of the torque to the applied
electric field in a noncollinear magnet, $\chi_{ijkl}^{\rm CIT2a}$:
\bege
\chi_{ijkl}^{\rm CIT2}=\chi_{ijkl}^{\rm CIT2a}+\chi^{\rm CIT2b}_{ijkl}.
\ee
In order to determine $\chi_{ijkl}^{\rm CIT2a}$,
we assume that the magnetization direction $\hat{\vn{M}}(\vn{r})$ oscillates
spatially as described by
\bege\label{eq_magnetization_texture_q}
\hat{\vn{M}}(\vn{r})
=\begin{pmatrix}
\eta\sin(\vn{q}\cdot \vn{r})\\
0\\
1
\end{pmatrix}
\frac{1}{\sqrt{1+\eta^2\sin^2(\vn{q}\cdot\vn{r})}},
\ee
where we will take the limit $\vn{q}\rightarrow 0$ at the end of the
calculation.
Since the spatial derivative of the magnetization direction is
\bege
\frac{\partial \hat{\vn{M}}(\vn{r})}{\partial r_i}
=
\begin{pmatrix}
\eta q_i \cos(\vn{q}\cdot\vn{r})\\
0\\
0
\end{pmatrix}+\mathcal{O}(\eta^3),
\ee
the chiral contribution to the CIT  
oscillates spatially proportional to $\cos(\vn{q}\cdot\vn{r})$.
In order to extract this spatially oscillating contribution we multiply
with $\cos(\vn{q} \cdot\vn{r})$ and integrate over the unit cell.
The resulting expression for $\chi^{\rm CIT2a}_{ijkl}$ is
\bege\label{eq_def_torkance_ij11}
\begin{aligned}
\chi^{\rm CIT2a}_{ijkl}&=-\frac{2e}{V\eta}\lim_{\vn{q}\rightarrow 0}
\lim_{\omega\rightarrow 0}\Biggl[\\
&\!\!\!\!\!\!\!\!\frac{1}{q_{l}^{\phantom{1}}}
\!\!\int\!\!
\cos(q_{l}^{\phantom{l}} r_{l}^{\phantom{l}})
\frac{{\rm Im}
\langle\langle \mathcal{T}_{i}(\vn{r});v_{j}(\vn{r}') \rangle\rangle^{\rm R}
(\hbar\omega)
}{\hbar\omega}
d^3 r d^3 r'\Biggr],
\end{aligned}
\ee
where $V$ is the volume of the unit 
cell, and the retarded torque-velocity correlation 
function 
$\langle\langle \mathcal{T}_{i}(\vn{r});v_{j}(\vn{r}') \rangle\rangle^{\rm R} (\hbar\omega)$
needs to be evaluated in the presence of the perturbation
\bege\label{eq_noco_perturbation_cit}
\delta H=
\mathcal{T}_{k}
\eta\sin(\vn{q}\cdot\vn{r})
\ee
due to the noncollinearity (the index $k$ in
Eq.~\eqref{eq_noco_perturbation_cit}
needs to match the index $k$ in $\chi^{\rm CIT2a}_{ijkl}$).

In Appendix~\ref{sec_appendix} we show 
that $\chi^{\rm CIT2a}_{ijkl}$ 
can be written as
\bege\label{eq_chiral_torkance_green}
\chi^{\rm CIT2a}_{ijkl}
=-\frac{2e}{\hbar}
{\rm Im}
\left[
\mathscr{W}^{(\rm surf)}_{ijkl}
\!+\!
\mathscr{W}^{(\rm sea)}_{ijkl}
\right],
\ee
where
\bege\label{eq_fermi_surface_torkance}
\begin{aligned}
\mathscr{W}^{(\rm surf)}_{ijkl}&=
\frac{1}{4 \pi \hbar}
\intkspa
\int d\,\mathcal{E}
f'(\mathcal{E})
{\rm Tr}
\Biggl[\\
&\mathcal{T}_{i}
G^{\rm R}_{\vn{k}}(\mathcal{E})
v_{l}
G^{\rm R}_{\vn{k}}(\mathcal{E})
v_{j}
G^{\rm A}_{\vn{k}}(\mathcal{E})
\mathcal{T}_{k}
G^{\rm A}_{\vn{k}}(\mathcal{E})
\\
+&
\mathcal{T}_{i}
G^{\rm R}_{\vn{k}}(\mathcal{E})
v_{j}
G^{\rm A}_{\vn{k}}(\mathcal{E})
v_{l}
G^{\rm A}_{\vn{k}}(\mathcal{E})
\mathcal{T}_{k}
G^{\rm A}_{\vn{k}}(\mathcal{E})
\\
-&
\mathcal{T}_{i}
G^{\rm R}_{\vn{k}}(\mathcal{E})
v_{j}
G^{\rm A}_{\vn{k}}(\mathcal{E})
\mathcal{T}_{k}
G^{\rm A}_{\vn{k}}(\mathcal{E})
v_{l}
G^{\rm A}_{\vn{k}}(\mathcal{E})\\
+&
\frac{\hbar}{m_e}\delta_{jl}
\mathcal{T}_{i}
G^{\rm R}_{\vn{k}}(\mathcal{E})
G^{\rm A}_{\vn{k}}(\mathcal{E})
\mathcal{T}_{k}
G^{\rm A}_{\vn{k}}(\mathcal{E})
\Biggr]\\
\end{aligned}
\ee
is a Fermi surface term ($f'(\mathcal{E})=df(\mathcal{E})/d\mathcal{E}$)
and
\bege\label{eq_fermi_sea_torkance}
\begin{aligned}
&\mathscr{W}^{(\rm sea)}_{ijkl}=
\frac{1}{4\pi\hbar^2}
\intkspa
\int d \mathcal{E} f(\mathcal{E})
\Biggl[\\
&-
{\rm Tr}
\left[\mathcal{T}_{i}Rv_{l}RRv_{j}R\mathcal{T}_{k}R\right]
-
{\rm Tr}\left[
\mathcal{T}_{i}Rv_{l}R\mathcal{T}_{k}RRv_{j}R
\right]
\\
&-
{\rm Tr}\left[
\mathcal{T}_{i}RRv_{l}Rv_{j}R\mathcal{T}_{k}R
\right]
-
{\rm Tr}\left[
\mathcal{T}_{i}RRv_{j}Rv_{l}R\mathcal{T}_{k}R
\right]\\
&+
{\rm Tr}\left[
\mathcal{T}_{i}RRv_{j}R\mathcal{T}_{k}Rv_{l}R
\right]
+
{\rm Tr}\left[
\mathcal{T}_{i}RR\mathcal{T}_{k}Rv_{j}Rv_{l}R
\right]\\
&
+
{\rm Tr}\left[
\mathcal{T}_{i}RR\mathcal{T}_{k}Rv_{l}Rv_{j}R
\right]
-
{\rm Tr}\left[
\mathcal{T}_{i}RRv_{l}R\mathcal{T}_{k}Rv_{j}R
\right]
\\
&
-
{\rm Tr}\left[
\mathcal{T}_{i}Rv_{l}RR\mathcal{T}_{k}Rv_{j}R
\right]
+
{\rm Tr}\left[
\mathcal{T}_{i}R\mathcal{T}_{k}RRv_{j}Rv_{l}R
\right]
\\
&
+
{\rm Tr}\left[
\mathcal{T}_{i}R\mathcal{T}_{k}RRv_{l}Rv_{j}R
\right]
+
{\rm Tr}\left[
\mathcal{T}_{i}R\mathcal{T}_{k}Rv_{l}RRv_{j}R
\right]\\
&-
\frac{\hbar}{m_e}\delta_{jl}
{\rm Tr}\left[
\mathcal{T}_{i}RRR\mathcal{T}_{k}R
\right]
-
\frac{\hbar}{m_e}\delta_{jl}
{\rm Tr}\left[
\mathcal{T}_{i}A AA\mathcal{T}_{k}A
\right]\\
&-
\frac{\hbar}{m_e}\delta_{jl}
{\rm Tr}\left[
\mathcal{T}_{i}A
A\mathcal{T}_{k}AA
\right]
\Biggr]\\
\end{aligned}
\ee
is a Fermi sea term.

\subsection{Inverse CIT in noncollinear magnets}
\label{sec_formalism_icit}
The chiral 
contribution $\vn{J}^{\rm ICIT2}$ (see Eq.~\eqref{eq_j2})
to the charge pumping
is described by the coefficients
\bege\label{eq_icit2_complete}
\chi^{\rm ICIT2}_{ijkl}=\chi^{\rm ICIT2a}_{ijkl}
+\chi^{\rm ICIT2b}_{ijkl}
+\chi^{\rm ICIT2c}_{ijkl},
\ee
where $\chi^{\rm ICIT2a}_{ijkl}$ describes the response to the
time-dependent magnetization gradient 
(see Eq.~\eqref{eq_chi_icit2a}, 
Eq.~\eqref{eq_convert_icit2a}, 
and Eq.~\eqref{eq_icit2a_convert_to_standard})
and $\chi^{\rm ICIT2c}_{ijkl}$ results from the spatial
gradient of DOM (see Eq.~\eqref{eq_icit2c}).
$\chi^{\rm ICIT2b}_{ijkl}$ describes the response to the
perturbation by magnetization dynamics
in a noncollinear magnet. In order to
derive an expression for $\chi^{\rm ICIT2b}_{ijkl}$
we assume that the magnetization oscillates
spatially as described by
Eq.~\eqref{eq_magnetization_texture_q}.
Since the corresponding response oscillates spatially
proportional to $\cos(\vn{q}\cdot\vn{r})$,
we multiply by $\cos(\vn{q}\cdot\vn{r})$ and integrate 
over the unit cell in order to extract $\chi^{\rm ICIT2b}_{ijkl}$
from the retarded velocity-torque
correlation function $\langle\langle
v_{i}(\vn{r});\mathcal{T}_{j}(\vn{r}')
\rangle\rangle^{\rm R}(\hbar\omega)$,
which is evaluated in the presence of the
perturbation Eq.~\eqref{eq_noco_perturbation_cit}. 
We obtain
\bege\label{eq_chiral_invsot_correlation}
\begin{aligned}
\chi^{\rm ICIT2b}_{ijkl}&=\frac{2e}{V\eta}\lim_{\vn{q}\rightarrow 0}
\lim_{\omega\rightarrow 0}\Biggl[\\
&\frac{1}{q_{l}}
\int
\cos(q_{l}^{\phantom{l}} r_{l}^{\phantom{l}})
\frac{
{\rm Im}
\langle\langle
v_{i}(\vn{r});\mathcal{T}_{j}(\vn{r}')
\rangle\rangle^{\rm R}(\hbar\omega)
}{\hbar\omega}
d^3 r d^3 r'\Biggr],
\end{aligned}
\ee
which can be written as (see Appendix~\ref{sec_appendix})
 \bege\label{eq_chiral_invsot_green}
\chi^{\rm ICIT2b}_{ijkl}
=\frac{2e}{\hbar}
{\rm Im}
\left[
\mathscr{V}^{(\rm surf)}_{ijkl}
\!+\!
\mathscr{V}^{(\rm sea)}_{ijkl}
\right],
\ee
where
\bege\label{eq_fermi_surface_isot}
\begin{aligned}
\mathscr{V}^{(\rm surf)}_{ijkl}&=
\frac{1}{4 \pi \hbar}
\intkspa
\int d\,\mathcal{E}
f'(\mathcal{E})
{\rm Tr}
\Bigl[\\
&v_{i}
G^{\rm R}_{\vn{k}}(\mathcal{E})
v_{l}
G^{\rm R}_{\vn{k}}(\mathcal{E})
\mathcal{T}_{j}
G^{\rm A}_{\vn{k}}(\mathcal{E})
\mathcal{T}_{k}
G^{\rm A}_{\vn{k}}(\mathcal{E})
\\
+&
v_{i}
G^{\rm R}_{\vn{k}}(\mathcal{E})
\mathcal{T}_{j}
G^{\rm A}_{\vn{k}}(\mathcal{E})
v_{l}
G^{\rm A}_{\vn{k}}(\mathcal{E})
\mathcal{T}_{k}
G^{\rm A}_{\vn{k}}(\mathcal{E})
\\
-&
v_{i}
G^{\rm R}_{\vn{k}}(\mathcal{E})
\mathcal{T}_{j}
G^{\rm A}_{\vn{k}}(\mathcal{E})
\mathcal{T}_{k}
G^{\rm A}_{\vn{k}}(\mathcal{E})
v_{l}
G^{\rm A}_{\vn{k}}(\mathcal{E})\Bigr]\\
\end{aligned}
\ee
is the Fermi surface term
and
\bege\label{eq_fermi_sea_isot}
\begin{aligned}
&\mathscr{V}^{(\rm sea)}_{ijkl}=
\frac{1}{4\pi\hbar^2}
\intkspa
\int d \mathcal{E} f(\mathcal{E})
{\rm Tr}\Bigl[\\
&-{\rm Tr}
\left[
v_{i}Rv_{l}RR\mathcal{T}_{j}R\mathcal{T}_{k}R
\right]
-{\rm Tr}
\left[
v_{i}Rv_{l}R\mathcal{T}_{k}RR\mathcal{T}_{j}R
\right]
\\
&-{\rm Tr}
\left[
v_{i}RRv_{l}R\mathcal{T}_{j}R\mathcal{T}_{k}R
\right]
-{\rm Tr}
\left[
v_{i}RR\mathcal{T}_{j}Rv_{l}R\mathcal{T}_{k}R
\right]
\\
&+{\rm Tr}
\left[
v_{i}RR\mathcal{T}_{j}R\mathcal{T}_{k}Rv_{l}R
\right]
+{\rm Tr}
\left[
v_{i}RR\mathcal{T}_{k}R\mathcal{T}_{j}Rv_{l}R
\right]
\\
&
+{\rm Tr}
\left[
v_{i}RR\mathcal{T}_{k}Rv_{l}R\mathcal{T}_{j}R
\right]
-{\rm Tr}
\left[
v_{i}RRv_{l}R\mathcal{T}_{k}R\mathcal{T}_{j}R
\right]
\\
&
-{\rm Tr}
\left[
v_{i}Rv_{l}RR\mathcal{T}_{k}R\mathcal{T}_{j}R
\right]
+{\rm Tr}
\left[
v_{i}R\mathcal{T}_{k}RR\mathcal{T}_{j}Rv_{l}R
\right]
\\
&
+{\rm Tr}
\left[
v_{i}R\mathcal{T}_{k}RRv_{l}R\mathcal{T}_{j}R
\right]
+{\rm Tr}
\left[
v_{i}R\mathcal{T}_{k}Rv_{l}RR\mathcal{T}_{j}R
\right]
\Bigr]\\
\end{aligned}
\ee
is the Fermi sea term.

In Eq.~\eqref{eq_chiral_invsot_correlation} 
we use the Kubo formula to describe the response to
magnetization dynamics combined with perturbation theory to include the
effect of noncollinearity. Thereby, the time-dependent perturbation
and the perturbation by the magnetization gradient are 
separated
and perturbations of the form of Eq.~\eqref{eq_pert_type1} or
Eq.~\eqref{eq_pert_type2} are not automatically included.
For example the flat cycloidal spin spiral 
\bege\label{eq_cycloid_propagating}
\hat{\vn{M}}(x,t)=
\begin{pmatrix}
\sin(qx-\omega t)\\
0\\
\cos(qx-\omega t)
\end{pmatrix}
\ee
moving in $x$ direction
with speed $\omega/q$ 
and the
helical spin spiral
\bege\label{eq_helix_propagating}
\hat{\vn{M}}(y,t)=
\begin{pmatrix}
\sin(qy-\omega t)\\
0\\
\cos(qy-\omega t)
\end{pmatrix}
\ee
moving in $y$ direction
with speed $\omega/q$ 
behave like Eq.~\eqref{eq_example2}
when $t$ and $\vn{r}$ are small.
Thus, these moving domain walls
correspond to the perturbation of the type 
of Eq.~\eqref{eq_example2} 
and the resulting contribution $\vn{J}^{\rm ICIT2a}$
from the time-dependent  magnetization gradient is not described
by Eq.~\eqref{eq_chiral_invsot_correlation} and
needs to be added, which we do by adding $\chi_{ijkl}^{\rm ICIT2a}$
in Eq.~\eqref{eq_icit2_complete}.

\subsection{Damping and gyromagnetism in noncollinear magnets }
\label{sec_formalism_damping_gyromag}
The chiral contribution Eq.~\eqref{eq_t_tt2} 
to the
torque-torque correlation function
is expressed in terms of the coefficient
\bege
\chi_{ijkl}^{\rm TT}=
\chi_{ijkl}^{\rm TT2a}+
\chi_{ijkl}^{\rm TT2b}+
\chi_{ijkl}^{\rm TT2c},
\ee
where $\chi_{ijkl}^{\rm TT2c}$
results from the spatial gradient of DDMI 
(see Eq.~\eqref{eq_chi_tt2c}),
$\chi_{ijkl}^{\rm TT2a}$
describes the response to a time-dependent
magnetization gradient in a collinear magnet, and
$\chi_{ijkl}^{\rm TT2b}$ describes the response
to magnetization dynamics in a noncollinear magnet.

In order to derive an expression 
for $\chi_{ijkl}^{\rm TT2b}$ we assume that the
magnetization oscillates spatially according to
Eq.~\eqref{eq_magnetization_texture_q}.
We multiply the retarded torque-torque 
correlation 
function $\langle\langle
\mathcal{T}_{i}(\vn{r});\mathcal{T}_{j}(\vn{r}') \rangle\rangle^{\rm R}(\hbar\omega)$ 
with $\cos(q_l^{\phantom{l}} r_l^{\phantom{l}})$
and integrate over the unit cell in order to extract the
part of the response that varies spatially 
proportional to $\cos(q_l^{\phantom{l}} r_l^{\phantom{l}})$.
We obtain:
\bege\label{eq_def_lambda_ij11}
\begin{aligned}
\chi^{\rm TT2b}_{ijkl}&=\frac{2}{V\eta}\lim_{q_l^{\phantom{l}} \rightarrow 0}
\lim_{\omega\rightarrow 0}\Biggl[\\
&\frac{1}{q_l^{\phantom{l}} }
\int 
\cos(q_l^{\phantom{l}}  r_l^{\phantom{l}})
\frac{{\rm Im}
\langle\langle
\mathcal{T}_{i}(\vn{r});\mathcal{T}_{j}(\vn{r}')
\rangle\rangle^{\rm R}
(\hbar\omega)
}{\hbar\omega}
d^3 r d^3 r'\Biggr].
\end{aligned}
\ee

In Appendix~\ref{sec_appendix} we discuss how to 
evaluate Eq.~\eqref{eq_def_lambda_ij11}
in first order perturbation theory with respect to the 
perturbation Eq.~\eqref{eq_noco_perturbation_cit} and show
that $\chi^{\rm TT2b}_{ijkl}$
can be expressed as 
\bege\label{eq_chiral_gyromag_green}
\chi^{\rm TT2b}_{ijkl}=\frac{2}{\hbar}
{\rm Im}
\left[
\mathscr{X}^{(\rm surf)}_{ijkl}
\!+\!
\mathscr{X}^{(\rm sea)}_{ijkl}
\right],
\ee
where
\bege\label{eq_fermi_surface}
\begin{aligned}
\mathscr{X}^{(\rm surf)}_{ijkl}&=
\frac{1}{4 \pi \hbar}
\intkspa
\int d\,\mathcal{E}
f'(\mathcal{E})
{\rm Tr}
\Biggl[\\
&\mathcal{T}_{i}
G^{\rm R}_{\vn{k}}(\mathcal{E})
v_{l}
G^{\rm R}_{\vn{k}}(\mathcal{E})
\mathcal{T}_{j}
G^{\rm A}_{\vn{k}}(\mathcal{E})
\mathcal{T}_{k}
G^{\rm A}_{\vn{k}}(\mathcal{E})
\\
+&
\mathcal{T}_{i}
G^{\rm R}_{\vn{k}}(\mathcal{E})
\mathcal{T}_{j}
G^{\rm A}_{\vn{k}}(\mathcal{E})
v_{l}
G^{\rm A}_{\vn{k}}(\mathcal{E})
\mathcal{T}_{k}
G^{\rm A}_{\vn{k}}(\mathcal{E})
\\
-&
\mathcal{T}_{i}
G^{\rm R}_{\vn{k}}(\mathcal{E})
\mathcal{T}_{j}
G^{\rm A}_{\vn{k}}(\mathcal{E})
\mathcal{T}_{k}
G^{\rm A}_{\vn{k}}(\mathcal{E})
v_{l}
G^{\rm A}_{\vn{k}}(\mathcal{E})
\Biggr]\\
\end{aligned}
\ee
is a Fermi surface term
and
\bege\label{eq_fermi_sea}
\begin{aligned}
\mathscr{X}^{(\rm sea)}_{ijkl}&=
\frac{1}{4\pi\hbar^2}
\intkspa
\int d \mathcal{E} f(\mathcal{E})
{\rm Tr}\Biggl[\\
&-(\mathcal{T}_{i}Rv_{l}RR\mathcal{T}_{j}R\mathcal{T}_{k}R)
-(\mathcal{T}_{i}Rv_{l}R\mathcal{T}_{k}RR\mathcal{T}_{j}R)
\\
&-(\mathcal{T}_{i}RRv_{l}R\mathcal{T}_{j}R\mathcal{T}_{k}R)
-(\mathcal{T}_{i}RR\mathcal{T}_{j}Rv_{l}R\mathcal{T}_{k}R)\\
&+(\mathcal{T}_{i}RR\mathcal{T}_{j}R\mathcal{T}_{k}Rv_{l}R)
+(\mathcal{T}_{i}RR\mathcal{T}_{k}R\mathcal{T}_{j}Rv_{l}R)\\
&
+(\mathcal{T}_{i}RR\mathcal{T}_{k}Rv_{l}R\mathcal{T}_{j}R)
-(\mathcal{T}_{i}RRv_{l}R\mathcal{T}_{k}R\mathcal{T}_{j}R)
\\
&
-(\mathcal{T}_{i}Rv_{l}RR\mathcal{T}_{k}R\mathcal{T}_{j}R)
+(\mathcal{T}_{i}R\mathcal{T}_{k}RR\mathcal{T}_{j}Rv_{l}R)
\\
&
+(\mathcal{T}_{i}R\mathcal{T}_{k}RRv_{l}R\mathcal{T}_{j}R)
+(\mathcal{T}_{i}R\mathcal{T}_{k}Rv_{l}RR\mathcal{T}_{j}R)
\Biggr]\\
\end{aligned}
\ee
is a Fermi sea term. 

The contribution $\chi_{ijkl}^{\rm TT2a}$ from the time-dependent gradients
is given by
\bege
\chi_{ijkl}^{\rm TT2a}=-\sum_{m}\chi_{iml}^{\rm TT2a}[1-\delta_{jm}]\delta_{jk},
\ee
where
\bege\label{eq_chi_tt2a}
\begin{aligned}
&\chi^{\rm TT2a}_{iml}=
\frac{i}{4\pi\hbar^2}
\intkspa
\int
d \mathcal{E} 
f(\mathcal{E})
{\rm Tr}\Bigl[\\
&
\mathcal{T}_{i}
Rv_{l}RR \mathcal{O}_{m}R
+
\mathcal{T}_{i}
RRv_{l}R \mathcal{O}_{m}R
+
\\
-&
\mathcal{T}_{i}
RR \mathcal{O}_{m}Rv_{l}R
-
\mathcal{T}_{i}
Rv_{l}R \mathcal{O}_{m}AA
\\
+&
\mathcal{T}_{i}
R \mathcal{O}_{m}Av_{l}AA
+
\mathcal{T}_{i}
R
\mathcal{O}_{m}
AAv_{l}A\\
-&
\mathcal{T}_{i}
Rv_{l}RR
\mathcal{O}_{m}
A
-
\mathcal{T}_{i}
RRv_{l}R \mathcal{O}_{m}A
\\
+&
\mathcal{T}_{i}
RR \mathcal{O}_{m}Av_{l}A
+
\mathcal{T}_{i}
Av_{l}A \mathcal{O}_{m}AA\\
-&
\mathcal{T}_{i}
A \mathcal{O}_{m}Av_{l}AA
-
\mathcal{T}_{i}
A \mathcal{O}_{m}AAv_{l}A
\Bigr],
\\
\end{aligned}
\ee
with $\mathcal{O}_{m}=\partial H/\partial \hat{M}_{m}$ (see Appendix~\ref{app_time_dependent_gradients}).

\section{Symmetry properties}
\label{sec_symmetry}
In this section we discuss the symmetry properties of
CIDMI, DDMI and DOM in the case of the magnetic Rashba model
\bege\label{eq_rashba_model}
H_{\vn{k}}(\vn{r})=\frac{\hbar^2}{2m_e}k^2+
\alpha (\vn{k}\times\hat{\vn{e}}_{z})\cdot\vn{\sigma}+
\frac{\Delta V}{2}\vn{\sigma}\cdot\hat{\vn{M}}(\vn{r}).
\ee
Additionally, we discuss the symmetry properties of
the currents and torques induced by time-dependent
magnetization gradients of the form of Eq.~\eqref{eq_example2}.

We consider mirror reflection $\mathcal{M}_{xz}$
at the $xz$ plane, 
mirror reflection $\mathcal{M}_{yz}$
at the $yz$ plane, 
and c2 rotation around the
$z$ axis. When $\Delta V=0$ these operations
leave Eq.~\eqref{eq_rashba_model} invariant,
but when $\Delta V \ne 0$ they modify the
magnetization direction $\hat{\vn{M}}$ in Eq.~\eqref{eq_rashba_model},
as shown in Table~\ref{tab_symmetry_properties}. 
At the same time, these operations affect the torque $\vn{T}$
and the current $\vn{J}$ driven by the time-dependent
magnetization gradients (see Table~\ref{tab_symmetry_properties}). 
In Table~\ref{tab_symmetry_properties2} and Table~\ref{tab_symmetry_properties3}
we show how $\hat{\vn{M}}\times \partial\hat{\vn{M}}/\partial r_k$
is affected by the symmetry operations.

\begin{figure}
\includegraphics[width=\linewidth]{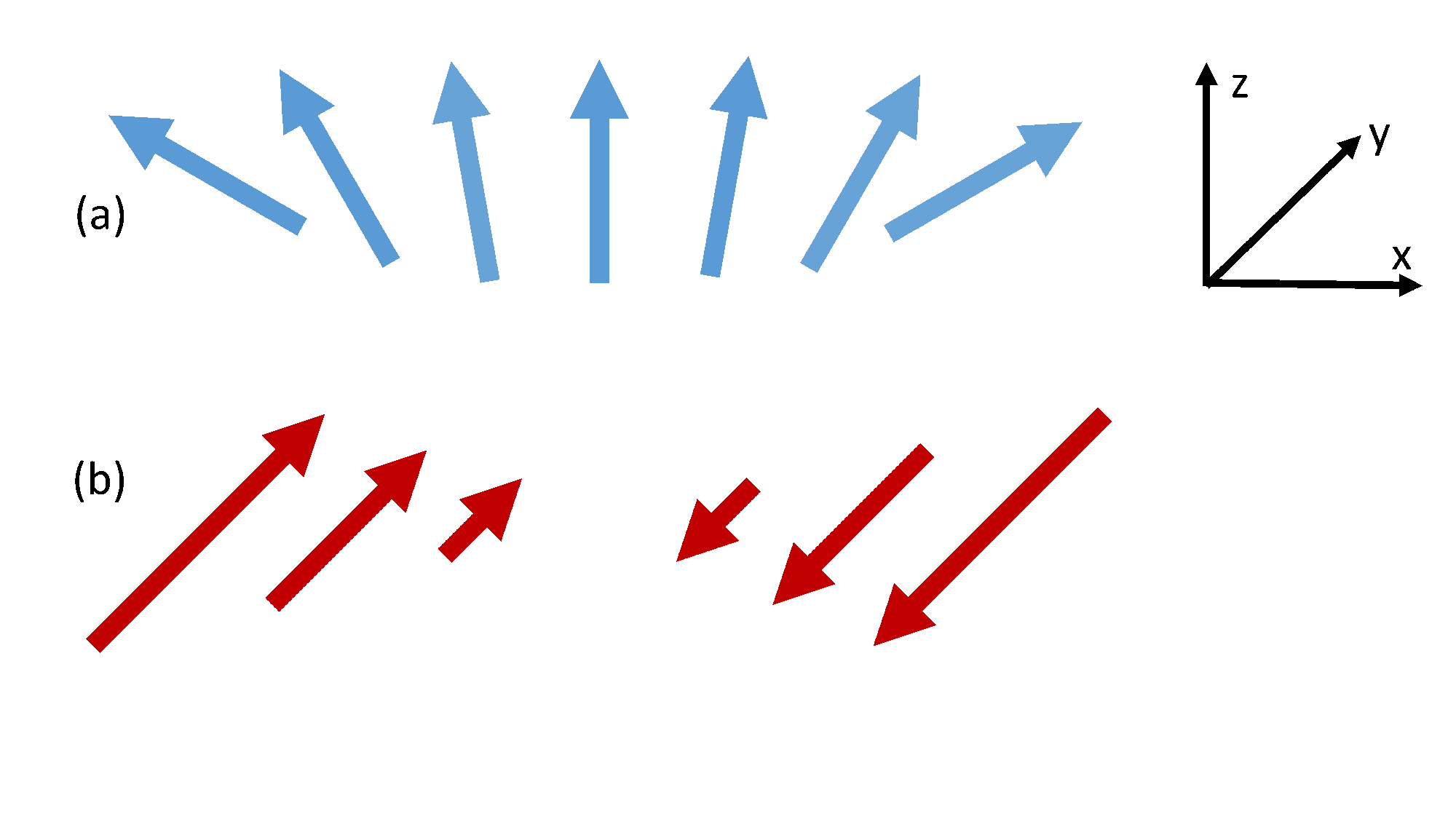}
\caption{\label{fig_cidmi_symmetry}
ICIDMI in a noncollinear magnet.
(a) Arrows illustrate the magnetization direction.
(b) Arrows illustrate the current $J_{y}$ induced by
a time-dependent magnetization gradient, which is
described by $\chi^{\rm ICIDMI}_{221}$.
When $\hat{\vn{M}}$ points in $z$ direction, $\chi^{\rm ICIDMI}_{221}$
and $J_{y}$ are zero. The sign of $\chi^{\rm ICIDMI}_{221}$
and of $J_{y}$ changes with the sign of $M_{x}$.
}
\end{figure}

A flat cycloidal spin spiral with spins rotating in the
$xz$ plane is mapped by a c2 rotation around
the $z$ axis onto the same spin spiral.
Similarly, a flat helical spin spiral with spins rotating
in the $yz$ plane is mapped by a c2 rotation around
the $z$ axis onto the same spin spiral.
Therefore, when $\hat{\vn{M}}$ points in $z$ direction,
a c2 rotation around the $z$ axis
does not change $\hat{\vn{M}}\times\partial\hat{\vn{M}}/\partial r_i$, but it
flips the in-plane current $\vn{J}$ and the
in-plane components of the torque, $T_{x}$ and $T_{y}$. 
Consequently, $\hat{\vn{M}}\times\partial^2\hat{\vn{M}}/\partial r_i\partial t$ 
does not induce currents or torques,  i.e., 
ICIDMI, CIDMI, IDDMI and DDMI are zero,
when $\hat{\vn{M}}$ points in $z$ direction. 
However, they become nonzero
when the magnetization has an in-plane component (see Fig.~\ref{fig_cidmi_symmetry}).

\begin{figure}
\includegraphics[width=\linewidth]{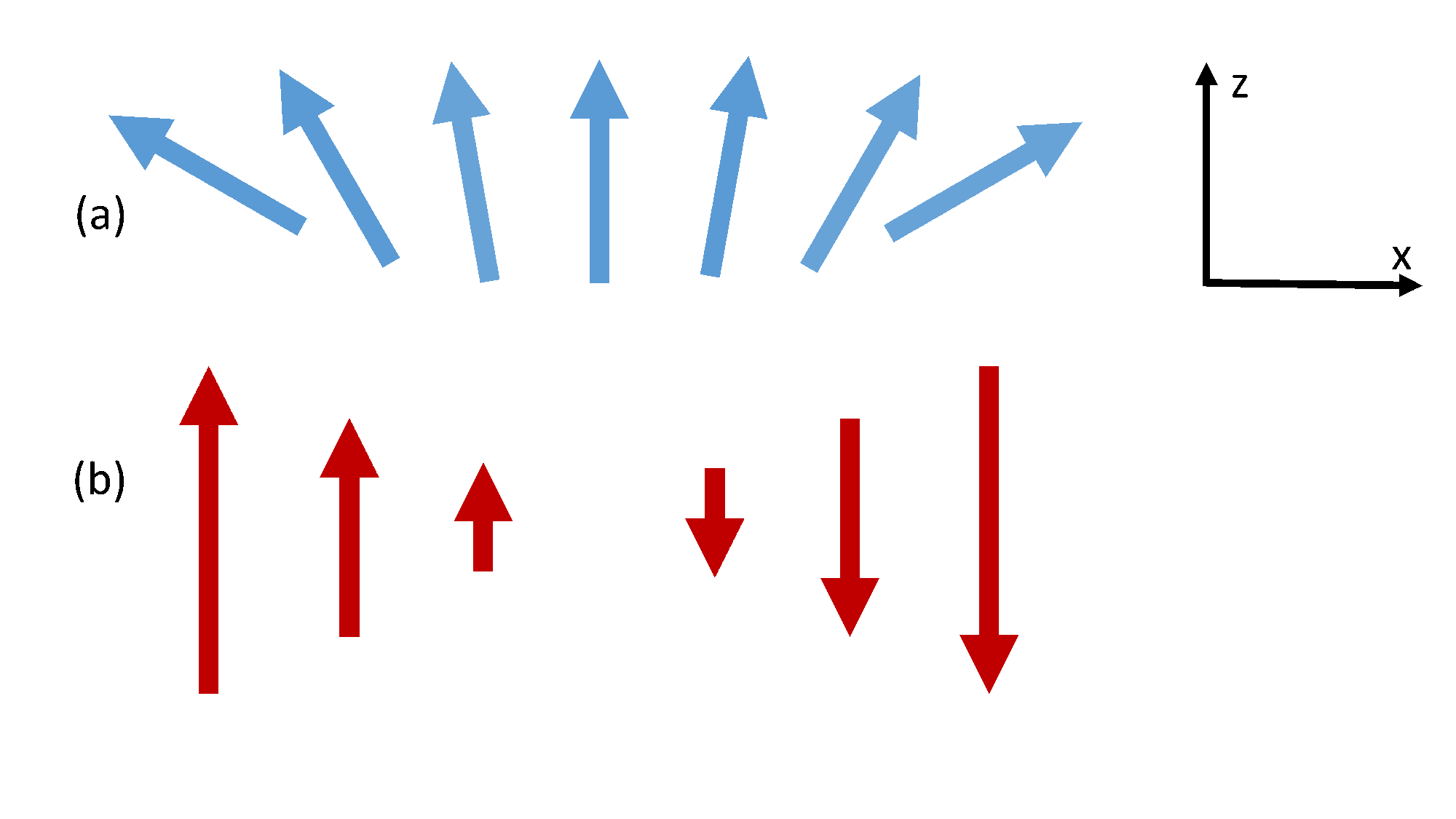}
\caption{\label{fig_dom_symmetry}
DOM in a noncollinear magnet.
(a) Arrows illustrate the magnetization direction.
(b) Arrows illustrate the orbital magnetization induced by 
magnetization dynamics (DOM).
When $\hat{\vn{M}}$ points in $z$ direction, DOM is zero.
The sign of DOM changes with the sign of $M_{x}$.
}
\end{figure}

Similarly, IDOM vanishes when the magnetization points in $z$ direction:
In that case Eq.~\eqref{eq_rashba_model}
is invariant under the c2 rotation. A time-dependent magnetic field along
$z$ direction is invariant under the c2 rotation as well. However,
$T_{x}$ and $T_{y}$ change sign under the c2 rotation. 
Consequently, symmetry forbids IDOM in this case.
However, when the magnetization has an in-plane
component, IDOM and DOM become nonzero (see Fig.~\ref{fig_dom_symmetry}).

That time-dependent magnetization gradients 
of the type of Eq.~\eqref{eq_example1}
do not induce in-plane currents and torques when $\hat{\vn{M}}$
points in $z$ direction can also be seen directly from Eq.~\eqref{eq_example1}:
The c2 rotation transforms $\vn{q}\rightarrow -\vn{q}$
and $M_{x}\rightarrow -M_{x}$. Since $\sin(\vn{q}\cdot\vn{r})$ is odd
in $\vn{r}$, Eq.~\eqref{eq_example1} is invariant under c2 rotation,
while the in-plane currents and torques induced by time-dependent
magnetization gradients change sign under c2 rotation.
In contrast, Eq.~\eqref{eq_example2} is not invariant under c2 rotation,
because $\sin(\vn{q}\cdot\vn{r}-\omega t)$ is not odd in $\vn{r}$
for $t>0$.
Consequently, time-dependent magnetization gradients 
of the type of Eq.~\eqref{eq_example2}
induce currents and torques also when $\hat{\vn{M}}$ points
locally into the $z$ direction. These currents and torques, 
which are described by Eq.~\eqref{eq_icit2a_convert_to_standard} 
and Eq.~\eqref{eq_chi_tt2a},
respectively, need to be added to the chiral ICIT and the chiral torque-torque
correlation. While CIDMI, DDMI, and DOM 
are zero when the 
magnetization points in $z$ direction, their gradients 
are not (see Fig.~\ref{fig_cidmi_symmetry} 
and Fig.~\ref{fig_dom_symmetry}).
Therefore, the gradients of CIDMI, DOM, and DDMI
contribute to CIT, to ICIT and to the torque-torque 
correlation, respectively, even when  $\hat{\vn{M}}$ points
locally into the $z$ direction. 

\begin{threeparttable}
\caption{Effect of mirror reflection $\mathcal{M}_{xz}$ at the $xz$ plane,
mirror reflection $\mathcal{M}_{yz}$ at the $yz$ plane,
and c2 rotation around the $z$ axis. The magnetization
$\vn{M}$ and the torque $\vn{T}$ transform like axial vectors,
while the current $\vn{J}$ transforms like a polar vector.}
\label{tab_symmetry_properties}
\begin{ruledtabular}
\begin{tabular}{c|c|c|c|c|c|c|c|c}
&$M_{x}$ &$M_{y}$ &$M_{z}$ &$J_{x}$ &$J_{y}$ &$T_{x}$ &$T_{y}$ &$T_{z}$\\
\hline
$\mathcal{M}_{xz}$
&$-M_{x}$ &$M_{y}$ &$-M_{z}$ &$J_{x}$ &$-J_{y}$ &$-T_{x}$ &$T_{y}$ &$-T_{z}$\\
\hline
$\mathcal{M}_{yz}$
&$M_{x}$ &$-M_{y}$ &$-M_{z}$ &$-J_{x}$ &$J_{y}$ &$T_{x}$ &$-T_{y}$ &$-T_{z}$\\
\hline
c2
& -$M_{x}$ &-$M_{y}$ &$M_{z}$ &-$J_{x}$ &$-J_{y}$ &$-T_{x}$ &$-T_{y}$ &$T_{z}$\\
\end{tabular}
\end{ruledtabular}
\end{threeparttable}

\begin{threeparttable}
\caption{Effect of symmetry operations on the magnetization
gradients. Magnetization gradients
are described by three indices $(ijk)$. The first index denotes the
magnetization direction at $\vn{r}=0$. The third index denotes
the direction along which the magnetization changes.
The second index denotes the direction 
of $\partial\hat{\vn{M}}/\partial r_k \delta r_k$.
The direction of $\hat{\vn{M}}\times \partial\hat{\vn{M}}/\partial r_k$
is specified by the number below the indices $(ijk)$.}
\label{tab_symmetry_properties2}
\begin{ruledtabular}
\begin{tabular}{c|c|c|c|c|c|c}
&(1,2,1) &(1,3,1) &(2,1,1) &(2,3,1) &(3,1,1) &(3,2,1)\\
&3 &-2 &-3 &1 &2 &-1\\
\hline
$\mathcal{M}_{xz}$
&(-1,2,1)&(-1,-3,1)&(2,-1,1)&(2,-3,1)&(-3,-1,1)&(-3,2,1)
\\
&-3 &-2 &3 &-1 &2 &1\\
\hline
$\mathcal{M}_{yz}$
&(1,2,1)&(1,3,1)&(-2,-1,1)&(-2,3,1)&(-3,-1,1)&(-3,2,1)
\\
&3  &-2  &-3  &-1  &2  &1  \\
\hline
c2
&(-1,2,1)&(-1,-3,1)&(-2,1,1)&(-2,-3,1)&(3,1,1)&(3,2,1)
\\
&-3  &-2  &3  &1  &2  &-1\\
\end{tabular}
\end{ruledtabular}
\end{threeparttable}

\begin{threeparttable}
\caption{Continuation of Table~\ref{tab_symmetry_properties2}}.
\label{tab_symmetry_properties3}
\begin{ruledtabular}
\begin{tabular}{c|c|c|c|c|c|c}
&(1,2,2) &(1,3,2) &(2,1,2) &(2,3,2) &(3,1,2) &(3,2,2)\\
&3 &-2 &-3 &1 &2 &-1\\
\hline
$\mathcal{M}_{xz}$
&(-1,-2,2)&(-1,3,2)&(2,1,2)&(2,3,2)&(-3,1,2)&(-3,-2,2)
\\
&3 &2 &-3 &1 &-2 &-1\\
\hline
$\mathcal{M}_{yz}$
&(1,-2,2)&(1,-3,2)&(-2,1,2)&(-2,-3,2)&(-3,1,2)&(-3,-2,2)
\\
&-3  &2  &3  &1  &-2  &-1  \\
\hline
c2
&(-1,2,2)&(-1,-3,2)&(-2,1,2)&(-2,-3,2)&(3,1,2)&(3,2,2)
\\
&-3  &-2  &3  &1  &2  &-1\\
\end{tabular}
\end{ruledtabular}
\end{threeparttable}

\subsection{Symmetry properties of ICIDMI and IDDMI}
\label{sec_symmetry_properties_cidmi_ddmi}
In the following we discuss how Table~\ref{tab_symmetry_properties},
Table~\ref{tab_symmetry_properties2}, and Table~\ref{tab_symmetry_properties3}
can be used to analyze the symmetry of ICIDMI and IDDMI.
According to Eq.~\eqref{eq_chi_icidmi_ijk}
the coefficient $\chi^{\rm ICIDMI}_{ijk}$ describes the
response of the current $J_{i}^{\rm ICIT2a}$
to the time-dependent magnetization gradient
$\hat{\vn{e}}_{j}\cdot[\hat{\vn{M}}\times\frac{\partial^{2} \hat{\vn{M}}}{\partial r_k \partial t}]$.
Since $\hat{\vn{M}}\times\frac{\partial^{2} \hat{\vn{M}}}{\partial r_k \partial t}=\frac{\partial}{\partial t}[\hat{\vn{M}}\times\frac{\partial \hat{\vn{M}}}{\partial r_k}]$ for
time-dependent magnetization gradients of the type Eq.~\eqref{eq_example1}
the symmetry properties of $\chi^{\rm ICIDMI}_{ijk}$ follow from 
the transformation behaviour
of $\hat{\vn{M}}\times\frac{\partial\hat{\vn{M}}}{\partial r_k}$ and $\vn{J}$
under symmetry operations.

We consider the case with
magnetization in $x$ direction. 
The 
component $\chi^{\rm ICIDMI}_{132}$
describes the current in $x$ direction
induced by the time-dependence of
a cycloidal magnetization gradient in $y$ 
direction (with spins rotating in the $xy$ plane).
$\mathcal{M}_{yz}$ flips 
both $\hat{\vn{M}}\times\frac{\partial\hat{\vn{M}}}{\partial y}$
and $J_{x}$, but it
preserves $\hat{\vn{M}}$.
$\mathcal{M}_{zx}$  
preserves $\hat{\vn{M}}\times\frac{\partial\hat{\vn{M}}}{\partial y}$
and $J_{x}$, but it flips $\hat{\vn{M}}$. 
A c2 rotation around the $z$ axis flips
$\hat{\vn{M}}\times\frac{\partial\hat{\vn{M}}}{\partial y}$,
$\hat{\vn{M}}$
and $J_{x}$. Consequently, $\chi^{\rm ICIDMI}_{132}(\hat{\vn{M}})$  
is allowed by symmetry and it is even in $\hat{\vn{M}}$.
The component $\chi^{\rm ICIDMI}_{122}$
describes the current in $x$ direction
induced by the time-dependence of
a helical magnetization gradient in $y$ 
direction (with spins rotating in the $xz$ plane). 
$\mathcal{M}_{yz}$ 
flips $\hat{\vn{M}}\times\frac{\partial\hat{\vn{M}}}{\partial y}$
and $J_{x}$, but it preserves $\hat{\vn{M}}$. 
$\mathcal{M}_{zx}$ 
flips $\hat{\vn{M}}\times\frac{\partial\hat{\vn{M}}}{\partial y}$
and $\hat{\vn{M}}$, 
but it preserves $J_{x}$. A c2 rotation
around the $z$ axis flips $J_{x}$ and $\hat{\vn{M}}$, but it
preserves $\hat{\vn{M}}\times\frac{\partial\hat{\vn{M}}}{\partial y}$.
Consequently, $\chi^{\rm ICIDMI}_{122}$ is allowed
by symmetry and it is odd in $\hat{\vn{M}}$.
The component
$\chi^{\rm ICIDMI}_{221}$ describes the current
in $y$ direction 
induced by the time-dependence of
a cycloidal magnetization gradient in $x$ direction (with spins
rotating in the $xz$ plane). 
$\mathcal{M}_{zx}$ preserves $\hat{\vn{M}}\times\frac{\partial\hat{\vn{M}}}{\partial x}$, 
but it flips $J_{y}$ and $\hat{\vn{M}}$.
$\mathcal{M}_{yz}$ preserves $\hat{\vn{M}}$, $J_{y}$, 
and $\hat{\vn{M}}\times\frac{\partial\hat{\vn{M}}}{\partial x}$.
The c2 rotation around the $z$ axis
preserves $\hat{\vn{M}}\times\frac{\partial\hat{\vn{M}}}{\partial x}$, 
but it flips $\hat{\vn{M}}$ and $J_{y}$.
Consequently, $\chi^{\rm ICIDMI}_{221}$ is allowed by symmetry 
and it is odd in $\hat{\vn{M}}$.
The component $\chi^{\rm ICIDMI}_{231}$ 
describes the current in $y$ direction 
induced by the time-dependence of
a cycloidal magnetization gradient in $x$ direction
(with spins rotating in the $xy$ plane).
$\mathcal{M}_{zx}$ 
flips $\hat{\vn{M}}\times\frac{\partial\hat{\vn{M}}}{\partial x}$, 
$\hat{\vn{M}}$, and $J_{y}$.
$\mathcal{M}_{yz}$ 
preserves $\hat{\vn{M}}\times\frac{\partial\hat{\vn{M}}}{\partial x}$,
$\hat{\vn{M}}$ and $J_{y}$.
The c2 rotation around the $z$ axis
flips $\hat{\vn{M}}\times\frac{\partial\hat{\vn{M}}}{\partial x}$,
$J_{y}$, and $\hat{\vn{M}}$.
Consequently, $\chi^{\rm ICIDMI}_{231}$
is allowed by symmetry and it is even in $\hat{\vn{M}}$.

These properties are
summarized in Table~\ref{table_cidmi_inplanex}.
Due to the relations between CIDMI and DOM 
(see Table~\ref{table_icidmi_idom_inplanex}
and Table~\ref{table_icidmi_idom_inplaney}),
they can be used for DOM as well.
When the magnetization lies at a general angle in the
$xz$ plane or in the $yz$ plane several additional components
of CIDMI and DOM are nonzero (see 
Table~\ref{table_icidmi_idom_inplanex} and
Table~\ref{table_icidmi_idom_inplaney}, respectively).

\begin{threeparttable}
\caption{Allowed components of
$\chi^{\rm ICIDMI}_{ijk}$ when $\hat{\vn{M}}$
points in $x$ direction. + components are even in $\hat{\vn{M}}$,
while - components are odd in $\hat{\vn{M}}$.}
\label{table_cidmi_inplanex}
\begin{ruledtabular}
\begin{tabular}{c|c|c|c}
132 &122 &221 &231 \\
\hline
+ &- &- &+ \\
\end{tabular}
\end{ruledtabular}
\end{threeparttable}

Similarly, one can analyze the symmetry of DDMI.
Table~\ref{table_ddmi_inplanex}
lists the components of 
DDMI, $\chi_{ijk}^{\rm DDMI}$,
which are allowed by symmetry 
when $\hat{\vn{M}}$
points in $x$ direction.

\begin{threeparttable}
\caption{Allowed components of
$\chi^{\rm DDMI}_{ijk}$ when $\hat{\vn{M}}$
points in $x$ direction.
+ components are even in $\hat{\vn{M}}$,
while - components are odd in $\hat{\vn{M}}$.
}
\label{table_ddmi_inplanex}
\begin{ruledtabular}
\begin{tabular}{c|c|c|c}
222 &232 &322 &332 \\
\hline
- &+ &+ &- \\
\end{tabular}
\end{ruledtabular}
\end{threeparttable}

\subsection{Response to time-dependent magnetization
gradients of the second type (Eq.~\eqref{eq_example2})}
According to Eq.~\eqref{eq_second_deriv_example2}
the time-dependent magnetization gradient is along the magnetization.
Therefore, in contrast to the discussion in section~\ref{sec_symmetry_properties_cidmi_ddmi}
we cannot 
use $\hat{\vn{M}}\times\frac{\partial^2\hat{\vn{M}}}{\partial r_k\partial t}$
in the symmetry analysis.
Eq.~\eqref{eq_icit2a_convert_to_standard}
and Eq.~\eqref{eq_convert_icit2a}
show that $\chi_{ijjl}^{\rm ICIT2a}$
describes the response of $J_{i}^{\rm ICIT2a}$
to $\hat{\vn{e}}_{j}\cdot
\left[\hat{\vn{M}}\times\frac{\partial \hat{\vn{M}}}{\partial t}
\right]
\hat{\vn{e}}_{j}\cdot
\left[
\hat{\vn{M}}\times\frac{\partial \hat{\vn{M}}}{\partial r_l}
\right]
$
while $\chi_{ijkl}^{\rm ICIT2a}=0$ for $j\ne k$.
According to Eq.~\eqref{eq_product_first_derivatives}
the symmetry properties of
$\left[\hat{\vn{M}}\times\frac{\partial \hat{\vn{M}}}{\partial t}
\right]\cdot
\left[
\hat{\vn{M}}\times\frac{\partial \hat{\vn{M}}}{\partial r_l}
\right]$
agree to the symmetry properties 
of $\hat{\vn{M}}\cdot\frac{\partial^2\hat{\vn{M}}}{\partial r_l\partial t}$.
Therefore, in order to understand the symmetry properties of
$\chi_{ijjl}^{\rm ICIT2a}$
we consider the transformation of $\vn{J}$ 
and $\hat{\vn{M}}\cdot\frac{\partial^2\hat{\vn{M}}}{\partial r_l\partial t}$
under symmetry operations.

We consider the case where $\hat{\vn{M}}$ points in $z$ direction.
$\chi_{1jj1}^{\rm ICIT2a}$ describes the current driven in $x$ direction,
when the magnetization varies in $x$ direction.
$\mathcal{M}_{xz}$ flips $\hat{\vn{M}}$, but preserves $J_{x}$
and $\hat{\vn{M}}\cdot\partial^2\hat{\vn{M}}/(\partial x \partial t)$.
$\mathcal{M}_{yz}$ flips $\hat{\vn{M}}$, $J_{x}$, 
and $\hat{\vn{M}}\cdot\partial^2\hat{\vn{M}}/(\partial x \partial t)$.
c2 rotation flips $\hat{\vn{M}}\cdot\partial^2\hat{\vn{M}}/(\partial x \partial t)$ 
and $J_{x}$, but preserves $\hat{\vn{M}}$.
Consequently, $\chi_{1jj1}^{\rm ICIT2a}$ is allowed by symmetry and 
it is even in $\hat{\vn{M}}$.

$\chi_{2jj1}^{\rm ICIT2a}$
describes the current flowing in $y$ direction,
when magnetization varies in $x$ direction.
$\mathcal{M}_{xz}$ flips $\hat{\vn{M}}$ and $J_{y}$, but preserves
$\hat{\vn{M}}\cdot\partial^2\hat{\vn{M}}/(\partial x \partial t)$.
$\mathcal{M}_{yz}$ flips $\hat{\vn{M}}$, 
and $\hat{\vn{M}}\cdot\partial^2\hat{\vn{M}}/(\partial x \partial t)$,
but preserves $J_{y}$.
c2 rotation flips $\hat{\vn{M}}\cdot\partial^2\hat{\vn{M}}/(\partial x \partial t)$ 
and $J_{y}$, but preserves $\hat{\vn{M}}$.
Consequently, $\chi_{2jj1}^{\rm ICIT2a}$ is allowed by symmetry and 
it is odd in $\hat{\vn{M}}$.

Similarly, one can show that  $\chi_{1jj2}^{\rm ICIT2a}$ is odd in $\hat{\vn{M}}$
and that $\chi_{2jj2}^{\rm ICIT2a}$ is even in $\hat{\vn{M}}$.

Analogously, one can investigate the symmetry properties
of $\chi_{ijjl}^{\rm TT2a}$. We find that
$\chi_{1jj1}^{\rm TT2a}$ and 
$\chi_{2jj2}^{\rm TT2a}$
are odd in $\hat{\vn{M}}$,
while $\chi_{2jj1}^{\rm TT2a}$
and $\chi_{1jj2}^{\rm TT2a}$ are even in $\hat{\vn{M}}$.

\section{Results}
\label{sec_results}
In the following sections we discuss the results for the 
direct and inverse chiral CIT 
and for the chiral torque-torque correlation 
in the two-dimensional (2d) Rashba model
Eq.~\eqref{eq_rashba_model},
and in the one-dimensional (1d) Rashba 
model~\cite{chigyromag}
\bege\label{eq_rashba_model_onedim}
H_{k_{x}}(x)=\frac{\hbar^2}{2m_e}k^2_{x}-
\alpha k_{x}\sigma_{y}+
\frac{\Delta V}{2}\vn{\sigma}\cdot\hat{\vn{M}}(x).
\ee
Additionally, we discuss the contributions of the
time-dependent magnetization gradients, and of 
DDMI, DOM and CIDMI to these effects.

While vertex corrections
to the chiral CIT and to 
the chiral torque-torque correlation
are important in the
Rashba model~\cite{chigyromag}, 
the purpose of this work is to show the
importance of the contributions
from time-dependent magnetization
gradients, DDMI, DOM and CIDMI. 
We therefore consider only the intrinsic contributions here,
i.e., we set
\bege
\gret_{\vn{k}}(\mathcal{E})=\hbar
\left[
\mathcal{E}-H_{\vn{k}}
+i\Gamma
\right]^{-1},
\ee
where $\Gamma$ is a constant
broadening, and we leave the study of vertex corrections
for future work. 

The results shown in the following sections
are obtained for
the model parameters $\Delta V=1$eV, 
$\alpha=$2eV\AA, and $\Gamma=0.1{\rm Ry}=1.361$eV, 
when the magnetization points in $z$ direction,
i.e., $\hat{\vn{M}}=\hat{\vn{e}}_{z}$.
The unit of $\chi_{ijkl}^{\rm CIT2}$
is charge times length in the 1d case
and charge in the 2d case. 
Therefore, in the 1d case we discuss the chiral torkance 
in units of $e a_{0}$, where $a_{0}$ is Bohr's radius.
In the 2d case we discuss the
chiral torkance in units of $e$.
The unit of $\chi_{ijkl}^{\rm TT2}$
is angular momentum in the 1d case
and angular momentum per length in the 2d case.
Therefore, we discuss $\chi_{ijkl}^{\rm TT2}$ in 
units of $\hbar$ in the 1d case, and in
units of $\hbar/a_{0}$ in the 2d case.



\subsection{Direct and inverse chiral CIT}
\label{sec_results_cit}

\begin{figure}
\includegraphics[width=\linewidth]{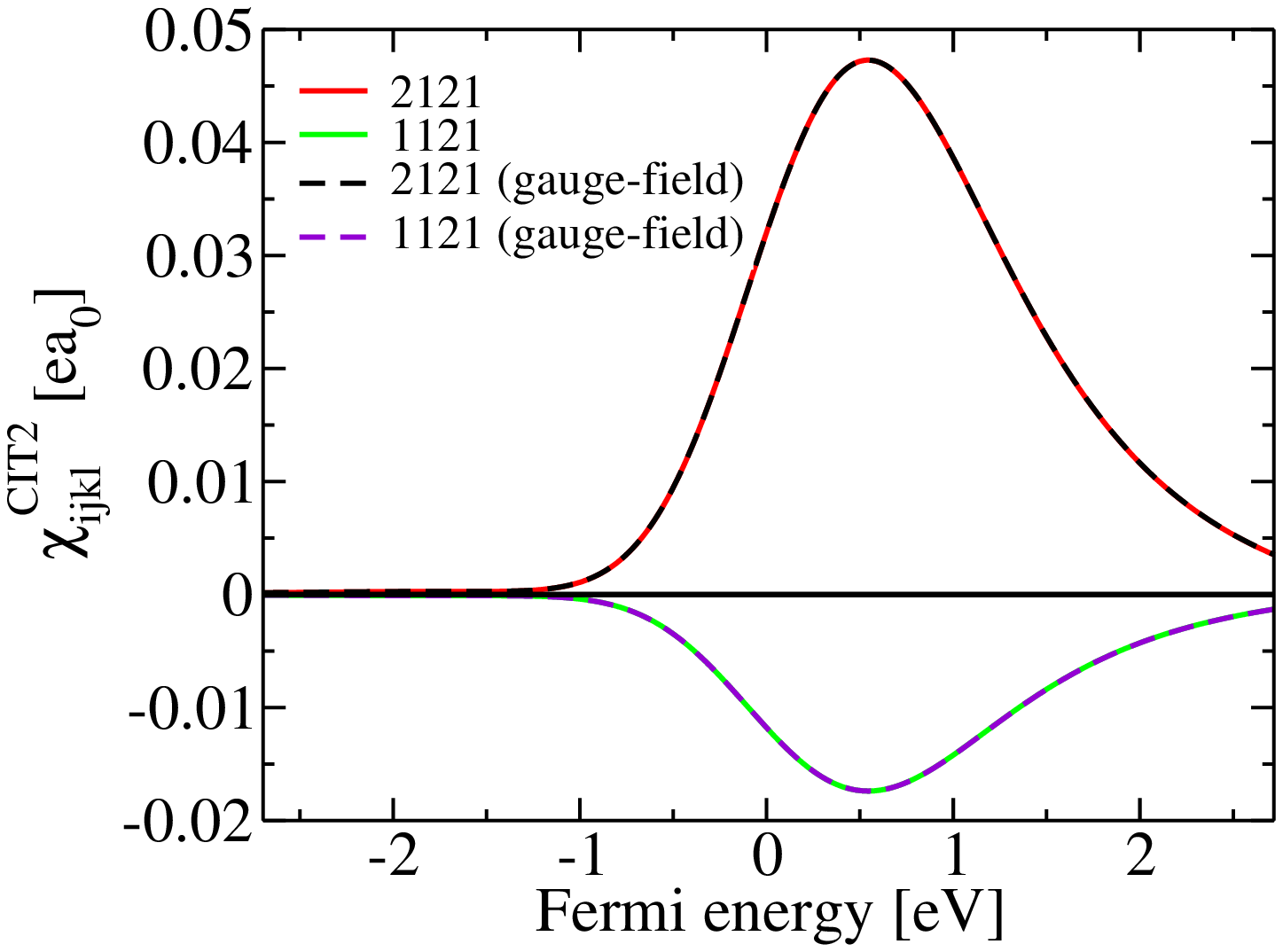}
\caption{\label{fig_onedim_gauge_field_vs_perturbation_cit}
Chiral CIT in the 1d
Rashba model for cycloidal gradients vs.\ Fermi
energy. General perturbation
theory (solid lines) agrees to the gauge-field 
approach (dashed lines). 
}
\end{figure}

In 
Fig.~\ref{fig_onedim_gauge_field_vs_perturbation_cit} 
we show the chiral CIT as a function of
the Fermi energy for cycloidal magnetization gradients
in the 1d Rashba model. 
The components $\chi_{2121}^{\rm CIT2}$
and $\chi_{1121}^{\rm CIT2}$ are labelled by
2121 and 1121, respectively.
The component 2121 of CIT describes the non-adiabatic torque,
while the component 1121 describes the adiabatic STT (modified
by SOI).
In the one-dimensional Rashba model, 
the contributions $\chi_{2121}^{\rm CIT2b}$ 
and $\chi_{1121}^{\rm CIT2b}$ (Eq.~\eqref{eq_cit2b})
from the CIDMI are zero 
when $\hat{\vn{M}}=\hat{\vn{e}}_{z}$ (not shown in the figure).
For cycloidal spin spirals,
it is possible to solve the 1d Rashba model
by a gauge-field approach~\cite{chigyromag},
which allows us to test the
perturbation theory, Eq.~\eqref{eq_chiral_torkance_green}.
For comparison we show in Fig.~\ref{fig_onedim_gauge_field_vs_perturbation_cit} 
the results obtained
from the gauge-field approach,
which agree to the
perturbation theory, Eq.~\eqref{eq_chiral_torkance_green}.
This demonstrates the validity of
Eq.~\eqref{eq_chiral_torkance_green}.

\begin{figure}
\includegraphics[width=\linewidth]{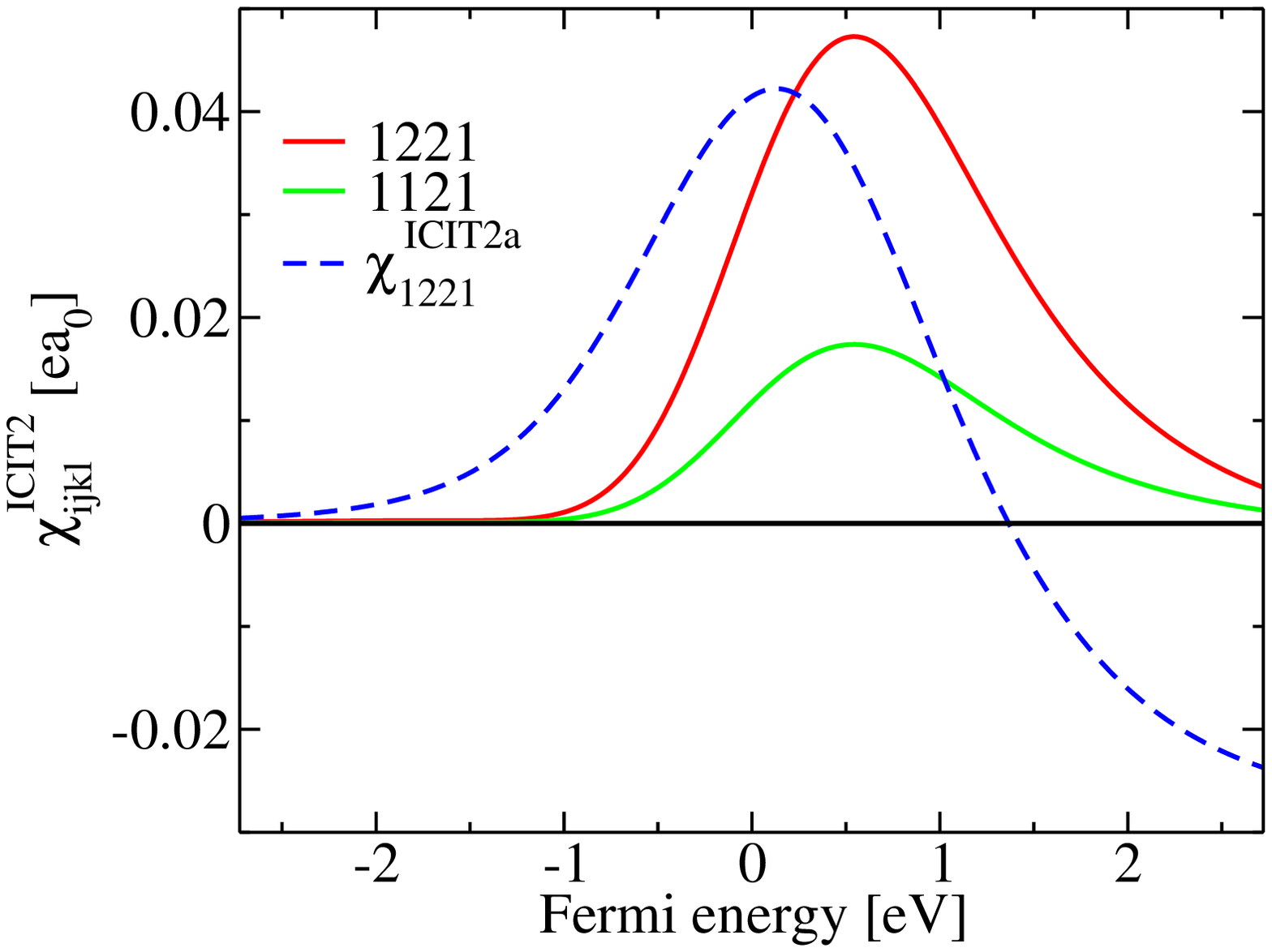}
\caption{\label{fig_onedim_gauge_field_vs_perturbation_icit}
Chiral ICIT in the 1d
Rashba model for cycloidal gradients
vs.\ Fermi energy. 
Dashed line: Contribution 
from the time-dependent gradient.
}
\end{figure}

In 
Fig.~\ref{fig_onedim_gauge_field_vs_perturbation_icit} 
we show the chiral ICIT in the 1d Rashba model. 
The components $\chi_{1221}^{\rm ICIT2}$
and $\chi_{1121}^{\rm ICIT2}$ are labelled by
1221 and 1121, respectively.
The contribution $\chi_{1221}^{\rm ICIT2a}$
from the time-dependent
gradient is of the same order of magnitude as the 
total $\chi_{1221}^{\rm ICIT2}$.
Comparison of Fig.~\ref{fig_onedim_gauge_field_vs_perturbation_cit} 
and Fig.~\ref{fig_onedim_gauge_field_vs_perturbation_icit} shows
that
CIT and ICIT satisfy the reciprocity relations Eq.~\eqref{eq_cit_icit_recipro},
that $\chi_{1121}^{\rm CIT2}$ is odd in $\hat{\vn{M}}$,
and that $\chi_{2121}^{\rm CIT2}$ is even in $\hat{\vn{M}}$,
i.e., $\chi_{2121}^{\rm CIT2}=\chi_{1221}^{\rm ICIT2}$ 
and $\chi_{1121}^{\rm CIT2}=-\chi_{1121}^{\rm ICIT2}$.
The contribution $\chi_{1221}^{\rm ICIT2a}$  from the 
time-dependent gradients is crucial to satisfy the
reciprocity relations between $\chi_{2121}^{\rm CIT2}$
and $\chi_{1221}^{\rm ICIT2}$.

\begin{figure}
\includegraphics[width=\linewidth]{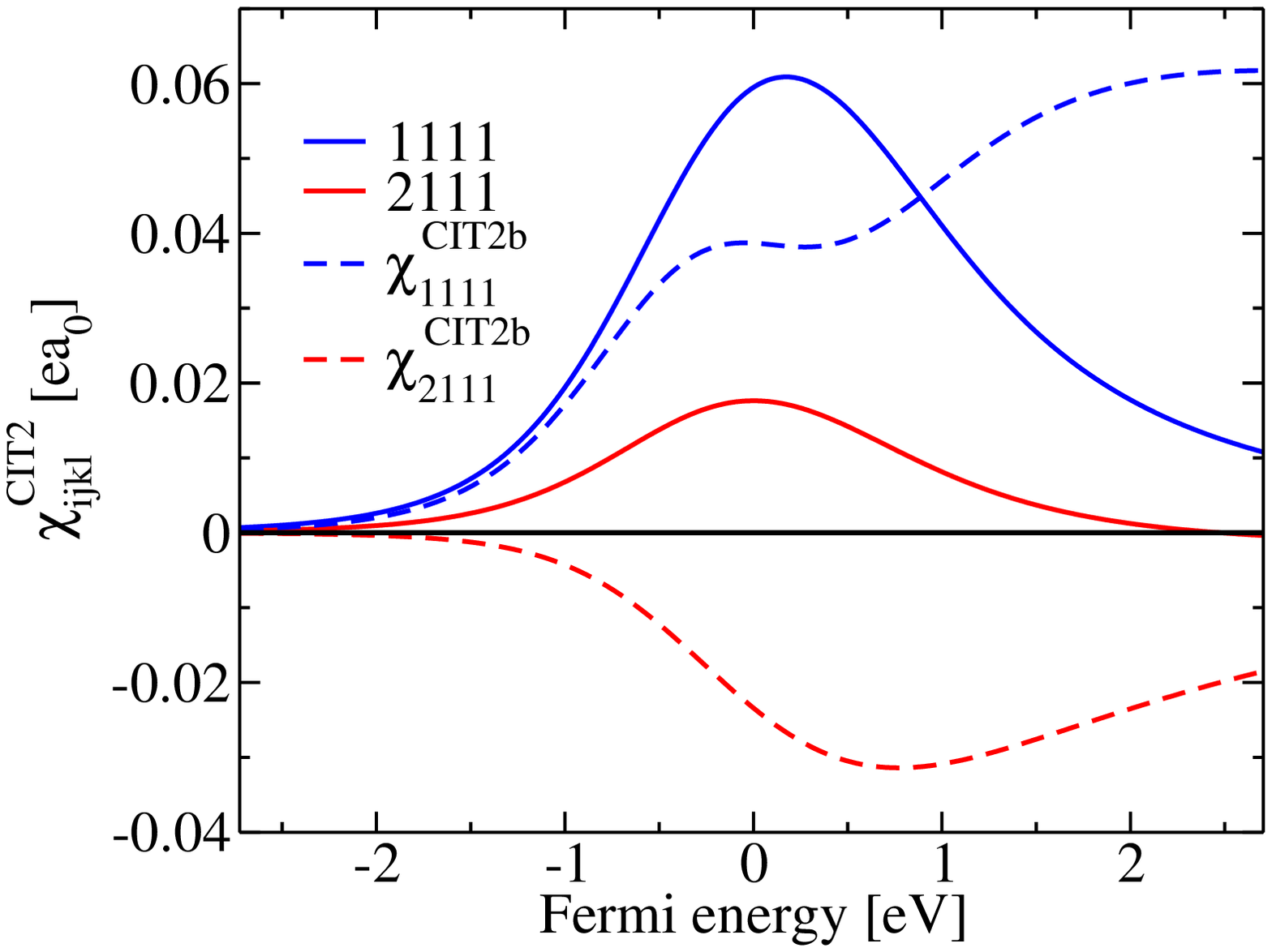}
\caption{\label{fig_onedim_cit_vs_fermi_helical}
Chiral CIT for helical gradients
in the 1d Rashba model vs.\ Fermi energy. 
Dashed lines: Contributions from CIDMI.
}
\end{figure}

\begin{figure}
\includegraphics[width=\linewidth]{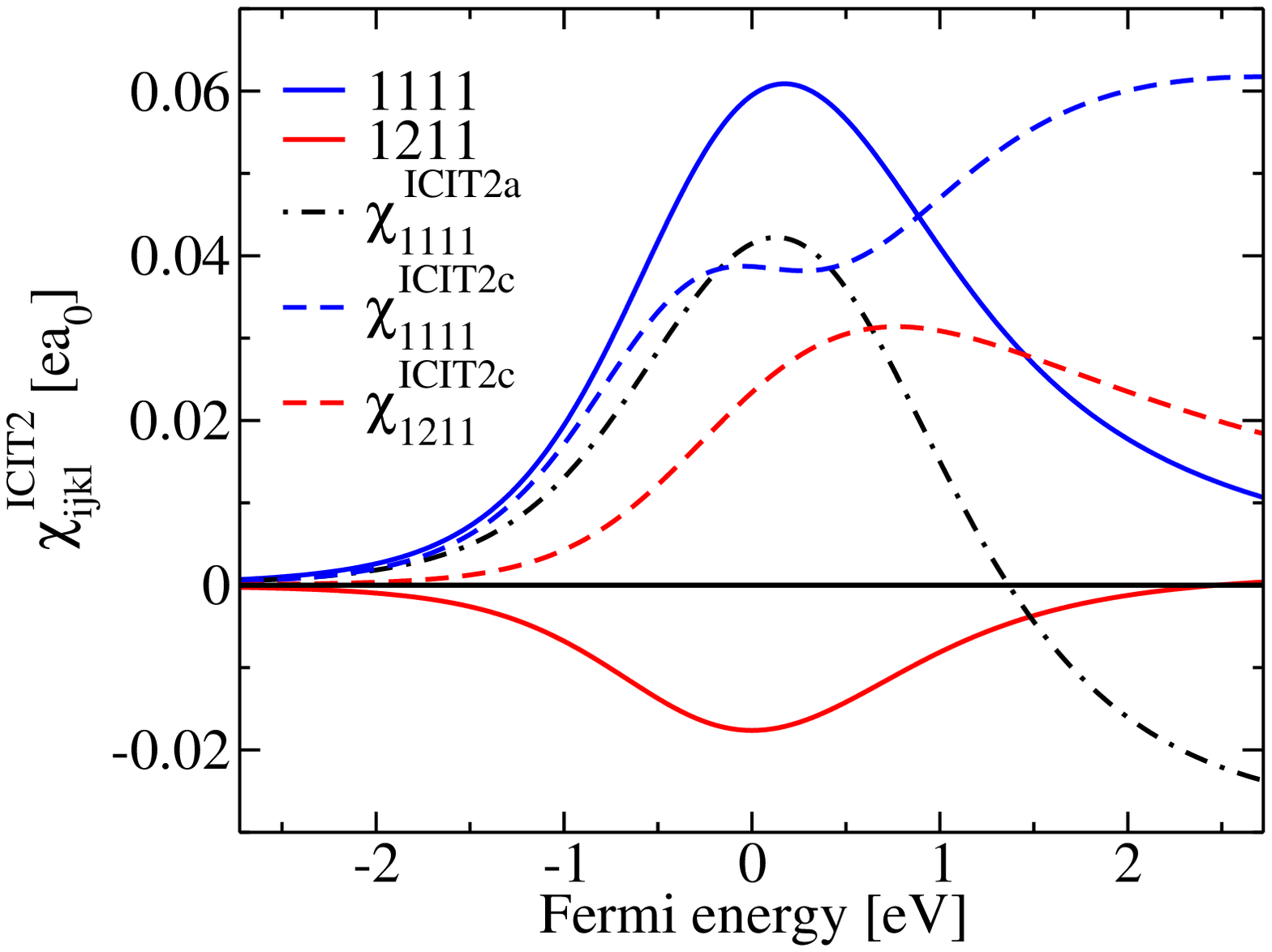}
\caption{\label{fig_onedim_icit_vs_fermi_helical}
Chiral ICIT for helical gradients
in the 1d Rashba model vs.\ Fermi energy. 
Dashed lines: Contributions from DOM.
Dashed-dotted line: Contribution from the
time-dependent magnetization gradient.
}
\end{figure}

In Fig.~\ref{fig_onedim_cit_vs_fermi_helical}
and Fig.~\ref{fig_onedim_icit_vs_fermi_helical}
we show the CIT and the ICIT, respectively, 
for helical gradients in the 1d Rashba model.
The components $\chi_{2111}^{\rm CIT2}$ 
and $\chi_{1111}^{\rm CIT2}$ are labelled
2111 and 1111, respectively, in Fig.~\ref{fig_onedim_cit_vs_fermi_helical},
while $\chi_{1211}^{\rm ICIT2}$ 
and $\chi_{1111}^{\rm ICIT2}$ are labelled
1211 and 1111, respectively, in Fig.~\ref{fig_onedim_icit_vs_fermi_helical}.
The contributions $\chi_{2111}^{\rm CIT2b}$
and $\chi_{1111}^{\rm CIT2b}$  from CIDMI
are of the
same order of magnitude as the total $\chi_{2111}^{\rm CIT2}$
and $\chi_{1111}^{\rm CIT2}$.
Similarly, the contributions $\chi_{1211}^{\rm ICIT2c}$
and $\chi_{1111}^{\rm ICIT2c}$  from DOM
are of the
same order of magnitude as the total $\chi_{1211}^{\rm ICIT2}$
and $\chi_{1111}^{\rm ICIT2}$.
Additionally, the contribution $\chi_{1111}^{\rm ICIT2a}$
from the time-dependent gradient is substantial.
Comparison of Fig.~\ref{fig_onedim_cit_vs_fermi_helical} and
Fig.~\ref{fig_onedim_icit_vs_fermi_helical}
shows that
CIT and ICIT satisfy the reciprocity relation Eq.~\eqref{eq_cit_icit_recipro},
that $\chi_{2111}^{\rm CIT2}$ is odd in $\hat{\vn{M}}$,
and that $\chi_{1111}^{\rm CIT2}$ is even in $\hat{\vn{M}}$,
i.e., $\chi_{1111}^{\rm CIT2}=\chi_{1111}^{\rm ICIT2}$ 
and $\chi_{2111}^{\rm CIT2}=-\chi_{1211}^{\rm ICIT2}$.
These reciprocity relations between CIT 
and ICIT are only satisfied when CIDMI, DOM, and
the response to time-dependent magnetization
gradients are included.
Additionally, the comparison between 
Fig.~\ref{fig_onedim_cit_vs_fermi_helical} and
Fig.~\ref{fig_onedim_icit_vs_fermi_helical}
shows that the contributions of CIDMI to 
CIT ($\chi_{1111}^{\rm CIT2b}$ and $\chi_{2111}^{\rm CIT2b}$)
are related to the contributions of 
DOM to ICIT ($\chi_{1111}^{\rm ICIT2c}$ and $\chi_{1211}^{\rm ICIT2c}$).
These relations between DOM and ICIT are expected from
Table~\ref{table_icidmi_idom_inplanex}.

\begin{figure}
\includegraphics[width=\linewidth]{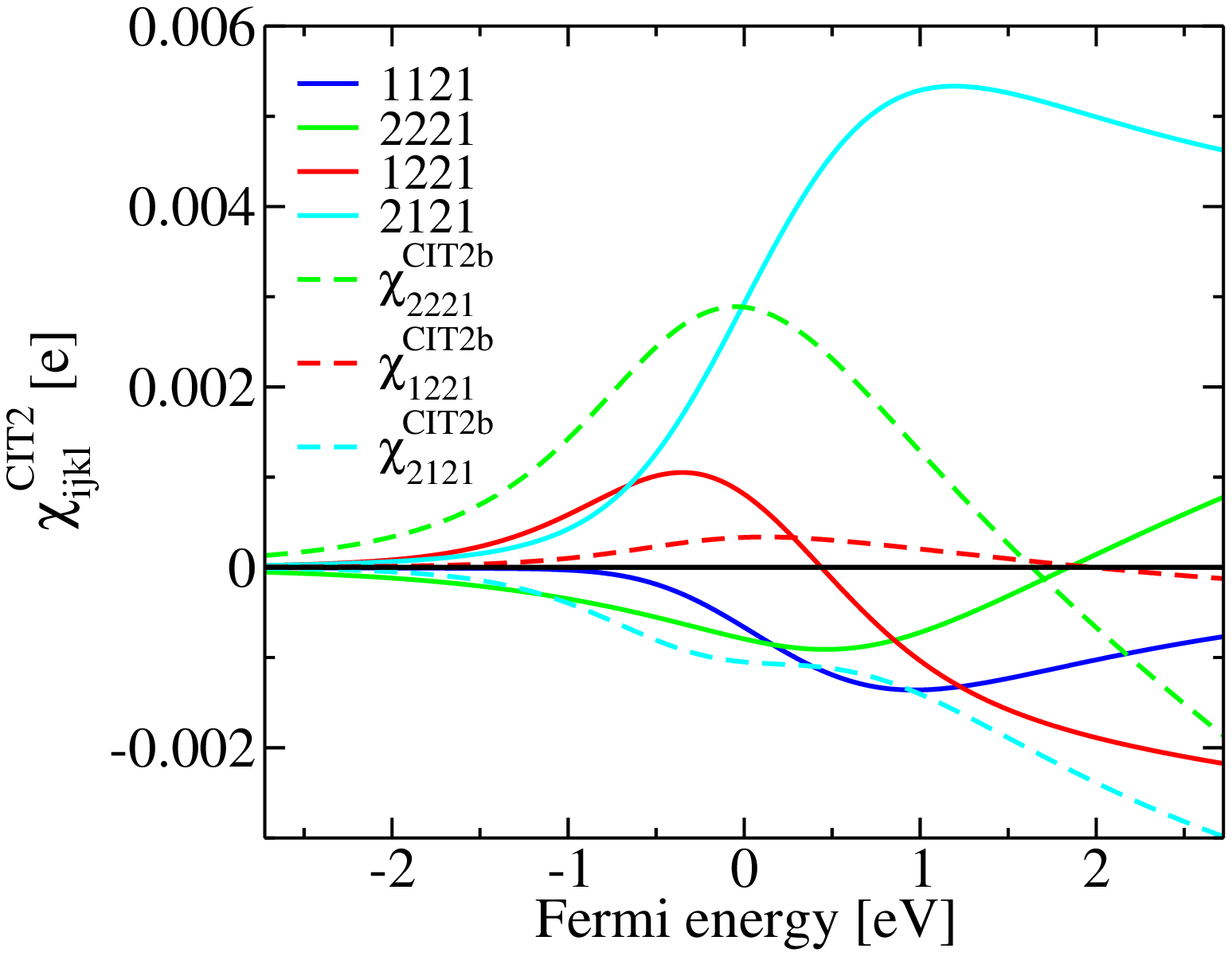}
\caption{\label{fig_twodim_cit_vs_fermi_cycloidal}
Chiral CIT for cycloidal gradients
in the 2d Rashba model vs.\ Fermi energy. 
Dashed lines: Contributions from CIDMI.
}
\end{figure}

\begin{figure}
\includegraphics[width=\linewidth]{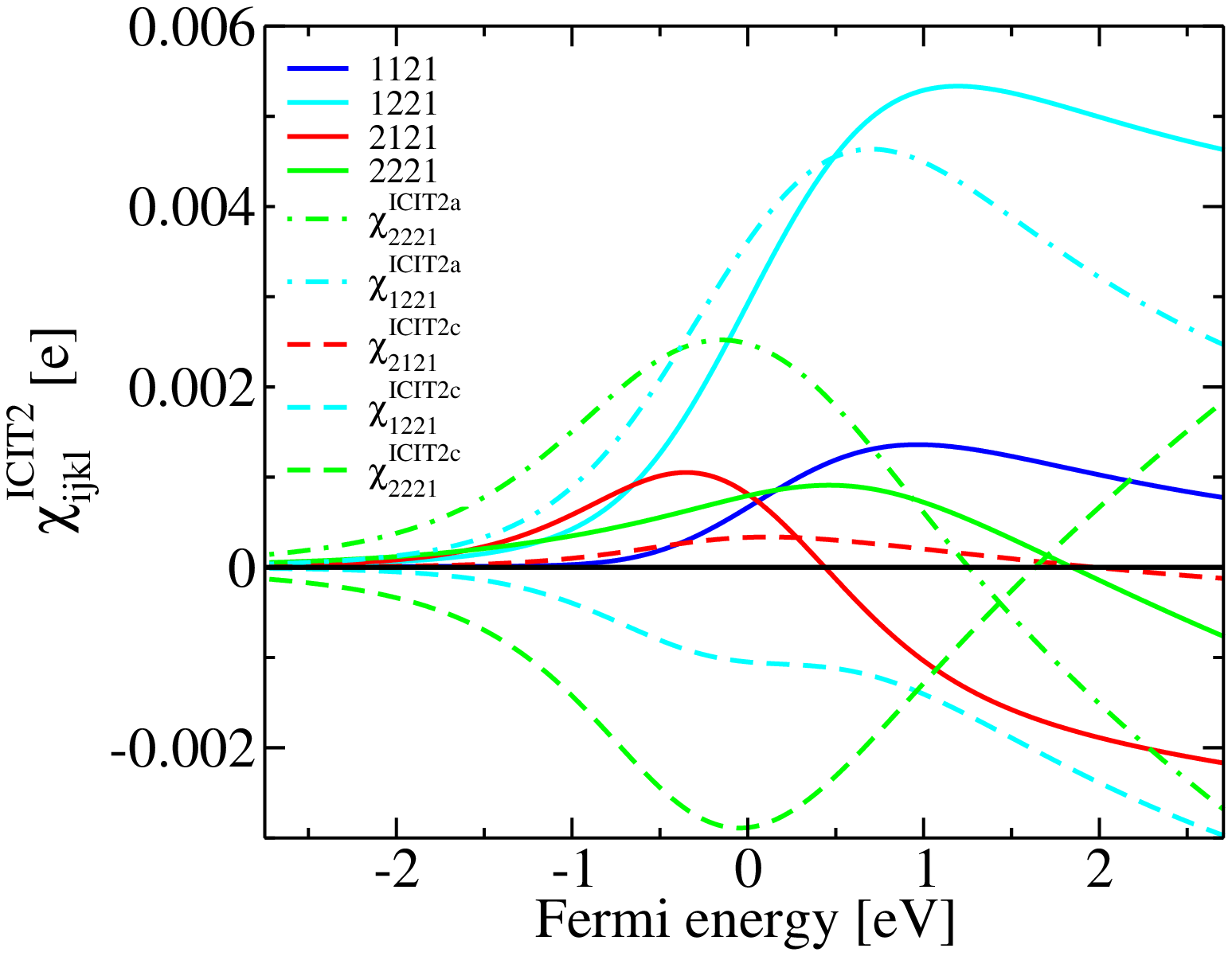}
\caption{\label{fig_twodim_icit_vs_fermi_cycloidal}
Chiral ICIT for cycloidal gradients
in the 2d Rashba model vs.\ Fermi energy. 
Dashed lines: Contributions from DOM.
Dashed-dotted lines: Contributions from the time-dependent gradients.
}
\end{figure}

In Fig.~\ref{fig_twodim_cit_vs_fermi_cycloidal}
and Fig.~\ref{fig_twodim_icit_vs_fermi_cycloidal}
we show the CIT and the ICIT, respectively, 
for cycloidal gradients in the 2d Rashba model.
In this case there are contributions from CIDMI and DOM 
in contrast to the 1d case 
with cycloidal gradients (Fig.~\ref{fig_onedim_gauge_field_vs_perturbation_cit}).
Comparison between Fig.~\ref{fig_twodim_cit_vs_fermi_cycloidal}
and Fig.~\ref{fig_twodim_icit_vs_fermi_cycloidal}
shows that $\chi_{1121}^{\rm CIT2}$
and $\chi_{2221}^{\rm CIT2}$ are odd in $\hat{\vn{M}}$,
that  $\chi_{1221}^{\rm CIT2}$
and $\chi_{2121}^{\rm CIT2}$ are even in $\hat{\vn{M}}$,
 and that
CIT and ICIT satisfy the reciprocity relation Eq.~\eqref{eq_cit_icit_recipro}
when the gradients of CIDMI and DOM are included,
i.e., $\chi_{1121}^{\rm CIT2}=-\chi_{1121}^{\rm ICIT2}$,
$\chi_{2221}^{\rm CIT2}=-\chi_{2221}^{\rm ICIT2}$,
$\chi_{1221}^{\rm CIT2}=\chi_{2121}^{\rm ICIT2}$,
and $\chi_{2121}^{\rm CIT2}=\chi_{1221}^{\rm ICIT2}$.
$\chi_{1121}^{\rm CIT2}$ describes the adiabatic STT with SOI,
while $\chi_{2121}^{\rm CIT2}$ describes the non-adiabatic STT.
Experimentally, it has been found that
CITs occur also when the electric field is applied parallel to
domain-walls (i.e., perpendicular to 
the $\vn{q}$-vector of spin spirals)~\cite{sots_perpendicular_to_wall}.
In our calculations, the components $\chi_{2221}^{\rm CIT2}$ 
and  $\chi_{1221}^{\rm CIT2}$ describe such a case, where the
applied electric field points in $y$ direction, while the magnetization
direction varies with the $x$ coordinate.

\begin{figure}
\includegraphics[width=\linewidth]{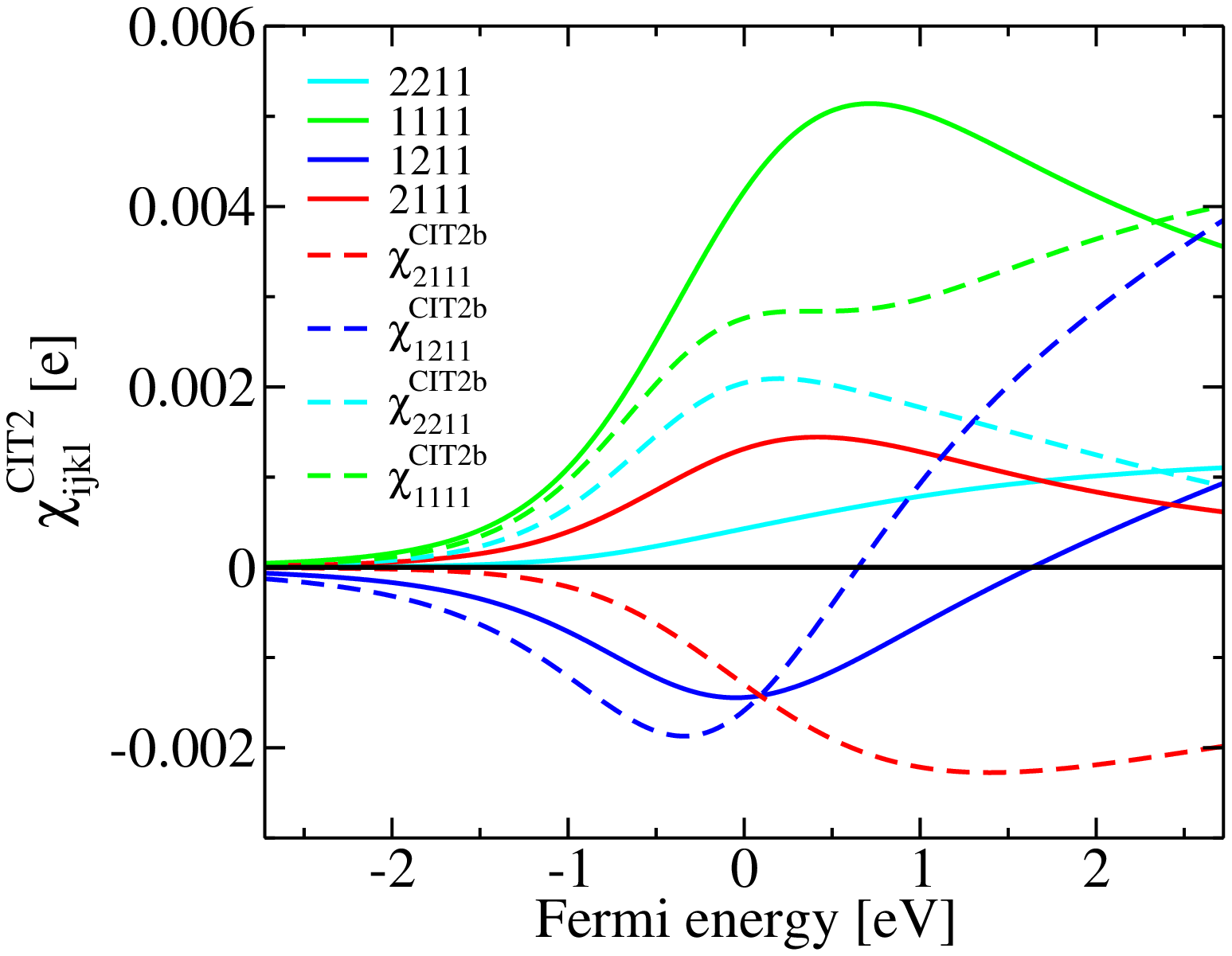}
\caption{\label{fig_twodim_cit_vs_fermi_helical}
Chiral CIT for helical gradients
in the 2d Rashba model vs.\ Fermi energy. 
Dashed lines: Contributions from CIDMI.
}
\end{figure}

\begin{figure}
\includegraphics[width=\linewidth]{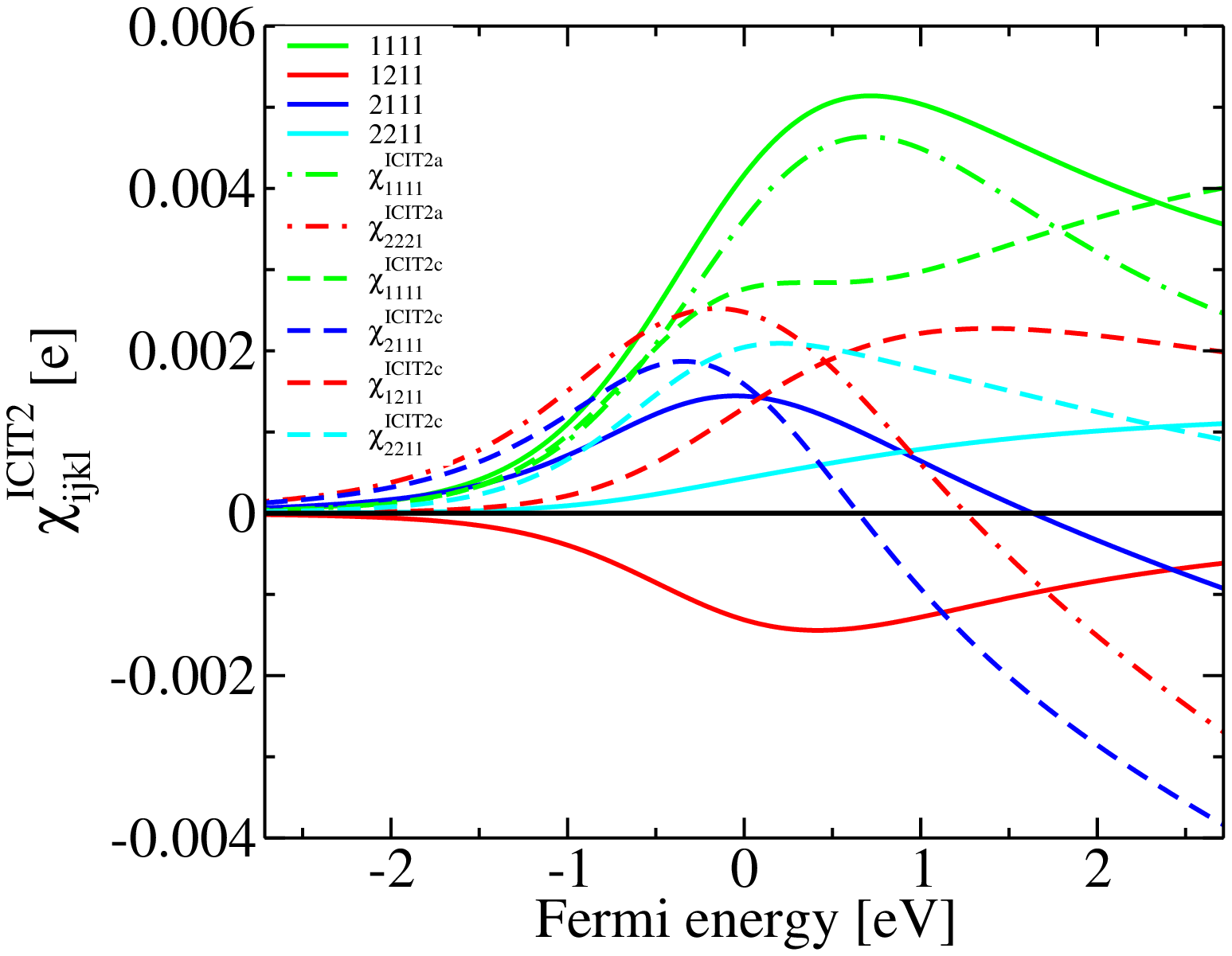}
\caption{\label{fig_twodim_icit_vs_fermi_helical}
Chiral ICIT for helical gradients
in the 2d Rashba model vs.\ Fermi energy. 
Dashed lines: Contributions from DOM.
Dashed-dotted lines: Contributions from the
time-dependent gradient.
}
\end{figure}

In Fig.~\ref{fig_twodim_cit_vs_fermi_helical}
and Fig.~\ref{fig_twodim_icit_vs_fermi_helical}
we show the chiral CIT and ICIT, respectively, 
for helical gradients in the 2d Rashba model.
The component $\chi_{2111}^{\rm CIT2}$ describes the
adiabatic STT with SOI and the 
component $\chi_{1111}^{\rm CIT2}$ describes the 
non-adiabatic STT. The components $\chi_{2211}^{\rm CIT2}$
and $\chi_{1211}^{\rm CIT2}$ describe the case when 
the applied electric field points in $y$ direction, i.e., perpendicular
to the direction along which the magnetization direction varies.
Comparison between Fig.~\ref{fig_twodim_cit_vs_fermi_helical}
and Fig.~\ref{fig_twodim_icit_vs_fermi_helical} shows that
$\chi_{1111}^{\rm CIT2}$
and $\chi_{2211}^{\rm CIT2}$ are even in $\hat{\vn{M}}$,
that $\chi_{1211}^{\rm CIT2}$ and $\chi_{2111}^{\rm CIT2}$
are odd in $\hat{\vn{M}}$ and that
CIT and ICIT satisfy the reciprocity relation Eq.~\eqref{eq_cit_icit_recipro}
when the gradients of CIDMI and DOM are included, i.e.,
$\chi_{1111}^{\rm CIT2}=\chi_{1111}^{\rm ICIT2}$,
$\chi_{2211}^{\rm CIT2}=\chi_{2211}^{\rm ICIT2}$,
$\chi_{1211}^{\rm CIT2}=-\chi_{2111}^{\rm ICIT2}$,
and $\chi_{2111}^{\rm CIT2}=-\chi_{1211}^{\rm ICIT2}$.

\subsection{Chiral torque-torque correlation}
\label{sec_results_gyrodamp}
\begin{figure}
\includegraphics[width=\linewidth]{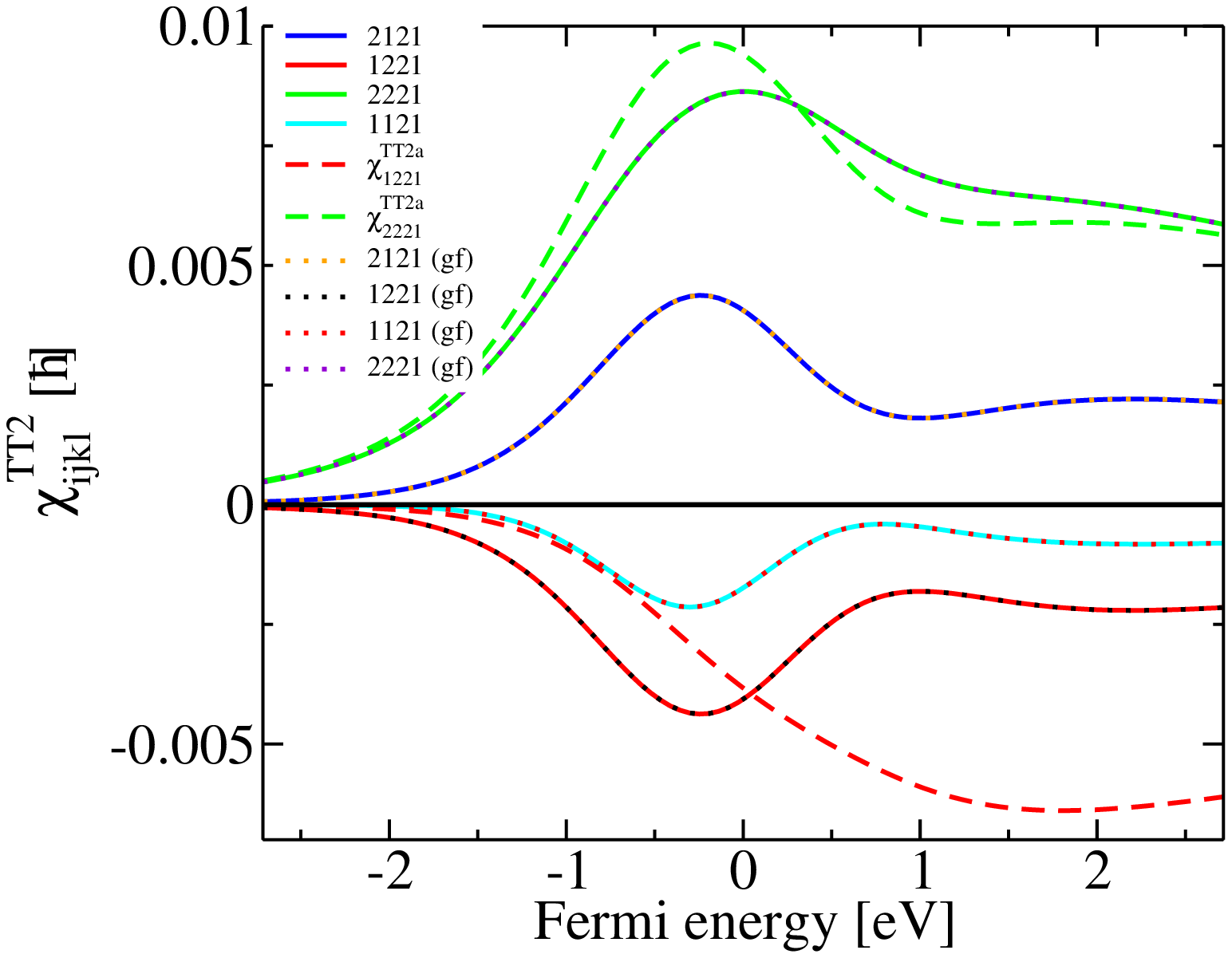}
\caption{\label{fig_onedim_gauge_field_vs_perturbation_gyro}
Chiral contribution to the torque-torque correlation for cycloidal gradients
in the 1d Rashba model vs.\ Fermi energy. Perturbation
theory (solid lines) agrees to the gauge-field (gf) 
approach (dotted lines). Dashed lines: Contribution 
from the time-dependent gradient.
}
\end{figure}

In Fig.~\ref{fig_onedim_gauge_field_vs_perturbation_gyro} we show the 
chiral contribution to the torque-torque correlation
in the 1d Rashba model for cycloidal gradients.
We compare the perturbation theory Eq.~\eqref{eq_chiral_gyromag_green}
plus Eq.~\eqref{eq_chi_tt2a}
to the gauge-field approach from Ref.~\cite{chigyromag}.
This comparison shows that perturbation theory provides the correct answer 
only 
when the contribution $\chi_{ijkl}^{\rm TT2a}$ (Eq.~\eqref{eq_chi_tt2a}) 
from the time-dependent gradients is taken into account.
The contributions $\chi_{1221}^{\rm TT2a}$
and $\chi_{2221}^{\rm TT2a}$ from the time-dependent
gradients are comparable in magnitude to the total values.
In the 1d Rashba model 
the DDMI-contribution in Eq.~\eqref{eq_chi_tt2c} 
is zero for cycloidal gradients (not shown in the figure).
The components $\chi^{\rm TT2}_{2121}$ 
and $\chi^{\rm TT2}_{1221}$ describe the chiral gyromagnetism
while the components $\chi^{\rm TT2}_{1121}$ 
and $\chi^{\rm TT2}_{2221}$ describe the chiral 
damping~\cite{chigyromag,chiral_damping_magnetic_domain_walls,phenomenology_chiral_damping}.
The components $\chi^{\rm TT2}_{2121}$ 
and $\chi^{\rm TT2}_{1221}$ are odd in $\hat{\vn{M}}$ and they satisfy 
the Onsager relation Eq.~\eqref{eq_torquetorque_onsa}, 
i.e., $\chi^{\rm TT2}_{2121}=-\chi^{\rm TT2}_{1221}$.

\begin{figure}
\includegraphics[width=\linewidth]{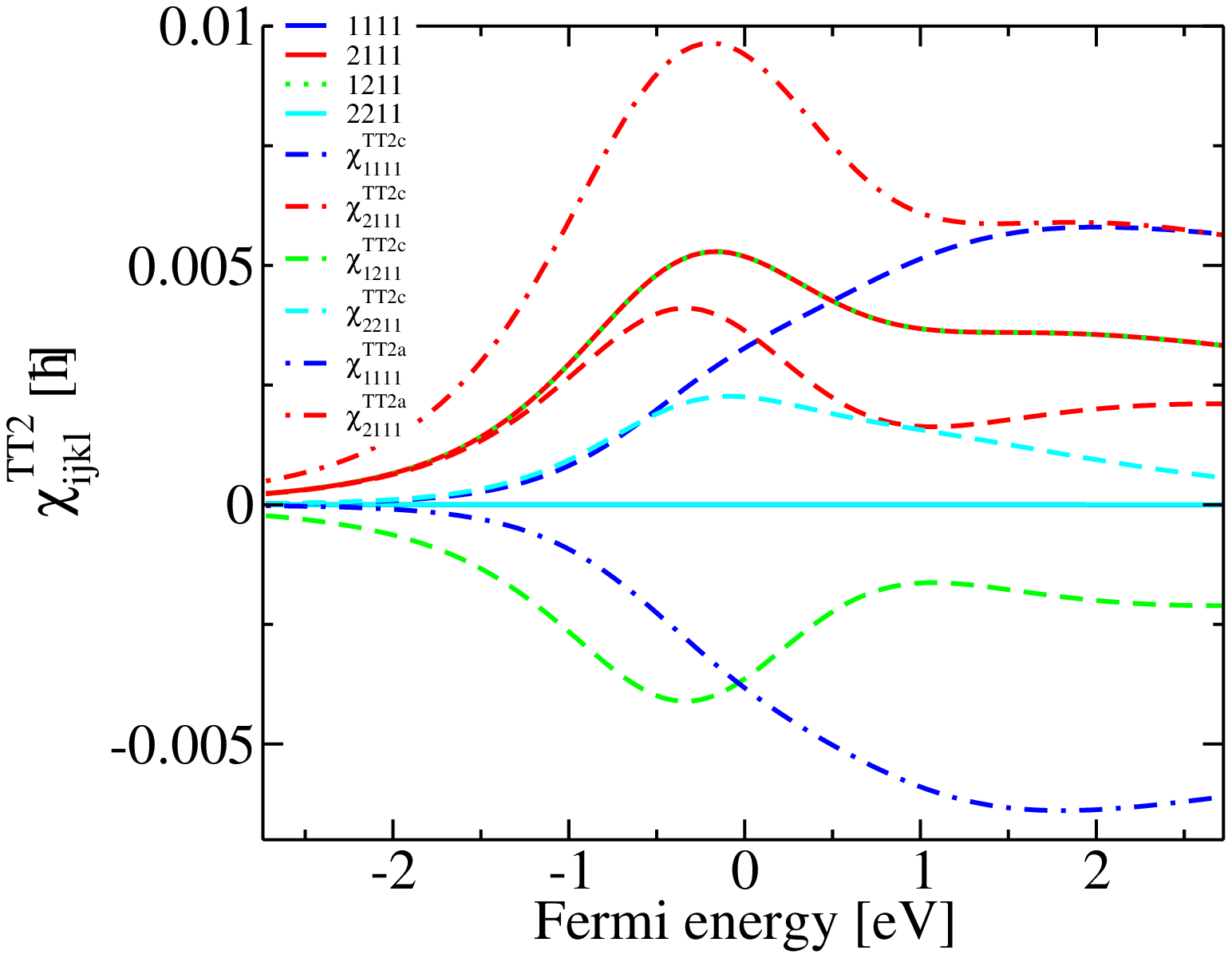}
\caption{\label{fig_onedim_lambda_vs_fermi_helical}
Chiral contribution to the torque-torque correlation for helical gradients
in the 1d Rashba model vs.\ Fermi energy.
Dashed lines: Contributions from DDMI. 
Dashed-dotted lines: Contributions from the time-dependent gradients.
}
\end{figure}

In Fig.~\ref{fig_onedim_lambda_vs_fermi_helical}
we show the chiral contributions to the torque-torque
correlation in the 1d Rashba model for
helical gradients.
In contrast to the cycloidal gradients 
(Fig.~\ref{fig_onedim_gauge_field_vs_perturbation_gyro})
there are contributions from the spatial gradients of 
DDMI (Eq.~\eqref{eq_chi_tt2c}) in this case.
The Onsager relation Eq.~\eqref{eq_torquetorque_onsa}
for the components $\chi_{2111}^{\rm TT2}$ 
and $\chi_{1211}^{\rm TT2}$
is satisfied only when these contributions from DDMI are taken
into account, which are of the same order of magnitude as
the total values.
The components $\chi_{2111}^{\rm TT2}$ 
and $\chi_{1211}^{\rm TT2}$ are even 
in $\hat{\vn{M}}$ and describe chiral damping,
while the components 
$\chi_{1111}^{\rm TT2}$ 
and $\chi_{2211}^{\rm TT2}$
are odd in $\hat{\vn{M}}$ and describe chiral gyromagnetism.
As a consequence of the Onsager relation
Eq.~\eqref{eq_torquetorque_onsa}
we obtain $\chi_{1111}^{\rm TT2}=\chi_{2211}^{\rm TT2}=0$
for the total components: Eq.~\eqref{eq_torquetorque_onsa} shows that
diagonal components of the torque-torque correlation function 
are zero unless they are even in $\hat{\vn{M}}$.
However, $\chi_{1111}^{\rm TT2a}$,  $\chi_{1111}^{\rm TT2c}$,
and $\chi_{1111}^{\rm TT2b}=-\chi_{1111}^{\rm TT2a}-\chi_{1111}^{\rm TT2c}$
are individually nonzero.
Interestingly, the off-diagonal components of the
torque-torque correlation describe
chiral damping for helical gradients, while for cycloidal
gradients the off-diagonal elements describe chiral gyromagnetism
and the diagonal elements describe chiral damping. 

\begin{figure}
\includegraphics[width=\linewidth]{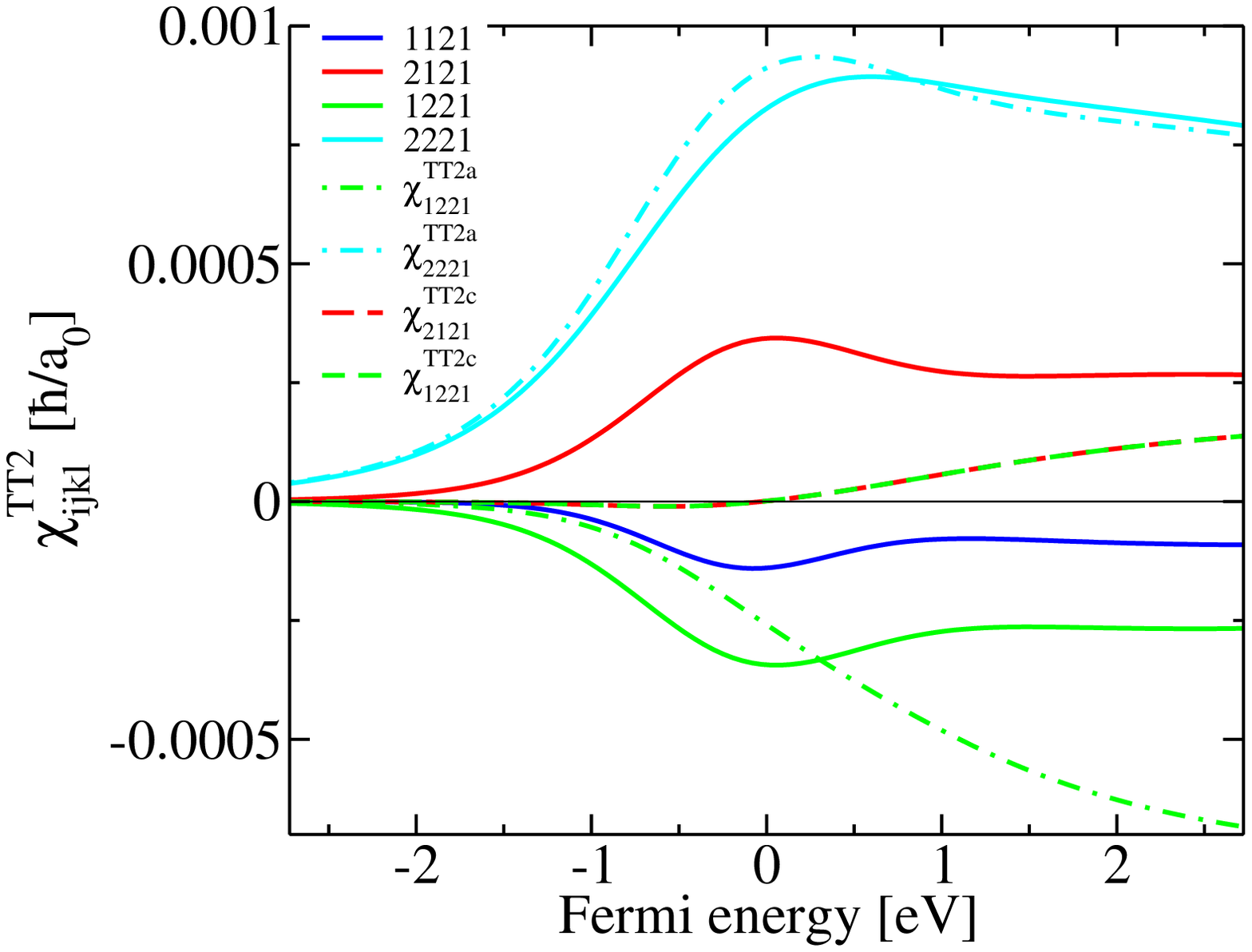}
\caption{\label{fig_twodim_lambda_vs_fermi_chiral}
Chiral contribution to the torque-torque correlation for cycloidal gradients
in the 2d Rashba model vs.\ Fermi energy. 
Dashed lines: Contributions from DDMI.
Dashed-dotted lines: Contributions from the time-dependent gradients.
}
\end{figure}

In Fig.~\ref{fig_twodim_lambda_vs_fermi_chiral}
we show the chiral contributions to the torque-torque
correlation in the 2d Rashba model
for cycloidal gradients.
In contrast to the 1d Rashba model
with cycloidal gradients (Fig.~\ref{fig_onedim_gauge_field_vs_perturbation_gyro}) the
contributions from DDMI $\chi_{ijkl}^{\rm TT2c}$ 
(Eq.~\eqref{eq_chi_tt2c}) are nonzero in this case.
Without these contributions from DDMI the Onsager relation~\eqref{eq_torquetorque_onsa}
$\chi_{2121}^{\rm TT2}=-\chi_{1221}^{\rm TT2}$
is violated. The DDMI contribution is of the same order of magnitude as
the total values.
The components $\chi_{2121}^{\rm TT2}$ 
and $\chi_{1221}^{\rm TT2}$ are odd in $\hat{\vn{M}}$ and describe chiral gyromagnetism,
while the components $\chi_{1121}^{\rm TT2}$ 
and $\chi_{2221}^{\rm TT2}$ are even in $\hat{\vn{M}}$ and describe chiral damping. 

\begin{figure}
\includegraphics[width=\linewidth]{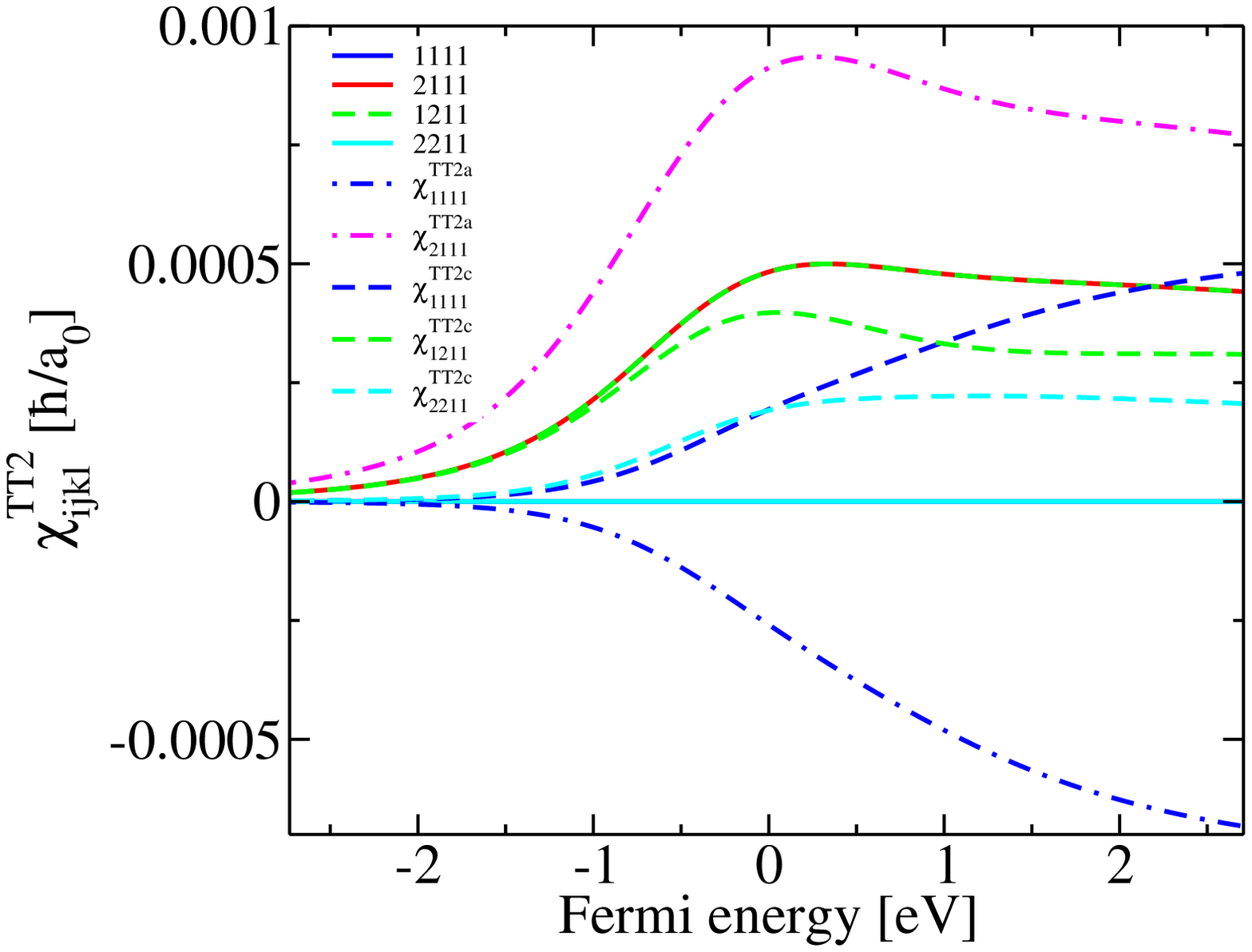}
\caption{\label{fig_twodim_lambda_vs_fermi_helical}
Chiral contribution to the torque-torque correlation for helical gradients
in the 2d Rashba model vs.\ Fermi energy. 
Dashed lines: Contributions from DDMI.
Dashed-dotted lines: Contributions from the time-dependent gradients.
}
\end{figure}

In Fig.~\ref{fig_twodim_lambda_vs_fermi_helical}
we show the chiral contributions to the torque-torque
correlation in the 2d Rashba model for
helical gradients.
The components $\chi_{1211}^{\rm TT2}$ 
and $\chi_{2111}^{\rm TT2}$ are even in $\hat{\vn{M}}$ and describe chiral damping, while the
components $\chi_{1111}^{\rm TT2}$ 
and $\chi_{2211}^{\rm TT2}$ are odd in $\hat{\vn{M}}$ and describe
chiral gyromagnetism.
The Onsager relation Eq.~\eqref{eq_torquetorque_onsa}
requires $\chi_{1111}^{\rm TT2}=\chi_{2211}^{\rm TT2}=0$
and $\chi_{2111}^{\rm TT2}=\chi_{1211}^{\rm TT2}$.
Without the contributions from DDMI 
these Onsager relations are violated.

\section{Summary}
\label{sec_summary}
Finding ways to tune
the Dzyaloshinskii-Moriya interaction (DMI)
by external means, such as an applied electric current,
holds much promise for applications
in which DMI determines the magnetic texture
of domain walls or skyrmions.
In order to derive an expression for current-induced
Dzyaloshinskii-Moriya interaction (CIDMI) we first identify
its inverse effect: 
When magnetic textures vary as a function of time, electric currents
are driven by various mechanisms, which can be distinguished
according to their different dependence on the time-derivative of magnetization,
$\partial \hat{\vn{M}}(\vn{r},t)/\partial t$, and on the spatial 
derivative $\partial \hat{\vn{M}}(\vn{r},t)/\partial \vn{r}$: One group of 
effects is proportional to $\partial \hat{\vn{M}}(\vn{r},t)/\partial t$,
a second group of effects is proportional to the 
product $\partial \hat{\vn{M}}(\vn{r},t)/\partial t \,\,\, \partial \hat{\vn{M}}(\vn{r},t)/\partial \vn{r}$,
and a third group is proportional to the second derivative
$\partial^2 \hat{\vn{M}}(\vn{r},t)/\partial \vn{r}\partial t$.   
We show that the response of the electric current
to the time-dependent magnetization 
gradient $\partial^2 \hat{\vn{M}}(\vn{r},t)/\partial \vn{r}\partial t$
contais the inverse of CIDMI. 
We establish the reciprocity relation between inverse and direct CIDMI
and thereby obtain an expression for CIDMI.
We find that CIDMI is related to the modification of orbital magnetism
induced by magnetization dynamics, 
which we call dynamical orbital magnetism (DOM).
We show that torques are generated by
time-dependent gradients of magnetization as well. The inverse
effect consists in the modification of DMI by
magnetization dynamics, which we call
dynamical DMI (DDMI). 

Additionally, we develop a formalism to calculate the chiral contributions to
the direct and inverse
current-induced torques (CITs) and to the torque-torque
correlation
in noncollinear magnets.
We show that the response to time-dependent magnetization
gradients contributes substantially
to these effects
and that the Onsager reciprocity relations are violated
when it is not taken into account.  
In noncollinear magnets CIDMI, DDMI and DOM depend on the
local magnetization direction.
We show that the resulting
spatial gradients of CIDMI, DDMI and DOM
have to be subtracted from the CIT, from the torque-torque 
correlation, and from the inverse CIT, respectively. 

We apply our formalism to study CITs and
the torque-torque correlation in
textured Rashba ferromagnets.
We find that the contribution of CIDMI to the
chiral CIT is of the order of magnitude of the total effect.
Similarly, we find that the contribution of  
DDMI to the chiral torque-torque correlation is of the
order of magnitude of the total effect.

\section*{Acknowledgments}
We acknowledge financial support from Leibniz Collaborative Excellence project OptiSPIN 
$-$ Optical Control of Nanoscale Spin Textures. We  acknowledge  funding  under SPP 2137 
``Skyrmionics" of the DFG. We gratefully acknowledge financial support from the European 
Research Council (ERC) under the European Union's Horizon 2020 research and innovation program 
(Grant No. 856538, project "3D MAGiC”). The work was also supported by the Deutsche 
Forschungsgemeinschaft (DFG, German Research Foundation) $-$ TRR 173 $-$ 268565370 (project 
A11). We gratefully acknowledge the J\"ulich Supercomputing Centre and RWTH Aachen 
University for providing computational resources under project No. jiff40.

\appendix
\section{Response to time-dependent gradients}
\label{app_time_dependent_gradients}
In this appendix we derive Eq.~\eqref{eq_chi_icit2a},
Eq.~\eqref{eq_icidmi},
Eq.~\eqref{eq_iddmi},
and Eq.~\eqref{eq_chi_tt2a},
which describe the response to
time-dependent magnetization gradients,
and Eq.~\eqref{eq_chi_idom},
which describes the response to
time-dependent magnetic fields.
We consider perturbations
of the form
\bege\label{eq_perturb_sinr_sint}
\delta H(\vn{r},t)=\mathfrak{B}
b\frac{1}{q\omega}
\sin(\vn{q}\cdot \vn{r})\sin(\omega t).
\ee
When we set $\mathfrak{B}=\frac{\partial H}{\partial \hat{M}_{k}}$
and $b=\frac{\partial^2 \hat{M}_{k}}{\partial r_{i}\partial t}$, 
Eq.~\eqref{eq_perturb_sinr_sint}
turns into Eq.~\eqref{eq_pert_general_tidegra}, while
when we set $\mathfrak{B}=-ev_{i}$
and $b=\frac{1}{2}\epsilon_{ijk}\frac{\partial B_{k}}{\partial t}$ we obtain
Eq.~\eqref{eq_pert_tidegra_bfi}.
We need to derive an expression for the
response $\delta A(\vn{r},t)$ of an 
observable $\mathfrak{A}$ to this perturbation, which varies
in time like $\cos(\omega t)$ and in space 
like $\cos(\vn{q}\cdot \vn{r})$, because
$\frac{\partial^2 \hat{\vn{M}}(\vn{r},t)}{\partial r_i\partial
  t}\propto \cos(\vn{q}\cdot \vn{r})\cos(\omega t)$.
Therefore, we use the Kubo linear response 
formalism to obtain the coefficient $\chi$ in
\bege
\delta A(\vn{r},t)=
\chi
\cos(\vn{q}\cdot \vn{r})\cos(\omega t),
\ee
which
is given by
\bege\label{eq_retarded_func}
\begin{aligned}
\chi=\frac{i}{ \hbar q\omega V}
\Bigl[
&\langle\langle
\mathfrak{A}\cos(\vn{q}\cdot \vn{r}),
\mathfrak{B}\sin(\vn{q}\cdot \vn{r})
\rangle\rangle^{\rm R}(\hbar\omega)\\
-
&\langle\langle
\mathfrak{A}\cos(\vn{q}\cdot \vn{r}),
\mathfrak{B}\sin(\vn{q}\cdot \vn{r})
\rangle\rangle^{\rm R}(-\hbar\omega)\Bigr],
\end{aligned}
\ee
where $\langle\langle
\mathfrak{A}\cos(\vn{q}\cdot \vn{r}),
\mathfrak{B}\sin(\vn{q}\cdot \vn{r})
\rangle\rangle^{\rm R}(\hbar\omega)$ is the
retarded function at frequency $\omega$
and $V$ is the volume of the unit cell.

The operator $\mathfrak{B}\sin(\vn{q}\cdot \vn{r})$ can be written as
\bege\label{eq_b_sin}
\mathfrak{B}\sin(\vn{q}\cdot \vn{r})
=\frac{1}{2i}\sum_{\vn{k}nm}
\left[
\mathfrak{B}^{(1)}_{\vn{k}nm}
c^{\dagger}_{\vn{k}_{+}n}
c^{\phantom{\dagger}}_{\vn{k}_{-}m}
-\mathfrak{B}^{(2)}_{\vn{k}nm}
c^{\dagger}_{\vn{k}_{-}n}
c^{\phantom{\dagger}}_{\vn{k}_{+}m}
\right],
\ee
where $\vn{k}_{+}=\vn{k}+\vn{q}/2$, $\vn{k}_{-}=\vn{k}-\vn{q}/2$, 
$c^{\dagger}_{\vn{k}_{+}n}$ is the creation operator of
an electron in 
state $|u_{\vn{k}_{+}n}^{\phantom{k}}\rangle$,
$c^{\phantom{\dagger}}_{\vn{k}_{-}m}$
is the annihilation operator of 
an electron in 
state $|u_{\vn{k}_{-}m}^{\phantom{k}}\rangle$,
\bege
\mathfrak{B}^{(1)}_{\vn{k}nm}=\frac{1}{2}\langle
u_{\vn{k}_{+}n}^{\phantom{k}}|
[\mathfrak{B}_{\vn{k}_{+}}
+\mathfrak{B}_{\vn{k}_{-}}
]
|
u_{\vn{k}_{-}m}^{\phantom{k}}
\rangle
\ee
and
\bege
\mathfrak{B}^{(2)}_{\vn{k}nm}=\frac{1}{2}\langle
u_{\vn{k}_{-}n}^{\phantom{k}}|
[
\mathfrak{B}_{\vn{k}_{+}}
+
\mathfrak{B}_{\vn{k}_{-}}
]
|
u_{\vn{k}_{+}m}^{\phantom{k}}
\rangle.
\ee
Similarly, 
\bege\label{eq_a_cos}
\mathfrak{A}
\cos(\vn{q}\cdot\vn{r})=\frac{1}{2}\sum_{\vn{k}nm}
\left[
\mathfrak{A}^{(1)}_{\vn{k}nm}
c^{\dagger}_{\vn{k}_{+}n}
c^{\phantom{\dagger}}_{\vn{k}_{-}m}
+
\mathfrak{A}^{(2)}_{\vn{k}nm}
c^{\dagger}_{\vn{k}_{-}n}
c^{\phantom{\dagger}}_{\vn{k}_{+}m}
\right],
\ee
where
\bege
\mathfrak{A}^{(1)}_{\vn{k}nm}=\frac{1}{2}\langle
u_{\vn{k}_{+}n}^{\phantom{k}}|
\left[
\mathfrak{A}_{\vn{k}_{+}}
+
\mathfrak{A}_{\vn{k}_{-}}
\right]
|
u_{\vn{k}_{-}m}^{\phantom{k}}
\rangle
\ee
and
\bege
\mathfrak{A}^{(2)}_{\vn{k}nm}=\frac{1}{2}\langle
u_{\vn{k}_{-}n}^{\phantom{k}}|
\left[
\mathfrak{A}_{\vn{k}_{+}}
+
\mathfrak{A}_{\vn{k}_{-}}
\right]
|
u_{\vn{k}_{+}m}^{\phantom{k}}
\rangle.
\ee

It is convenient to obtain the retarded response
function in Eq.~\eqref{eq_retarded_func} from the corresponding
Matsubara function in imaginary time $\tau$
\bege\label{eq_correlation_matsubaratime}
\begin{aligned}
&\frac{1}{V}\langle\langle
\mathfrak{A}\cos(\vn{q}\cdot\vn{r})
,
\mathfrak{B}\sin(\vn{q}\cdot\vn{r})
\rangle\rangle^{\rm M}(\tau)=\\
&=\frac{1}{4i}
\intkspa
\sum_{nm}\sum_{n'm'}
\Bigl [
\mathfrak{A}^{(1)}_{\vn{k}nm}
\mathfrak{B}^{(2)}_{\vn{k}n'm'}
\mathscr{Z}^{(1)}_{\vn{k}nmn'm'}(\tau)\\
&\quad\quad\quad\quad\quad-
\mathfrak{A}^{(2)}_{\vn{k}nm}
\mathfrak{B}^{(1)}_{\vn{k}n'm'}
\mathscr{Z}^{(2)}_{\vn{k}nmn'm'}(\tau)
\Bigr],
\end{aligned}
\ee
where $d=1,2$ or 3 is the dimension,
\bege
\begin{aligned}
\mathscr{Z}^{(1)}_{\vn{k}nmn'm'}(\tau)&=\langle
T_{\tau}
\crea{\vn{k}_{+}n}(\tau)
\annihi{\vn{k}_{-}m}(\tau)
\crea{\vn{k}_{-}n'}(0)
\annihi{\vn{k}_{+}m'}(0)
\rangle\\
&=-\gmat_{m'n}(\vn{k}_{+},-\tau)
\gmat_{mn'}(\vn{k}_{-},\tau),
\end{aligned}
\ee
\bege
\begin{aligned}
\mathscr{Z}^{(2)}_{\vn{k}nmn'm'}(\tau)&=\langle
T_{\tau}
\crea{\vn{k}_{-}n}(\tau)
\annihi{\vn{k}_{+}m}(\tau)
\crea{\vn{k}_{+}n'}(0)
\annihi{\vn{k}_{-}m'}(0)
\rangle\\
&=-\gmat_{m'n}(\vn{k}_{-},-\tau)
\gmat_{mn'}(\vn{k}_{+},\tau),
\end{aligned}
\ee
and
\bege
\gmat_{mn'}(\vn{k}_{+},\tau)=-
\langle
T_{\tau}
\annihi{\vn{k}_{+}m}(\tau)
\crea{\vn{k}_{+}n'}(0)
\rangle
\ee
is the single-particle Matsubara function.
The Fourier transform of Eq.~\eqref{eq_correlation_matsubaratime}
is given by
\bege\label{eq_tidegra_fourier}
\begin{aligned}
&\frac{1}{V}\langle\langle
\mathfrak{A}\cos(\vn{q}\cdot\vn{r})
,
\mathfrak{B}\sin(\vn{q}\cdot\vn{r})
\rangle\rangle^{\rm M}(i\mathcal{E}_{N})=\\
&=\frac{i}{4\hbar\beta}\intkspa\sum_{nm}\sum_{n'm'}\sum_{p}
\Bigl [\\
&\mathfrak{A}^{(1)}_{\vn{k}nm}
\mathfrak{B}^{(2)}_{\vn{k}n'm'}
\gmat_{m'n}(\vn{k}_{+},i\mathcal{E}_{p})
\gmat_{mn'}(\vn{k}_{-},i\mathcal{E}_{p}+i\mathcal{E}_{N})
\\
-
&\mathfrak{A}^{(2)}_{\vn{k}nm}
\mathfrak{B}^{(1)}_{\vn{k}n'm'}
\gmat_{m'n}(\vn{k}_{-},i\mathcal{E}_{p})
\gmat_{mn'}(\vn{k}_{+},i\mathcal{E}_{p}+i\mathcal{E}_{N})
\Bigr],\\
\end{aligned}
\ee
where $\mathcal{E}_{N}=2\pi N/\beta$ 
and $\mathcal{E}_{p}=(2p+1)\pi /\beta$ are
bosonic and fermionic
Matsubara energy points, respectively, and
$\beta=1/(k_{\rm B}T)$ is the inverse temperature.

In order to carry out the Matsubara 
summation over $\mathcal{E}_{p}$ we make use of
\bege\label{eq_greengreen_matsubarasum}
\begin{aligned}
&\frac{1}{\beta}\sum_{p}
\gmat_{mn'}(i\mathcal{E}_{p}+i\mathcal{E}_{N})
\gmat_{m'n}(i\mathcal{E}_{p})=\\
&=\frac{i}{2\pi}\int
d \mathcal{E}' 
f(\mathcal{E}')\gmat_{mn'}(\mathcal{E}'+i\mathcal{E}_N)
\gmat_{m'n}(\mathcal{E}'+i\delta)\\
&+\frac{i}{2\pi}\int
d \mathcal{E}' 
f(\mathcal{E}')\gmat_{mn'}(\mathcal{E}'+i\delta)
\gmat_{m'n}(\mathcal{E}'-i\mathcal{E}_N)\\
&-\frac{i}{2\pi}\int
d \mathcal{E}' 
f(\mathcal{E}')\gmat_{mn'}(\mathcal{E}'+i\mathcal{E}_N)
\gmat_{m'n}(\mathcal{E}'-i\delta)\\
&-\frac{i}{2\pi}\int
d \mathcal{E}' 
f(\mathcal{E}')\gmat_{mn'}(\mathcal{E}'-i\delta)
\gmat_{m'n}(\mathcal{E}'-i\mathcal{E}_N),
\end{aligned}
\ee
where $\delta$ is a positive infinitesimal.
The retarded function $\langle\langle
\mathfrak{A}\cos(\vn{q}\cdot \vn{r}),
\mathfrak{B}\sin(\vn{q}\cdot \vn{r})
\rangle\rangle^{\rm R}(\omega)$
is obtained from the Matsubara function
$\langle\langle
\mathfrak{A}\cos(\vn{q}\cdot \vn{r}),
\mathfrak{B}\sin(\vn{q}\cdot \vn{r})
\rangle\rangle^{\rm M}(i\mathcal{E}_{N})$
by the analytic continuation $i\mathcal{E}_{N}\rightarrow \hbar\omega$
to real frequencies.
The right-hand side of
Eq.~\eqref{eq_greengreen_matsubarasum} 
has the following analytic continuation to real frequencies:
\bege
\begin{aligned}
\frac{i}{2\pi}&\int
d \mathcal{E}' 
f(\mathcal{E}')\gret_{mn'}(\mathcal{E}'+\hbar\omega)
\gret_{m'n}(\mathcal{E}')\\
+\frac{i}{2\pi}&\int
d \mathcal{E}' 
f(\mathcal{E}')\gret_{mn'}(\mathcal{E}')
\gadv_{m'n}(\mathcal{E}'-\hbar\omega)\\
-\frac{i}{2\pi}&\int
d \mathcal{E}' 
f(\mathcal{E}')\gret_{mn'}(\mathcal{E}'+\hbar\omega)
\gadv_{m'n}(\mathcal{E}')\\
-\frac{i}{2\pi}&\int
d \mathcal{E}' 
f(\mathcal{E}')\gadv_{mn'}(\mathcal{E}')
\gadv_{m'n}(\mathcal{E}'-\hbar\omega).
\end{aligned}
\ee
Therefore, we obtain
\bege\label{eq_chi_from_z}
\begin{aligned}
\chi=\frac{-i}{8\pi \hbar^2 q\omega}\intkspa[
&Z_{\vn{k}}(q,\omega)-Z_{\vn{k}}(-q,\omega)\\
-
&Z_{\vn{k}}(q,-\omega)+Z_{\vn{k}}(-q,-\omega)
],
\end{aligned}
\ee
where
\bege
\begin{aligned}
&Z_{\vn{k}}(q,\omega)=\\
&=\int
d \mathcal{E}' 
f(\mathcal{E}')
{\rm Tr}\left[
\mathfrak{A}_{\vn{k}}
\gret_{\vn{k}_{-}}(\mathcal{E}'+\hbar\omega)
\mathfrak{B}_{\vn{k}}
\gret_{\vn{k}_{+}}(\mathcal{E}')
\right]
\\
&+\int
d \mathcal{E}' 
f(\mathcal{E}')
{\rm Tr}\left[
\mathfrak{A}_{\vn{k}}
\gret_{\vn{k}_{-}}(\mathcal{E}')
\mathfrak{B}_{\vn{k}}
\gadv_{\vn{k}_{+}}(\mathcal{E}'-\hbar\omega)
\right]
\\
&-\int
d \mathcal{E}' 
f(\mathcal{E}')
{\rm Tr}\left[
\mathfrak{A}_{\vn{k}}
\gret_{\vn{k}_{-}}(\mathcal{E}'+\hbar\omega)
\mathfrak{B}_{\vn{k}}
\gadv_{\vn{k}_{+}}(\mathcal{E}')
\right]
\\
&-\int
d \mathcal{E}' 
f(\mathcal{E}')
{\rm Tr}\left[
\mathfrak{A}_{\vn{k}}
\gadv_{\vn{k}_{-}}(\mathcal{E}')
\mathfrak{B}_{\vn{k}}
\gadv_{\vn{k}_{+}}(\mathcal{E}'-\hbar\omega)
\right].\\
\end{aligned}
\ee
We consider the limit $\lim_{q\rightarrow 0}\lim_{\omega\rightarrow 0}\chi$.
In this limit Eq.~\eqref{eq_chi_from_z} may be rewritten
as
\bege
\chi=\frac{-i}{2\pi\hbar^2}\intkspa
\left.
\frac{\partial^2 Z_{\vn{k}}(q,\omega)}{\partial q\partial\omega}
\right|_{q=\omega=0}
.
\ee
The frequency derivative of $Z_{\vn{k}}(q,\omega)$
is given by
\bege\label{eq_enediff_zq}
\begin{aligned}
\frac{1}{\hbar}
\left.
\frac{\partial Z_{\vn{k}}}{\partial \omega}
\right|_{\omega=0}
=&\int
d \mathcal{E}' 
f(\mathcal{E}')
{\rm Tr}\left[
\mathfrak{A}_{\vn{k}}
\frac{\partial\gret_{\vn{k}_{-}}(\mathcal{E}')}{\partial \mathcal{E}'}
\mathfrak{B}_{\vn{k}}
\gret_{\vn{k}_{+}}(\mathcal{E}')
\right]
\\
-&\int
d \mathcal{E}' 
f(\mathcal{E}')
{\rm Tr}\left[
\mathfrak{A}_{\vn{k}}
\gret_{\vn{k}_{-}}(\mathcal{E}')
\mathfrak{B}_{\vn{k}}
\frac{\partial
\gadv_{\vn{k}_{+}}(\mathcal{E}')}
{\partial \mathcal{E}'}
\right]
\\
-&\int
d \mathcal{E}' 
f(\mathcal{E}')
{\rm Tr}\left[
\mathfrak{A}_{\vn{k}}
\frac{\partial
\gret_{\vn{k}_{-}}(\mathcal{E}')}
{\partial \mathcal{E}'}
\mathfrak{B}_{\vn{k}}
\gadv_{\vn{k}_{+}}(\mathcal{E}')
\right]
\\
+&\int
d \mathcal{E}' 
f(\mathcal{E}')
{\rm Tr}\left[
\mathfrak{A}_{\vn{k}}
\gadv_{\vn{k}_{-}}(\mathcal{E}')
\mathfrak{B}_{\vn{k}}
\frac{\partial
\gadv_{\vn{k}_{+}}(\mathcal{E}')}
{\partial \mathcal{E}'}
\right].\\
\end{aligned}
\ee
Using $\partial\gret(\mathcal{E})/\partial\mathcal{E}=-\gret(\mathcal{E})\gret(\mathcal{E})/\hbar$ we
obtain
\bege\label{eq_enediff_zq2}
\begin{aligned}
\left.
\frac{\partial Z_{\vn{k}}}{\partial \omega}
\right|_{\omega=0}=
-&\int
d \mathcal{E}' 
f(\mathcal{E}')
{\rm Tr}\left[
\mathfrak{A}_{\vn{k}}
\gret_{\vn{k}_{-}}
\gret_{\vn{k}_{-}}
\mathfrak{B}_{\vn{k}}
\gret_{\vn{k}_{+}}
\right]
\\
+&\int
d \mathcal{E}' 
f(\mathcal{E}')
{\rm Tr}\left[
\mathfrak{A}_{\vn{k}}
\gret_{\vn{k}_{-}}
\mathfrak{B}_{\vn{k}}
\gadv_{\vn{k}_{+}}
\gadv_{\vn{k}_{+}}
\right]
\\
+&\int
d \mathcal{E}' 
f(\mathcal{E}')
{\rm Tr}\left[
\mathfrak{A}_{\vn{k}}
\gret_{\vn{k}_{-}}
\gret_{\vn{k}_{-}}
\mathfrak{B}_{\vn{k}}
\gadv_{\vn{k}_{+}}
\right]
\\
-&\int
d \mathcal{E}' 
f(\mathcal{E}')
{\rm Tr}\left[
\mathfrak{A}_{\vn{k}}
\gadv_{\vn{k}_{-}}
\mathfrak{B}_{\vn{k}}
\gadv_{\vn{k}_{+}}
\gadv_{\vn{k}_{+}}
\right].\\
\end{aligned}
\ee
Making use of
\bege
\lim_{q\rightarrow 0}
\frac{\partial \gret_{\vn{k}_{+}}}
{\partial q}
=\frac{1}{2}
\gret_{\vn{k}}
\frac{\vn{v}\cdot\vn{q}}{q}
\gret_{\vn{k}}
\ee
we finally obtain
\bege\label{eq_qdiffenediff_zq}
\begin{aligned}
&
\!\!\!\!\!\!\!\!\!\!\!\!\!\!\!\!
\chi=\frac{-i}{2\pi\hbar^2}
\intkspa
\lim_{q\rightarrow 0}
\lim_{\omega\rightarrow 0}
\frac{\partial^2 Z(q,\omega)}{\partial q\partial \omega}=\\
&
\!\!\!\!\!\!\!\!\!\!\!\!
=\frac{-i}{4\pi\hbar^2}
\frac{\vn{q}}{q}\cdot
\intkspa
\int
d \mathcal{E} 
f(\mathcal{E})
{\rm Tr}\Bigl[\\
&
\mathfrak{A}_{\vn{k}}
R\vn{v}RR \mathfrak{B}_{\vn{k}}R
+
\mathfrak{A}_{\vn{k}}
RR\vn{v}R \mathfrak{B}_{\vn{k}}R
\\
-&
\mathfrak{A}_{\vn{k}}
RR \mathfrak{B}_{\vn{k}}R\vn{v}R
-
\mathfrak{A}_{\vn{k}}
R\vn{v}R \mathfrak{B}_{\vn{k}}AA
\\
+&
\mathfrak{A}_{\vn{k}}
R \mathfrak{B}_{\vn{k}}A\vn{v}AA
+
\mathfrak{A}_{\vn{k}}
R
\mathfrak{B}_{\vn{k}}
AA\vn{v}A\\
-&
\mathfrak{A}_{\vn{k}}
R\vn{v}RR
\mathfrak{B}_{\vn{k}}
A
-
\mathfrak{A}_{\vn{k}}
RR\vn{v}R \mathfrak{B}_{\vn{k}}A
\\
+&
\mathfrak{A}_{\vn{k}}
RR \mathfrak{B}_{\vn{k}}A\vn{v}A
\\
+&
\mathfrak{A}_{\vn{k}}
A\vn{v}A \mathfrak{B}_{\vn{k}}AA
-
\mathfrak{A}_{\vn{k}}
A \mathfrak{B}_{\vn{k}}A\vn{v}AA
\\
-&
\mathfrak{A}_{\vn{k}}
A \mathfrak{B}_{\vn{k}}AA\vn{v}A
\Bigr],
\\
\end{aligned}
\ee
where we use the
abbreviations $R=\gret_{\vn{k}}(\mathcal{E})$
and $A=\gadv_{\vn{k}}(\mathcal{E})$.
When we substitute $\mathfrak{B}=\frac{\partial H}{\partial\hat{M}_{j}}$,
$\mathfrak{A}=-ev_{i}$, and $\vn{q}=q_{k}\hat{\vn{e}}_{k}$, 
we obtain Eq.~\eqref{eq_chi_icit2a}.
When we substitute $\mathfrak{B}=\mathcal{T}_{j}$,
$\mathfrak{A}=-ev_{i}$, and $\vn{q}=q_{k}\hat{\vn{e}}_{k}$, 
we obtain Eq.~\eqref{eq_icidmi}.
When we substitute $\mathfrak{A}=-\mathcal{T}_{i}$, 
$\mathfrak{B}=\mathcal{T}_{j}$, and $\vn{q}=q_{k}\hat{\vn{e}}_{k}$, 
we obtain Eq.~\eqref{eq_iddmi}.
When we substitute $\mathfrak{B}=-ev_{j}$,
$\mathfrak{A}=-\mathcal{T}_{i}$, and $\vn{q}=q_{k}\hat{\vn{e}}_{k}$, 
we obtain
Eq.~\eqref{eq_chi_idom}.
When we substitute $\mathfrak{B}=\frac{\partial H}{\partial\hat{M}_{j}}$,
$\mathfrak{A}=-\mathcal{T}_{i}$, and $\vn{q}=q_{k}\hat{\vn{e}}_{k}$, 
we obtain Eq.~\eqref{eq_chi_tt2a}.

\section{Perturbation theory for the chiral contributions
to CIT and to the torque-torque correlation}
\label{sec_appendix}
In this appendix we derive
expressions for the retarded
function
\bege
\langle\langle
\mathfrak{A}
\cos(\vn{q}\cdot\vn{r});
\mathfrak{C}
\rangle\rangle^{\rm R}(\hbar\omega)
\ee
within first-order perturbation theory with respect to the
perturbation 
\bege\label{eq_delta_h}
\delta H=\mathfrak{B}\eta\sin(\vn{q}\cdot \vn{r}),
\ee
which may arise e.g.\ from the spatial oscillation of the magnetization direction.
As usual, it is convenient to obtain the retarded response function 
from the corresponding Matsubara function
\bege\label{eq_chigyromag_matsubara}
\begin{aligned}
&
\langle\langle
\cos(\vn{q}\cdot \vn{r})\mathfrak{A};\mathfrak{C}
\rangle\rangle^{\rm M}(\tau)=-\langle
T_{\tau}
\cos(\vn{q}\cdot\vn{r})\mathfrak{A}(\tau)
\mathfrak{C}(0)
\rangle.
\end{aligned}
\ee

The starting point for the perturbative expansion is the
equation
\bege\label{eq_matsubara_function_generalized}
\begin{aligned}
&-\langle
T_{\tau}
\cos(\vn{q} \cdot\vn{r})\mathfrak{A}(\tau_1)
\mathfrak{C}(0)
\rangle=\\
&=-
\frac{
\Trace
\left[
e^{-\beta H}
T_{\tau}
\cos(\vn{q} \cdot\vn{r})\mathfrak{A}(\tau_1)
\mathfrak{C}(0)
\right]
}{\Trace
\left[
e^{-\beta H}
\right]
}=\\
&=-
\frac{
\Trace
\left\{
e^{-\beta H_0}
T_{\tau}\left[U
\cos(\vn{q}\cdot \vn{r})
\mathfrak{A}(\tau_1)
\mathfrak{C}(0)
\right]
\right\}
}{\Trace
\left[
e^{-\beta H_0}U
\right]
},
\end{aligned}
\ee
where $H_0$ is the unperturbed Hamiltonian
and we consider the first order in the perturbation $\delta H$:
\bege
U^{(1)}=-\frac{1}{\hbar}\int_{0}^{\hbar\beta}\!\!\!\!d\,\tau_{1}
{\rm T}_{\tau}
\{
e^{\tau_1 H_0 /\hbar}
\delta H
e^{-\tau_1 H_0 /\hbar}
\}.
\ee
The essential difference between Eq.~\eqref{eq_retarded_func}
and Eq.~\eqref{eq_matsubara_function_generalized}
is that in Eq.~\eqref{eq_retarded_func} the 
operator $\mathfrak{B}$ enters together with the 
factor $\sin(\vn{q}\cdot\vn{r})\sin(\omega t)$ (see Eq.~\eqref{eq_perturb_sinr_sint}),
while in Eq.~\eqref{eq_matsubara_function_generalized} only the
factor $\sin(\vn{q}\cdot\vn{r})$ is connected to $\mathfrak{B}$
in Eq.~\eqref{eq_delta_h}, while the factor $\sin(\omega t)$
is coupled to the additional operator $\mathfrak{C}$.

We use Eq.~\eqref{eq_b_sin}
and Eq.~\eqref{eq_a_cos}
in order to express $\mathfrak{A}\cos(\vn{q}\cdot\vn{r})$
and $\mathfrak{B}\sin(\vn{q}\cdot\vn{r})$
in terms of annihilation and creation operators.
In terms of the correlators
\bege
\begin{aligned}
&\mathscr{Z}^{(3)}_{\vn{k}nmn'm'n''m''}(\tau,\tau_1)=\\
&\langle
T_{\tau}
\crea{\vn{k}_{-}n}(\tau)
\annihi{\vn{k}_{+}m}(\tau)
\crea{\vn{k}_{+}n'}(\tau_1)
\annihi{\vn{k}_{-}m'}(\tau_1)
\crea{\vn{k}_{-}n''}
\annihi{\vn{k}_{-}m''}
\rangle\\
\end{aligned}
\ee
and
\bege
\begin{aligned}
&\mathscr{Z}^{(4)}_{\vn{k}nmn'm'n''m''}(\tau,\tau_1)=\\
&\langle
T_{\tau}
\crea{\vn{k}_{-}n}(\tau)
\annihi{\vn{k}_{+}m}(\tau)
\crea{\vn{k}_{+}n'}(\tau_1)
\annihi{\vn{k}_{-}m'}(\tau_1)
\crea{\vn{k}_{+}n''}
\annihi{\vn{k}_{+}m''}
\rangle\\
\end{aligned}
\ee
and
\bege
\begin{aligned}
&\mathscr{Z}^{(5)}_{\vn{k}nmn'm'n''m''}(\tau,\tau_1)=\\
&\langle
T_{\tau}
\crea{\vn{k}_{+}n}(\tau)
\annihi{\vn{k}_{-}m}(\tau)
\crea{\vn{k}_{-}n'}(\tau_1)
\annihi{\vn{k}_{+}m'}(\tau_1)
\crea{\vn{k}_{+}n''}
\annihi{\vn{k}_{+}m''}
\rangle\\
\end{aligned}
\ee
and
\bege
\begin{aligned}
&\mathscr{Z}^{(6)}_{\vn{k}nmn'm'n''m''}(\tau,\tau_1)=\\
&\langle
T_{\tau}
\crea{\vn{k}_{+}n}(\tau)
\annihi{\vn{k}_{-}m}(\tau)
\crea{\vn{k}_{-}n'}(\tau_1)
\annihi{\vn{k}_{+}m'}(\tau_1)
\crea{\vn{k}_{-}n''}
\annihi{\vn{k}_{-}m''}
\rangle\\
\end{aligned}
\ee
Eq.~\eqref{eq_matsubara_function_generalized} can be written as
\bege\label{eq_pi_general_imagtau}
\begin{aligned}
\langle\langle&
\cos(\vn{q}\cdot\vn{r})
\mathfrak{A};\mathfrak{C}
\rangle\rangle^{\rm M}
(\tau_1)=\\
=&\frac{\eta V}{4i\hbar}
\intkspa
\int_{0}^{\hbar\beta}
d\tau
\sum_{nm}
\sum_{n'm'}
\sum_{n''m''}
\Biggl[\\
-&\mathfrak{B}^{(2)}_{\vn{k}nm}
\mathfrak{A}^{(1)}_{\vn{k}n'm'}
\mathfrak{C}_{\vn{k}_{-}n''m''}
\mathscr{Z}^{(3)}_{\vn{k}nmn'm'n''m''}(\tau,\tau_1)\\
-&\mathfrak{B}^{(2)}_{\vn{k}nm}
\mathfrak{A}^{(1)}_{\vn{k}n'm'}
\mathfrak{C}_{\vn{k}_{+}n''m''}
\mathscr{Z}^{(4)}_{\vn{k}nmn'm'n''m''}(\tau,\tau_1)\\
+&\mathfrak{B}^{(1)}_{\vn{k}nm}
\mathfrak{A}^{(2)}_{\vn{k}n'm'}
\mathfrak{C}_{\vn{k}_{+}n''m''}
\mathscr{Z}^{(5)}_{\vn{k}nmn'm'n''m''}(\tau,\tau_1)\\
+&\mathfrak{B}^{(1)}_{\vn{k}nm}
\mathfrak{A}^{(2)}_{\vn{k}n'm'}
\mathfrak{C}_{\vn{k}_{-}n''m''}
\mathscr{Z}^{(6)}_{\vn{k}nmn'm'n''m''}(\tau,\tau_1)
\Biggr]
\end{aligned}
\ee
within first-order perturbation theory,
where we 
defined $\mathfrak{C}_{\vn{k}_{-}n''m''}=\langle u_{\vn{k}_{-}n''}|\mathfrak{C}|u_{\vn{k}_{-}m''} \rangle$
and $\mathfrak{C}_{\vn{k}_{+}n''m''}=\langle u_{\vn{k}_{+}n''}|\mathfrak{C}|u_{\vn{k}_{+}m''} \rangle$.

Note that $\mathscr{Z}^{(5)}$ can be obtained from $\mathscr{Z}^{(3)}$
by replacing $\vn{k}_{-}$ by $\vn{k}_{+}$ and $\vn{k}_{+}$ by $\vn{k}_{-}$.
Similarly, $\mathscr{Z}^{(6)}$ can be obtained from $\mathscr{Z}^{(4)}$
by replacing $\vn{k}_{-}$ by $\vn{k}_{+}$ and $\vn{k}_{+}$ by $\vn{k}_{-}$.
Therefore, we write down only the equations for $\mathscr{Z}^{(3)}$
and $\mathscr{Z}^{(4)}$ in the following.
Using Wick's theorem we find
\bege
\begin{aligned}
&\mathscr{Z}^{(3)}_{\vn{k}nmn'm'n''m''}(\tau,\tau_1)=\\
=-&G^{\rm M}_{m'n}(\vn{k}_{-},\tau_1-\tau)
G^{\rm M}_{mn'}(\vn{k}_{+},\tau-\tau_1)
G^{\rm M}_{m''n''}(\vn{k}_{-},0)\\
+&G^{\rm M}_{mn'}(\vn{k}_{+},\tau-\tau_1)
G^{\rm M}_{m''n}(\vn{k}_{-},-\tau)
G^{\rm M}_{m'n''}(\vn{k}_{-},\tau_1)\\
\end{aligned}
\ee
and
\bege
\begin{aligned}
&\mathscr{Z}^{(4)}_{\vn{k}nmn'm'n''m''}(\tau,\tau_1)=\\
=-&G^{\rm M}_{mn'}(\vn{k}_{+},\tau-\tau_1)
G^{\rm M}_{m'n}(\vn{k}_{-},\tau_1-\tau)
G^{\rm M}_{m''n''}(\vn{k}_{+},0)\\
+&G^{\rm M}_{mn''}(\vn{k}_{+},\tau)
G^{\rm M}_{m'n}(\vn{k}_{-},\tau_1-\tau)
G^{\rm M}_{m''n'}(\vn{k}_{+},-\tau_1).\\
\end{aligned}
\ee
The Fourier transform 
\bege
\begin{aligned}
&\langle\langle
\cos(\vn{q}\cdot\vn{r})
\mathfrak{A};\mathfrak{C}
\rangle\rangle^{\rm M}
(i\mathcal{E}_{N})=\\
=&\int_{0}^{\hbar\beta}\!\!\!\!d\,\tau_1\,\,
e^{\frac{i}{\hbar}\mathcal{E}_N\tau_1}
\langle\langle
\cos(\vn{q}\cdot\vn{r})
\mathfrak{A};\mathfrak{C}
\rangle\rangle^{\rm M}
(\tau_1)
\end{aligned}
\ee
of Eq.~\eqref{eq_pi_general_imagtau}
can be written as
\bege\label{eq_pi_general_imagenerg}
\begin{aligned}
&
\langle\langle
\cos(\vn{q}\cdot\vn{r})
\mathfrak{A};\mathfrak{C}
\rangle\rangle^{\rm M}
(i\mathcal{E}_{N})=\\
&=\frac{\eta V}{4i\hbar}
\intkspa
\sum_{nm}
\sum_{n'm'}
\sum_{n''m''}
\Biggl[\\
-&\mathfrak{B}^{(2)}_{\vn{k}nm}
\mathfrak{A}^{(1)}_{\vn{k}n'm'}
\mathfrak{C}_{\vn{k}_{-}n''m''}
\mathscr{Z}^{(3a)}_{\vn{k}nmn'm'n''m''}(i\mathcal{E}_N)\\
-&\mathfrak{B}^{(2)}_{\vn{k}nm}
\mathfrak{A}^{(1)}_{\vn{k}n'm'}
\mathfrak{C}_{\vn{k}_{+}n''m''}
\mathscr{Z}^{(4a)}_{\vn{k}nmn'm'n''m''}(i\mathcal{E}_N)\\
+&\mathfrak{B}^{(1)}_{\vn{k}nm}
\mathfrak{A}^{(2)}_{\vn{k}n'm'}
\mathfrak{C}_{\vn{k}_{+}n''m''}
\mathscr{Z}^{(5a)}_{\vn{k}nmn'm'n''m''}(i\mathcal{E}_N)\\
+&\mathfrak{B}^{(1)}_{\vn{k}nm}
\mathfrak{A}^{(2)}_{\vn{k}n'm'}
\mathfrak{C}_{\vn{k}_{-}n''m''}
\mathscr{Z}^{(6a)}_{\vn{k}nmn'm'n''m''}(i\mathcal{E}_N)
\Biggr]
\end{aligned}
\ee
in terms of the integrals
\bege\label{eq_z3a}
\begin{aligned}
&\mathscr{Z}^{(3a)}_{\vn{k}nmn'm'n''m''}(i\mathcal{E}_N)
=\int_{0}^{\hbar\beta}\!\!\!\!d\,\tau
\int_{0}^{\hbar\beta}\!\!\!\!d\,\tau_1\,\,
e^{\frac{i}{\hbar}\mathcal{E}_N\tau_1}\times\\
&\times
G^{\rm M}_{mn'}(\vn{k}_{+},\tau-\tau_1)
G^{\rm M}_{m''n}(\vn{k}_{-},-\tau)
G^{\rm M}_{m'n''}(\vn{k}_{-},\tau_1)=\\
&=
\frac{1}{\hbar\beta}
\sum_{p}
G^{\rm M}_{\vn{k}_{+}mn'}(i\mathcal{E}_{p})
G^{\rm M}_{\vn{k}_{-}m''n}(i\mathcal{E}_{p})
G^{\rm M}_{\vn{k}_{-}m'n''}(i\mathcal{E}_{p}+i\mathcal{E}_{N})
\end{aligned}
\ee
and
\bege\label{eq_z4a}
\begin{aligned}
&\mathscr{Z}^{(4a)}_{\vn{k}nmn'm'n''m''}(i\mathcal{E}_N)=
\int_{0}^{\hbar\beta}\!\!\!\!d\,\tau
\int_{0}^{\hbar\beta}\!\!\!\!d\,\tau_1\,\,
e^{\frac{i}{\hbar}\mathcal{E}_N\tau_1}\times\\
&\times
G^{\rm M}_{mn''}(\vn{k}_{+},\tau)
G^{\rm M}_{m'n}(\vn{k}_{-},\tau_1-\tau)
G^{\rm M}_{m''n'}(\vn{k}_{+},-\tau_1)=\\
&=
\frac{1}{\hbar\beta}
\sum_{p}
G^{\rm M}_{\vn{k}_{+}mn''}(i\mathcal{E}_{p})
G^{\rm M}_{\vn{k}_{-}m'n}(i\mathcal{E}_{p})
G^{\rm M}_{\vn{k}_{+}m''n'}(i\mathcal{E}_{p}-i\mathcal{E}_{N}),
\end{aligned}
\ee
where $\mathcal{E}_{N}=2\pi N/\beta$
is a bosonic Matsubara energy point and we used
\bege
G^{\rm M}(\tau)
=\frac{1}{\hbar\beta}
\sum_{p=-\infty}^{\infty}\!\!\!\!
e^{-i\mathcal{E}_{p}\tau/\hbar}
G^{\rm M}(i\mathcal{E}_{p}),
\ee
where $\mathcal{E}_{p}=(2p+1)\pi/\beta$ is a fermionic Matsubara point.
Again $\mathscr{Z}^{(5a)}$ is obtained from $\mathscr{Z}^{(3a)}$
by replacing $\vn{k}_{-}$ by $\vn{k}_{+}$ and $\vn{k}_{+}$ by
$\vn{k}_{-}$ and $\mathscr{Z}^{(6a)}$ is obtained 
from $\mathscr{Z}^{(4a)}$
in the same way.

Summation over
Matsubara points $\mathcal{E}_{p}$ 
in Eq.~\eqref{eq_z3a}
and in Eq.~\eqref{eq_z4a} 
and analytic 
continuation $i\mathcal{E}_{N}\rightarrow\hbar\omega$ 
yields
\bege
\begin{aligned}
&2\pi i\hbar 
\mathscr{Z}^{(3a)}_{\vn{k}nmn'm'n''m''}(\hbar\omega)=\\
&-
\int\! d\,\mathcal{E}
f(\mathcal{E})
G^{\rm R}_{\vn{k}_{+}mn'}(\mathcal{E})
G^{\rm R}_{\vn{k}_{-}m''n}(\mathcal{E})
G^{\rm R}_{\vn{k}_{-}m'n''}(\mathcal{E}\!+\!\hbar\omega)\\
&+
\int\! d\,\mathcal{E}
f(\mathcal{E})
G^{\rm A}_{\vn{k}_{+}mn'}(\mathcal{E})
G^{\rm A}_{\vn{k}_{-}m''n}(\mathcal{E})
G^{\rm R}_{\vn{k}_{-}m'n''}(\mathcal{E}\!+\!\hbar\omega)\\
&-
\int\! d\,\mathcal{E}
f(\mathcal{E})
G^{\rm A}_{\vn{k}_{+}mn'}(\mathcal{E}\!-\!\hbar\omega)
G^{\rm A}_{\vn{k}_{-}m''n}(\mathcal{E}\!-\!\hbar\omega)
G^{\rm R}_{\vn{k}_{-}m'n''}(\mathcal{E})\\
&+
\int\! d\,\mathcal{E}
f(\mathcal{E})
G^{\rm A}_{\vn{k}_{+}mn'}(\mathcal{E}\!-\!\hbar\omega)
G^{\rm A}_{\vn{k}_{-}m''n}(\mathcal{E}\!-\!\hbar\omega)
G^{\rm A}_{\vn{k}_{-}m'n''}(\mathcal{E})\\
\end{aligned}
\ee
and
\bege
\begin{aligned}
&
2\pi i\hbar
\mathscr{Z}^{(4a)}_{\vn{k}nmn'm'n''m''}(\hbar\omega)=\\
&-
\int d\,\mathcal{E}
f(\mathcal{E})
G^{\rm R}_{\vn{k}_{+}mn''}(\mathcal{E})
G^{\rm R}_{\vn{k}_{-}m'n}(\mathcal{E})
G^{\rm A}_{\vn{k}_{+}m''n'}(\mathcal{E}\!-\!\hbar\omega)\\
&+
\int d\,\mathcal{E}
f(\mathcal{E})
G^{\rm A}_{\vn{k}_{+}mn''}(\mathcal{E})
G^{\rm A}_{\vn{k}_{-}m'n}(\mathcal{E})
G^{\rm A}_{\vn{k}_{+}m''n'}(\mathcal{E}\!-\!\hbar\omega)\\
&-
\int d\,\mathcal{E}
f(\mathcal{E})
G^{\rm R}_{\vn{k}_{+}mn''}(\mathcal{E}\!+\!\hbar\omega)
G^{\rm R}_{\vn{k}_{-}m'n}(\mathcal{E}\!+\!\hbar\omega)
G^{\rm R}_{\vn{k}_{+}m''n'}(\mathcal{E})\\
&+
\int d\,\mathcal{E}
f(\mathcal{E})
G^{\rm R}_{\vn{k}_{+}mn''}(\mathcal{E}\!+\!\hbar\omega)
G^{\rm R}_{\vn{k}_{-}m'n}(\mathcal{E}\!+\!\hbar\omega)
G^{\rm A}_{\vn{k}_{+}m''n'}(\mathcal{E}).\\
\end{aligned}
\ee
In the next step we take the 
limit $\omega\rightarrow 0$ (see Eq.~\eqref{eq_def_torkance_ij11},
Eq.~\eqref{eq_chiral_invsot_correlation}, and
Eq.~\eqref{eq_def_lambda_ij11}):
\bege
\begin{aligned}
&-\frac{1}{V}
\lim_{\omega\rightarrow 0}
\frac{{\rm Im}
\langle\langle
\mathfrak{A}
\cos(\vn{q}\cdot\vn{r})
;\mathfrak{C}
\rangle\rangle^{\rm R}
(\hbar\omega)
}
{\hbar\omega}=\\
&=\frac{\eta}{4\hbar}{\rm Im}
\left[
\mathscr{Y}^{(3)}+\mathscr{Y}^{(4)}-
\mathscr{Y}^{(5)}-\mathscr{Y}^{(6)}
\right],
\end{aligned}
\ee
where we defined
\bege
\begin{aligned}
\mathscr{Y}^{(3)}=&
\frac{1}{i\hbar}\intkspa
\sum_{nm}
\sum_{n'm'}
\sum_{n''m''}
\mathfrak{B}^{(2)}_{\vn{k}nm}
\mathfrak{A}^{(1)}_{\vn{k}n'm'}
\mathfrak{C}_{\vn{k}_{-}n''m''}\times\\
&\times
\left.
\frac{\partial
\mathscr{Z}^{(3a)}_{\vn{k}nmn'm'n''m''}(\hbar\omega)
}
{
\partial \omega
}
\right|_{\omega=0}
,\\
\mathscr{Y}^{(4)}=&
\frac{1}{i\hbar}\intkspa
\sum_{nm}
\sum_{n'm'}
\sum_{n''m''}
\mathfrak{B}^{(2)}_{\vn{k}nm}
\mathfrak{A}^{(1)}_{\vn{k}n'm'}
\mathfrak{C}_{\vn{k}_{+}n''m''}\times\\
&\times
\left.
\frac{\partial
\mathscr{Z}^{(4a)}_{\vn{k}nmn'm'n''m''}(\hbar\omega)
}
{
\partial \omega
}
\right|_{\omega=0}
,\\
\mathscr{Y}^{(5)}=&
\frac{1}{i\hbar}\intkspa
\sum_{nm}
\sum_{n'm'}
\sum_{n''m''}
\mathfrak{B}^{(1)}_{\vn{k}nm}
\mathfrak{A}^{(2)}_{\vn{k}n'm'}
\mathfrak{C}_{\vn{k}_{+}n''m''}\times\\
&\times
\left.
\frac{\partial
\mathscr{Z}^{(5a)}_{\vn{k}nmn'm'n''m''}(\hbar\omega)
}
{
\partial \omega
}
\right|_{\omega=0}
,\\
\mathscr{Y}^{(6)}=&
\frac{1}{i\hbar}\intkspa
\sum_{nm}
\sum_{n'm'}
\sum_{n''m''}
\mathfrak{B}^{(1)}_{\vn{k}nm}
\mathfrak{A}^{(2)}_{\vn{k}n'm'}
\mathfrak{C}_{\vn{k}_{-}n''m''}\times\\
&\times
\left.
\frac{\partial
\mathscr{Z}^{(6a)}_{\vn{k}nmn'm'n''m''}(\hbar\omega)
}
{
\partial \omega
}
\right|_{\omega=0},\\
\end{aligned}
\ee
which can be expressed 
as $\mathscr{Y}^{(3)}=\mathscr{Y}^{(3a)}+\mathscr{Y}^{(3b)}$
and $\mathscr{Y}^{(4)}=\mathscr{Y}^{(4a)}+\mathscr{Y}^{(4b)}$,
where
\bege\label{eq_y3a}
\begin{aligned}
2\pi \hbar
\mathscr{Y}^{(3a)}&=\frac{1}{\hbar}\intkspa\int d\,\mathcal{E}
f(\mathcal{E})\times\\
\times{\rm Tr}
\Biggl[
&\mathfrak{A}_{\vn{k}}
G^{\rm R}_{\vn{k}_{-}}(\mathcal{E})
\mathfrak{C}_{\vn{k}_{-}}
G^{\rm A}_{\vn{k}_{-}}(\mathcal{E})
\mathfrak{B}_{\vn{k}}
G^{\rm A}_{\vn{k}_{+}}(\mathcal{E})
G^{\rm A}_{\vn{k}_{+}}(\mathcal{E})
\\
+
&\mathfrak{A}_{\vn{k}}
G^{\rm R}_{\vn{k}_{-}}(\mathcal{E})
G^{\rm R}_{\vn{k}_{-}}(\mathcal{E})
\mathfrak{C}_{\vn{k}_{-}}
G^{\rm A}_{\vn{k}_{-}}(\mathcal{E})
\mathfrak{B}_{\vn{k}}
G^{\rm A}_{\vn{k}_{+}}(\mathcal{E})
\\
+
&\mathfrak{A}_{\vn{k}}
G^{\rm R}_{\vn{k}_{-}}(\mathcal{E})
\mathfrak{C}_{\vn{k}_{-}}
G^{\rm A}_{\vn{k}_{-}}(\mathcal{E})
G^{\rm A}_{\vn{k}_{-}}(\mathcal{E})
\mathfrak{B}_{\vn{k}}
G^{\rm A}_{\vn{k}_{+}}(\mathcal{E})
\Biggr]
\\
=&\!\intkspa\!
\int\! d\,\mathcal{E}
f'(\mathcal{E})\times\\
&\times{\rm Tr}
\!\left[
\mathfrak{A}_{\vn{k}}
G^{\rm R}_{\vn{k}_{-}}(\mathcal{E})
\mathfrak{C}_{\vn{k}_{-}}
G^{\rm A}_{\vn{k}_{-}}(\mathcal{E})
\mathfrak{B}_{\vn{k}}
G^{\rm A}_{\vn{k}_{+}}(\mathcal{E})
\right]
\\
\end{aligned}
\ee
and
\bege\label{eq_y3b}
\begin{aligned}
&2\pi \hbar
\mathscr{Y}^{(3b)}=-\frac{1}{\hbar}\intkspa
\int d\,\mathcal{E}
f(\mathcal{E})\times\\
\times{\rm Tr}
\Biggl[
&\mathfrak{A}_{\vn{k}}
G^{\rm A}_{\vn{k}_{-}}(\mathcal{E})
\mathfrak{C}_{\vn{k}_{-}}
G^{\rm A}_{\vn{k}_{-}}(\mathcal{E})
\mathfrak{B}_{\vn{k}}
G^{\rm A}_{\vn{k}_{+}}(\mathcal{E})
G^{\rm A}_{\vn{k}_{+}}(\mathcal{E})
\\
+&
\mathfrak{A}_{\vn{k}}
G^{\rm R}_{\vn{k}_{-}}(\mathcal{E})
G^{\rm R}_{\vn{k}_{-}}(\mathcal{E})
\mathfrak{C}_{\vn{k}_{-}}
G^{\rm R}_{\vn{k}_{-}}(\mathcal{E})
\mathfrak{B}_{\vn{k}}
G^{\rm R}_{\vn{k}_{+}}(\mathcal{E})
\\
+&
\mathfrak{A}_{\vn{k}}
G^{\rm A}_{\vn{k}_{-}}(\mathcal{E})
\mathfrak{C}_{\vn{k}_{-}}
G^{\rm A}_{\vn{k}_{-}}(\mathcal{E})
G^{\rm A}_{\vn{k}_{-}}(\mathcal{E})
\mathfrak{B}_{\vn{k}}
G^{\rm A}_{\vn{k}_{+}}(\mathcal{E})
\Biggr].
\\
\end{aligned}
\ee

Similarly,
\bege\label{eq_y4a}
\begin{aligned}
2\pi \hbar
\mathscr{Y}^{(4a)}
&=\frac{1}{\hbar}\intkspa
\int d\,\mathcal{E}
f(\mathcal{E})\times\\
\times{\rm Tr}
\Biggl[
&\mathfrak{A}_{\vn{k}}
G^{\rm R}_{\vn{k}_{-}}(\mathcal{E})
\mathfrak{B}_{\vn{k}}
G^{\rm R}_{\vn{k}_{+}}(\mathcal{E})
\mathfrak{C}_{\vn{k}_{+}}
G^{\rm A}_{\vn{k}_{+}}(\mathcal{E})
G^{\rm A}_{\vn{k}_{+}}(\mathcal{E})
\\
-&
\mathfrak{A}_{\vn{k}}
G^{\rm R}_{\vn{k}_{-}}(\mathcal{E})
G^{\rm R}_{\vn{k}_{-}}(\mathcal{E})
\mathfrak{B}_{\vn{k}}
G^{\rm R}_{\vn{k}_{+}}(\mathcal{E})
\mathfrak{C}_{\vn{k}_{+}}
G^{\rm A}_{\vn{k}_{+}}(\mathcal{E})
\\
-&
\mathfrak{A}_{\vn{k}}
G^{\rm R}_{\vn{k}_{-}}(\mathcal{E})
\mathfrak{B}_{\vn{k}}
G^{\rm R}_{\vn{k}_{+}}(\mathcal{E})
G^{\rm R}_{\vn{k}_{+}}(\mathcal{E})
\mathfrak{C}_{\vn{k}_{+}}
G^{\rm A}_{\vn{k}_{+}}(\mathcal{E})
\Biggr]\\
=
&\intkspa
\int d\,\mathcal{E}
f'(\mathcal{E})\times\\
&\times{\rm Tr}
\left[
\mathfrak{A}_{\vn{k}}
G^{\rm R}_{\vn{k}_{-}}(\mathcal{E})
\mathfrak{B}_{\vn{k}}
G^{\rm R}_{\vn{k}_{+}}(\mathcal{E})
\mathfrak{C}_{\vn{k}_{+}}
G^{\rm A}_{\vn{k}_{+}}(\mathcal{E})
\right]
\\
\end{aligned}
\ee
and
\bege\label{eq_y4b}
\begin{aligned}
&2\pi \hbar
\mathscr{Y}^{(4b)}=-
\frac{1}{\hbar}\intkspa
\int d\,\mathcal{E}
f(\mathcal{E})\times\\
\times{\rm Tr}
\Biggl[
&\mathfrak{A}_{\vn{k}}
G^{\rm A}_{\vn{k}_{-}}(\mathcal{E})
\mathfrak{B}_{\vn{k}}
G^{\rm A}_{\vn{k}_{+}}(\mathcal{E})
\mathfrak{C}_{\vn{k}_{+}}
G^{\rm A}_{\vn{k}_{+}}(\mathcal{E})
G^{\rm A}_{\vn{k}_{+}}(\mathcal{E})
\\
+&
\mathfrak{A}_{\vn{k}}
G^{\rm R}_{\vn{k}_{-}}(\mathcal{E})
G^{\rm R}_{\vn{k}_{-}}(\mathcal{E})
\mathfrak{B}_{\vn{k}}
G^{\rm R}_{\vn{k}_{+}}(\mathcal{E})
\mathfrak{C}_{\vn{k}_{+}}
G^{\rm R}_{\vn{k}_{+}}(\mathcal{E})
\\
+&
\mathfrak{A}_{\vn{k}}
G^{\rm R}_{\vn{k}_{-}}(\mathcal{E})
\mathfrak{B}_{\vn{k}}
G^{\rm R}_{\vn{k}_{+}}(\mathcal{E})
G^{\rm R}_{\vn{k}_{+}}(\mathcal{E})
\mathfrak{C}_{\vn{k}_{+}}
G^{\rm R}_{\vn{k}_{+}}(\mathcal{E})
\Biggr].
\\
\end{aligned}
\ee
We call $\mathscr{Y}^{(3a)}$ and $\mathscr{Y}^{(4a)}$ Fermi surface
terms
and $\mathscr{Y}^{(3b)}$ and $\mathscr{Y}^{(4b)}$ Fermi sea terms. 
Again $\mathscr{Y}^{(5)}$ is obtained from $\mathscr{Y}^{(3)}$
by replacing $\vn{k}_{-}$ by $\vn{k}_{+}$ and $\vn{k}_{+}$ by
$\vn{k}_{-}$ and $\mathscr{Y}^{(6)}$ is obtained
from $\mathscr{Y}^{(4)}$
in the same way.

Finally, we take the limit $\vn{q} \rightarrow 0$:
\bege
\begin{aligned}
&\Lambda=-\frac{2}{\hbar V\eta}{\rm Im}
\lim_{\vn{q}\rightarrow 0}
\lim_{\omega\rightarrow 0}
\frac{\partial
}
{
\partial \omega
}
\frac{\partial
}
{
\partial q_i
}
\langle\langle
\mathfrak{A}\cos(\vn{q}\cdot\vn{r});
\mathfrak{C}
\rangle\rangle^{\rm R}(\hbar\omega)
\\
&=\frac{1}{2\hbar}
\lim_{\vn{q}\rightarrow 0}
\frac{\partial}{\partial q_i}
{\rm Im}
\left[
\mathscr{Y}^{(3)}+\mathscr{Y}^{(4)}-\mathscr{Y}^{(5)}-\mathscr{Y}^{(6)}
\right]\\
&=\frac{1}{2\hbar}
{\rm Im}
\left[
\mathscr{X}^{(3)}+\mathscr{X}^{(4)}-\mathscr{X}^{(5)}-\mathscr{X}^{(6)}
\right],\\
\end{aligned}
\ee
where we defined 
\bege
\mathscr{X}^{(j)}=\left.
\frac{\partial}{\partial q_i}
\right|_{\vn{q}=0}\mathscr{Y}^{(j)}
\ee
for $j=3,4,5,6$.
Since $\mathscr{Y}^{(4)}$ and $\mathscr{Y}^{(6)}$
are related by the interchange of $\vn{k}_{-}$ and $\vn{k}_{+}$
it follows that $\mathscr{X}^{(6)}=-\mathscr{X}^{(4)}$.
Similarly, since $\mathscr{Y}^{(3)}$ and $\mathscr{Y}^{(5)}$
are related by the interchange of $\vn{k}_{-}$ and $\vn{k}_{+}$
it follows that $\mathscr{X}^{(5)}=-\mathscr{X}^{(3)}$.
Consequently, we need
\bege\label{eq_lambda}
\Lambda=
\frac{1}{\hbar}
{\rm Im}
\left[
\mathscr{X}^{(3a)}
\!+\!
\mathscr{X}^{(3b)}
\!+\!
\mathscr{X}^{(4a)}
\!+\!
\mathscr{X}^{(4b)}
\right],
\ee
where $\mathscr{X}^{(3a)}$ and $\mathscr{X}^{(4a)}$ are the Fermi surface
terms
and $\mathscr{X}^{(3b)}$ and $\mathscr{X}^{(4b)}$ are the Fermi sea terms. 
The Fermi surface terms are given by
\bege\label{eq_x_3a}
\begin{aligned}
&\mathscr{X}^{(3a)}=
\frac{-1}{4 \pi\hbar}
\intkspa
\int d\,\mathcal{E}
f'(\mathcal{E})
{\rm Tr}
\Biggl[\\
&\mathfrak{A}_{\vn{k}}
G^{\rm R}_{\vn{k}}(\mathcal{E})
v _{\vn{k}}
G^{\rm R}_{\vn{k}}(\mathcal{E})
\mathfrak{C}_{\vn{k}}
G^{\rm A}_{\vn{k}}(\mathcal{E})
\mathfrak{B}_{\vn{k}}
G^{\rm A}_{\vn{k}}(\mathcal{E})
\\
+&
\mathfrak{A}_{\vn{k}}
G^{\rm R}_{\vn{k}}(\mathcal{E})
\mathfrak{C}_{\vn{k}}
G^{\rm A}_{\vn{k}}(\mathcal{E})
v _{\vn{k}}
G^{\rm A}_{\vn{k}}(\mathcal{E})
\mathfrak{B}_{\vn{k}}
G^{\rm A}_{\vn{k}}(\mathcal{E})
\\
-&
\mathfrak{A}_{\vn{k}}
G^{\rm R}_{\vn{k}}(\mathcal{E})
\mathfrak{C}_{\vn{k}}
G^{\rm A}_{\vn{k}}(\mathcal{E})
\mathfrak{B}_{\vn{k}}
G^{\rm A}_{\vn{k}}(\mathcal{E})
v _{\vn{k}}
G^{\rm A}_{\vn{k}}(\mathcal{E})
\\
+&
\mathfrak{A}_{\vn{k}}
G^{\rm R}_{\vn{k}}(\mathcal{E})
\frac{\partial
\mathfrak{C}_{\vn{k}}
}{\partial k}
G^{\rm A}_{\vn{k}}(\mathcal{E})
\mathfrak{B}_{\vn{k}}
G^{\rm A}_{\vn{k}}(\mathcal{E})
\Biggr]\\
\end{aligned}
\ee
and
\bege\label{eq_4a_3a}
\mathscr{X}^{(4a)}=
-\left[\mathscr{X}^{(3a)}\right]^{*}.
\ee
The Fermi sea terms are given by
\bege\label{eq_x3b}
\begin{aligned}
\mathscr{X}^{(3b)}&=\frac{-1}{4\pi\hbar^2}
\intkspa
\int d \mathcal{E} f(\mathcal{E}) {\rm Tr}
\Biggl[\\
&-(\mathfrak{A}RvRR\mathfrak{C}R\mathfrak{B}R)
+(\mathfrak{A}A\mathfrak{C}AA\mathfrak{B}AvA)
\\
&-(\mathfrak{A}RRvR\mathfrak{C}R\mathfrak{B}R)
-(\mathfrak{A}RR\mathfrak{C}RvR\mathfrak{B}R)\\
&+(\mathfrak{A}RR\mathfrak{C}R\mathfrak{B}RvR)
-(\mathfrak{A}AvA\mathfrak{C}A\mathfrak{B}AA)\\
&-(\mathfrak{A}A\mathfrak{C}AvA\mathfrak{B}AA)
+(\mathfrak{A}A\mathfrak{C}A\mathfrak{B}AvAA)\\
&+(\mathfrak{A}A\mathfrak{C}A\mathfrak{B}AAvA)
-(\mathfrak{A}AvA\mathfrak{C}AA\mathfrak{B}A)\\
&-(\mathfrak{A}A\mathfrak{C}AvAA\mathfrak{B}A)
-(\mathfrak{A}A\mathfrak{C}AAvA\mathfrak{B}A)\\
&-(\mathfrak{A}RR\frac{\partial\mathfrak{C}}{\partial k}R\mathfrak{B}R)
-(\mathfrak{A}A\frac{\partial\mathfrak{C}}{\partial k}AA\mathfrak{B}A)\\
&-(\mathfrak{A}A\frac{\partial\mathfrak{C}}{\partial k}A\mathfrak{B}AA)
\Biggr]\\
\end{aligned}
\ee
and
\bege\label{eq_4b_3b}
\mathscr{X}^{(4b)}=
-\left[\mathscr{X}^{(3b)}\right]^{*}.
\ee
In Eq.~\eqref{eq_x3b}
we use the 
abbreviations $R=G^{\rm R}_{\vn{k}}(\mathcal{E})$,
$A=G^{\rm A}_{\vn{k}}(\mathcal{E})$,
$\mathfrak{A}=\mathfrak{A}_{\vn{k}}$,
$\mathfrak{B}=\mathfrak{B}_{\vn{k}}$,
$\mathfrak{C}=\mathfrak{C}_{\vn{k}}$.
It is important to note that $\mathfrak{C}_{\vn{k}_{-}}$
and $\mathfrak{C}_{\vn{k}_{+}}$
depend on $q$ through $\vn{k}_{-}=\vn{k}-\vn{q}/2$
and $\vn{k}_{+}=\vn{k}+\vn{q}/2$ .
The $q$ derivative therefore generates the
additional terms with $\partial \mathfrak{C}_{\vn{k}}/\partial k$
in Eq.~\eqref{eq_x_3a}
and Eq.~\eqref{eq_x3b}.
In contrast, $\mathfrak{A}_{\vn{k}}$ and  $\mathfrak{B}_{\vn{k}}$ do not
depend linearly on $q$.


Eq.~\eqref{eq_lambda} simplifies 
due to the relations Eq.~\eqref{eq_4a_3a}
and Eq.~\eqref{eq_4b_3b} as follows:
\bege
\Lambda=
\frac{2}{\hbar}
{\rm Im}
\left[
\mathscr{X}^{(3a)}
\!+\!
\mathscr{X}^{(3b)}
\right].
\ee

In order to obtain the expression for 
the chiral contribution to the torque-torque
correlation we
choose the operators as follows:
\bege\label{eq_gyro_choose_operators}
\begin{aligned}
&\mathfrak{B}\rightarrow\mathcal{T}_{k}\\
&\mathfrak{A}\rightarrow-\mathcal{T}_{i}\\
&\mathfrak{C}\rightarrow\mathcal{T}_{j}\\
&\frac{\partial\mathfrak{C}}{\partial k}=0\\
&v\rightarrow v_{l}.
\end{aligned}
\ee
This leads to Eq.~\eqref{eq_chiral_gyromag_green}, 
Eq.~\eqref{eq_fermi_surface} 
and Eq.~\eqref{eq_fermi_sea}
of the main text.

In order to obtain the
expression for the chiral contribution to the CIT,
we set
\bege\label{eq_cit_choose_operators}
\begin{aligned}
&\mathfrak{B}\rightarrow\mathcal{T}_{k}\\
&\mathfrak{A}\rightarrow-\mathcal{T}_{i}\\
&\mathfrak{C}\rightarrow -e v_{j}\\
&\frac{\partial\mathfrak{C}}{\partial k}\rightarrow -\frac{e\hbar}{m}\delta_{jl}\\
&v\rightarrow v_{l}.
\end{aligned}
\ee
This leads to Eq.~\eqref{eq_chiral_torkance_green},
Eq.~\eqref{eq_fermi_surface_torkance}
and Eq.~\eqref{eq_fermi_sea_torkance}.

In order to obtain the expression for the
chiral contribution to the ICIT,
we set
\bege\label{eq_icit_choose_operators}
\begin{aligned}
&\mathfrak{B}\rightarrow\mathcal{T}_{k}\\
&\mathfrak{A}\rightarrow -e v_{i}\\
&\mathfrak{C}\rightarrow \mathcal{T}_{j}\\
&\frac{\partial\mathfrak{C}}{\partial k}\rightarrow 0\\
&v\rightarrow v_{l}.
\end{aligned}
\ee
This leads to Eq.~\eqref{eq_chiral_invsot_green},
Eq.~\eqref{eq_fermi_surface_isot}
and Eq.~\eqref{eq_fermi_sea_isot}.

\bibliography{nedmitorno}

\begin{thebibliography}{41}
\expandafter\ifx\csname natexlab\endcsname\relax\def\natexlab#1{#1}\fi
\expandafter\ifx\csname bibnamefont\endcsname\relax
  \def\bibnamefont#1{#1}\fi
\expandafter\ifx\csname bibfnamefont\endcsname\relax
  \def\bibfnamefont#1{#1}\fi
\expandafter\ifx\csname citenamefont\endcsname\relax
  \def\citenamefont#1{#1}\fi
\expandafter\ifx\csname url\endcsname\relax
  \def\url#1{\texttt{#1}}\fi
\expandafter\ifx\csname urlprefix\endcsname\relax\def\urlprefix{URL }\fi
\providecommand{\bibinfo}[2]{#2}
\providecommand{\eprint}[2][]{\url{#2}}

\bibitem[{\citenamefont{Nawaoka et~al.}(2015)\citenamefont{Nawaoka, Miwa,
  Shiota, Mizuochi, and Suzuki}}]{voltage_induction_dmi_AuFeMgO}
\bibinfo{author}{\bibfnamefont{K.}~\bibnamefont{Nawaoka}},
  \bibinfo{author}{\bibfnamefont{S.}~\bibnamefont{Miwa}},
  \bibinfo{author}{\bibfnamefont{Y.}~\bibnamefont{Shiota}},
  \bibinfo{author}{\bibfnamefont{N.}~\bibnamefont{Mizuochi}}, \bibnamefont{and}
  \bibinfo{author}{\bibfnamefont{Y.}~\bibnamefont{Suzuki}},
  \bibinfo{journal}{Applied Physics Express} \textbf{\bibinfo{volume}{8}},
  \bibinfo{pages}{063004} (\bibinfo{year}{2015}).

\bibitem[{\citenamefont{{Yang} et~al.}(2018)\citenamefont{{Yang}, {Boulle},
  {Cros}, {Fert}, and
  {Chshiev}}}]{controlling_DMI_layer_stacking_capping_electric_field}
\bibinfo{author}{\bibfnamefont{H.}~\bibnamefont{{Yang}}},
  \bibinfo{author}{\bibfnamefont{O.}~\bibnamefont{{Boulle}}},
  \bibinfo{author}{\bibfnamefont{V.}~\bibnamefont{{Cros}}},
  \bibinfo{author}{\bibfnamefont{A.}~\bibnamefont{{Fert}}}, \bibnamefont{and}
  \bibinfo{author}{\bibfnamefont{M.}~\bibnamefont{{Chshiev}}},
  \bibinfo{journal}{Scientific Reports} \textbf{\bibinfo{volume}{8}},
  \bibinfo{pages}{12356} (\bibinfo{year}{2018}).

\bibitem[{\citenamefont{Srivastava et~al.}(2018)\citenamefont{Srivastava,
  Schott, Juge, K{\v r}i{\v z}{\'a}kov{\'a}, Belmeguenai, Roussign{\'e},
  Bernand-Mantel, Ranno, Pizzini, Ch{\'e}rif
  et~al.}}]{large_voltage_tuning_dmi_dynamic_control_skyrmion_chirality}
\bibinfo{author}{\bibfnamefont{T.}~\bibnamefont{Srivastava}},
  \bibinfo{author}{\bibfnamefont{M.}~\bibnamefont{Schott}},
  \bibinfo{author}{\bibfnamefont{R.}~\bibnamefont{Juge}},
  \bibinfo{author}{\bibfnamefont{V.}~\bibnamefont{K{\v r}i{\v
  z}{\'a}kov{\'a}}},
  \bibinfo{author}{\bibfnamefont{M.}~\bibnamefont{Belmeguenai}},
  \bibinfo{author}{\bibfnamefont{Y.}~\bibnamefont{Roussign{\'e}}},
  \bibinfo{author}{\bibfnamefont{A.}~\bibnamefont{Bernand-Mantel}},
  \bibinfo{author}{\bibfnamefont{L.}~\bibnamefont{Ranno}},
  \bibinfo{author}{\bibfnamefont{S.}~\bibnamefont{Pizzini}},
  \bibinfo{author}{\bibfnamefont{S.-M.} \bibnamefont{Ch{\'e}rif}},
  \bibnamefont{et~al.}, \bibinfo{journal}{Nano Letters}
  \textbf{\bibinfo{volume}{18}}, \bibinfo{pages}{4871} (\bibinfo{year}{2018}).

\bibitem[{\citenamefont{Mikhaylovskiy et~al.}(2015)\citenamefont{Mikhaylovskiy,
  Hendry, Secchi, Mentink, Eckstein, Wu, Pisarev, Kruglyak, Katsnelson, Rasing
  et~al.}}]{ultrafast_modification_exchange_interaction}
\bibinfo{author}{\bibfnamefont{R.~V.} \bibnamefont{Mikhaylovskiy}},
  \bibinfo{author}{\bibfnamefont{E.}~\bibnamefont{Hendry}},
  \bibinfo{author}{\bibfnamefont{A.}~\bibnamefont{Secchi}},
  \bibinfo{author}{\bibfnamefont{J.~H.} \bibnamefont{Mentink}},
  \bibinfo{author}{\bibfnamefont{M.}~\bibnamefont{Eckstein}},
  \bibinfo{author}{\bibfnamefont{A.}~\bibnamefont{Wu}},
  \bibinfo{author}{\bibfnamefont{R.~V.} \bibnamefont{Pisarev}},
  \bibinfo{author}{\bibfnamefont{V.~V.} \bibnamefont{Kruglyak}},
  \bibinfo{author}{\bibfnamefont{M.~I.} \bibnamefont{Katsnelson}},
  \bibinfo{author}{\bibfnamefont{T.}~\bibnamefont{Rasing}},
  \bibnamefont{et~al.}, \bibinfo{journal}{Nature Communications}
  \textbf{\bibinfo{volume}{6}}, \bibinfo{pages}{8190} (\bibinfo{year}{2015}).

\bibitem[{\citenamefont{Freimuth et~al.}(2018)\citenamefont{Freimuth, Bl\"ugel,
  and Mokrousov}}]{sotcocucotrila}
\bibinfo{author}{\bibfnamefont{F.}~\bibnamefont{Freimuth}},
  \bibinfo{author}{\bibfnamefont{S.}~\bibnamefont{Bl\"ugel}}, \bibnamefont{and}
  \bibinfo{author}{\bibfnamefont{Y.}~\bibnamefont{Mokrousov}},
  \bibinfo{journal}{Phys. Rev. B} \textbf{\bibinfo{volume}{98}},
  \bibinfo{pages}{024419} (\bibinfo{year}{2018}).

\bibitem[{\citenamefont{Freimuth
  et~al.}(2017{\natexlab{a}})\citenamefont{Freimuth, Bl\"ugel, and
  Mokrousov}}]{spicudmi}
\bibinfo{author}{\bibfnamefont{F.}~\bibnamefont{Freimuth}},
  \bibinfo{author}{\bibfnamefont{S.}~\bibnamefont{Bl\"ugel}}, \bibnamefont{and}
  \bibinfo{author}{\bibfnamefont{Y.}~\bibnamefont{Mokrousov}},
  \bibinfo{journal}{Phys. Rev. B} \textbf{\bibinfo{volume}{96}},
  \bibinfo{pages}{054403} (\bibinfo{year}{2017}{\natexlab{a}}).

\bibitem[{\citenamefont{Kikuchi et~al.}(2016)\citenamefont{Kikuchi, Koretsune,
  Arita, and Tatara}}]{dmi_doppler_shift}
\bibinfo{author}{\bibfnamefont{T.}~\bibnamefont{Kikuchi}},
  \bibinfo{author}{\bibfnamefont{T.}~\bibnamefont{Koretsune}},
  \bibinfo{author}{\bibfnamefont{R.}~\bibnamefont{Arita}}, \bibnamefont{and}
  \bibinfo{author}{\bibfnamefont{G.}~\bibnamefont{Tatara}},
  \bibinfo{journal}{Phys. Rev. Lett.} \textbf{\bibinfo{volume}{116}},
  \bibinfo{pages}{247201} (\bibinfo{year}{2016}).

\bibitem[{\citenamefont{Karnad et~al.}(2018)\citenamefont{Karnad, Freimuth,
  Martinez, Lo~Conte, Gubbiotti, Schulz, Senz, Ocker, Mokrousov, and
  Kl\"aui}}]{cidmi_karnad}
\bibinfo{author}{\bibfnamefont{G.~V.} \bibnamefont{Karnad}},
  \bibinfo{author}{\bibfnamefont{F.}~\bibnamefont{Freimuth}},
  \bibinfo{author}{\bibfnamefont{E.}~\bibnamefont{Martinez}},
  \bibinfo{author}{\bibfnamefont{R.}~\bibnamefont{Lo~Conte}},
  \bibinfo{author}{\bibfnamefont{G.}~\bibnamefont{Gubbiotti}},
  \bibinfo{author}{\bibfnamefont{T.}~\bibnamefont{Schulz}},
  \bibinfo{author}{\bibfnamefont{S.}~\bibnamefont{Senz}},
  \bibinfo{author}{\bibfnamefont{B.}~\bibnamefont{Ocker}},
  \bibinfo{author}{\bibfnamefont{Y.}~\bibnamefont{Mokrousov}},
  \bibnamefont{and} \bibinfo{author}{\bibfnamefont{M.}~\bibnamefont{Kl\"aui}},
  \bibinfo{journal}{Phys. Rev. Lett.} \textbf{\bibinfo{volume}{121}},
  \bibinfo{pages}{147203} (\bibinfo{year}{2018}).

\bibitem[{\citenamefont{Kato et~al.}(2019)\citenamefont{Kato, Kawaguchi, Lau,
  Kikuchi, Nakatani, and Hayashi}}]{cidmi_hayashi}
\bibinfo{author}{\bibfnamefont{N.}~\bibnamefont{Kato}},
  \bibinfo{author}{\bibfnamefont{M.}~\bibnamefont{Kawaguchi}},
  \bibinfo{author}{\bibfnamefont{Y.-C.} \bibnamefont{Lau}},
  \bibinfo{author}{\bibfnamefont{T.}~\bibnamefont{Kikuchi}},
  \bibinfo{author}{\bibfnamefont{Y.}~\bibnamefont{Nakatani}}, \bibnamefont{and}
  \bibinfo{author}{\bibfnamefont{M.}~\bibnamefont{Hayashi}},
  \bibinfo{journal}{Phys. Rev. Lett.} \textbf{\bibinfo{volume}{122}},
  \bibinfo{pages}{257205} (\bibinfo{year}{2019}).

\bibitem[{\citenamefont{Freimuth
  et~al.}(2014{\natexlab{a}})\citenamefont{Freimuth, Bl\"ugel, and
  Mokrousov}}]{mothedmisot}
\bibinfo{author}{\bibfnamefont{F.}~\bibnamefont{Freimuth}},
  \bibinfo{author}{\bibfnamefont{S.}~\bibnamefont{Bl\"ugel}}, \bibnamefont{and}
  \bibinfo{author}{\bibfnamefont{Y.}~\bibnamefont{Mokrousov}},
  \bibinfo{journal}{Journal of physics: Condensed matter}
  \textbf{\bibinfo{volume}{26}}, \bibinfo{pages}{104202}
  (\bibinfo{year}{2014}{\natexlab{a}}).

\bibitem[{\citenamefont{Freimuth et~al.}(2013)\citenamefont{Freimuth, Bamler,
  Mokrousov, and Rosch}}]{phase_space_berry}
\bibinfo{author}{\bibfnamefont{F.}~\bibnamefont{Freimuth}},
  \bibinfo{author}{\bibfnamefont{R.}~\bibnamefont{Bamler}},
  \bibinfo{author}{\bibfnamefont{Y.}~\bibnamefont{Mokrousov}},
  \bibnamefont{and} \bibinfo{author}{\bibfnamefont{A.}~\bibnamefont{Rosch}},
  \bibinfo{journal}{Phys. Rev. B} \textbf{\bibinfo{volume}{88}},
  \bibinfo{pages}{214409} (\bibinfo{year}{2013}).

\bibitem[{\citenamefont{Shi et~al.}(2007)\citenamefont{Shi, Vignale, Xiao, and
  Niu}}]{shi_quantum_theory_orbital_mag}
\bibinfo{author}{\bibfnamefont{J.}~\bibnamefont{Shi}},
  \bibinfo{author}{\bibfnamefont{G.}~\bibnamefont{Vignale}},
  \bibinfo{author}{\bibfnamefont{D.}~\bibnamefont{Xiao}}, \bibnamefont{and}
  \bibinfo{author}{\bibfnamefont{Q.}~\bibnamefont{Niu}},
  \bibinfo{journal}{Phys. Rev. Lett.} \textbf{\bibinfo{volume}{99}},
  \bibinfo{pages}{197202} (\bibinfo{year}{2007}).

\bibitem[{\citenamefont{Thonhauser et~al.}(2005)\citenamefont{Thonhauser,
  Ceresoli, Vanderbilt, and Resta}}]{om_insulators_mlwfs}
\bibinfo{author}{\bibfnamefont{T.}~\bibnamefont{Thonhauser}},
  \bibinfo{author}{\bibfnamefont{D.}~\bibnamefont{Ceresoli}},
  \bibinfo{author}{\bibfnamefont{D.}~\bibnamefont{Vanderbilt}},
  \bibnamefont{and} \bibinfo{author}{\bibfnamefont{R.}~\bibnamefont{Resta}},
  \bibinfo{journal}{Phys. Rev. Lett.} \textbf{\bibinfo{volume}{95}},
  \bibinfo{pages}{137205} (\bibinfo{year}{2005}).

\bibitem[{\citenamefont{Ceresoli et~al.}(2006)\citenamefont{Ceresoli,
  Thonhauser, Vanderbilt, and Resta}}]{om_crystals_mlwfs}
\bibinfo{author}{\bibfnamefont{D.}~\bibnamefont{Ceresoli}},
  \bibinfo{author}{\bibfnamefont{T.}~\bibnamefont{Thonhauser}},
  \bibinfo{author}{\bibfnamefont{D.}~\bibnamefont{Vanderbilt}},
  \bibnamefont{and} \bibinfo{author}{\bibfnamefont{R.}~\bibnamefont{Resta}},
  \bibinfo{journal}{Phys. Rev. B} \textbf{\bibinfo{volume}{74}},
  \bibinfo{pages}{024408} (\bibinfo{year}{2006}).

\bibitem[{\citenamefont{Malashevich et~al.}(2010)\citenamefont{Malashevich,
  Souza, Coh, and Vanderbilt}}]{theory_OM_response}
\bibinfo{author}{\bibfnamefont{A.}~\bibnamefont{Malashevich}},
  \bibinfo{author}{\bibfnamefont{I.}~\bibnamefont{Souza}},
  \bibinfo{author}{\bibfnamefont{S.}~\bibnamefont{Coh}}, \bibnamefont{and}
  \bibinfo{author}{\bibfnamefont{D.}~\bibnamefont{Vanderbilt}},
  \bibinfo{journal}{New Journal of Physics} \textbf{\bibinfo{volume}{12}},
  \bibinfo{pages}{053032} (\bibinfo{year}{2010}).

\bibitem[{\citenamefont{Zhong et~al.}(2016)\citenamefont{Zhong, Moore, and
  Souza}}]{gyrotropic_magnetic_effect_magnetic_moment_fermi_surface}
\bibinfo{author}{\bibfnamefont{S.}~\bibnamefont{Zhong}},
  \bibinfo{author}{\bibfnamefont{J.~E.} \bibnamefont{Moore}}, \bibnamefont{and}
  \bibinfo{author}{\bibfnamefont{I.}~\bibnamefont{Souza}},
  \bibinfo{journal}{Phys. Rev. Lett.} \textbf{\bibinfo{volume}{116}},
  \bibinfo{pages}{077201} (\bibinfo{year}{2016}).

\bibitem[{\citenamefont{Manchon et~al.}(2019)\citenamefont{Manchon,
  \ifmmode~\check{Z}\else \v{Z}\fi{}elezn\'y, Miron, Jungwirth, Sinova,
  Thiaville, Garello, and Gambardella}}]{sot_review}
\bibinfo{author}{\bibfnamefont{A.}~\bibnamefont{Manchon}},
  \bibinfo{author}{\bibfnamefont{J.}~\bibnamefont{\ifmmode~\check{Z}\else
  \v{Z}\fi{}elezn\'y}}, \bibinfo{author}{\bibfnamefont{I.~M.}
  \bibnamefont{Miron}},
  \bibinfo{author}{\bibfnamefont{T.}~\bibnamefont{Jungwirth}},
  \bibinfo{author}{\bibfnamefont{J.}~\bibnamefont{Sinova}},
  \bibinfo{author}{\bibfnamefont{A.}~\bibnamefont{Thiaville}},
  \bibinfo{author}{\bibfnamefont{K.}~\bibnamefont{Garello}}, \bibnamefont{and}
  \bibinfo{author}{\bibfnamefont{P.}~\bibnamefont{Gambardella}},
  \bibinfo{journal}{Rev. Mod. Phys.} \textbf{\bibinfo{volume}{91}},
  \bibinfo{pages}{035004} (\bibinfo{year}{2019}).

\bibitem[{\citenamefont{Freimuth et~al.}(2016)\citenamefont{Freimuth, Bl\"ugel,
  and Mokrousov}}]{itsot}
\bibinfo{author}{\bibfnamefont{F.}~\bibnamefont{Freimuth}},
  \bibinfo{author}{\bibfnamefont{S.}~\bibnamefont{Bl\"ugel}}, \bibnamefont{and}
  \bibinfo{author}{\bibfnamefont{Y.}~\bibnamefont{Mokrousov}},
  \bibinfo{journal}{J. Phys.: Condens. matter} \textbf{\bibinfo{volume}{28}},
  \bibinfo{pages}{316001} (\bibinfo{year}{2016}).

\bibitem[{\citenamefont{Qin et~al.}(2011)\citenamefont{Qin, Niu, and
  Shi}}]{energy_magnetization_and_thermal_hall_effect}
\bibinfo{author}{\bibfnamefont{T.}~\bibnamefont{Qin}},
  \bibinfo{author}{\bibfnamefont{Q.}~\bibnamefont{Niu}}, \bibnamefont{and}
  \bibinfo{author}{\bibfnamefont{J.}~\bibnamefont{Shi}},
  \bibinfo{journal}{Phys. Rev. Lett.} \textbf{\bibinfo{volume}{107}},
  \bibinfo{pages}{236601} (\bibinfo{year}{2011}).

\bibitem[{\citenamefont{Xiao et~al.}(2006)\citenamefont{Xiao, Yao, Fang, and
  Niu}}]{berry_anomalous_thermoelectric_xiao}
\bibinfo{author}{\bibfnamefont{D.}~\bibnamefont{Xiao}},
  \bibinfo{author}{\bibfnamefont{Y.}~\bibnamefont{Yao}},
  \bibinfo{author}{\bibfnamefont{Z.}~\bibnamefont{Fang}}, \bibnamefont{and}
  \bibinfo{author}{\bibfnamefont{Q.}~\bibnamefont{Niu}},
  \bibinfo{journal}{Phys. Rev. Lett.} \textbf{\bibinfo{volume}{97}},
  \bibinfo{pages}{026603} (\bibinfo{year}{2006}).

\bibitem[{\citenamefont{Cooper et~al.}(1997)\citenamefont{Cooper, Halperin, and
  Ruzin}}]{thermoelectric_response_magnetic_field}
\bibinfo{author}{\bibfnamefont{N.~R.} \bibnamefont{Cooper}},
  \bibinfo{author}{\bibfnamefont{B.~I.} \bibnamefont{Halperin}},
  \bibnamefont{and} \bibinfo{author}{\bibfnamefont{I.~M.} \bibnamefont{Ruzin}},
  \bibinfo{journal}{Phys. Rev. B} \textbf{\bibinfo{volume}{55}},
  \bibinfo{pages}{2344} (\bibinfo{year}{1997}).

\bibitem[{\citenamefont{Hals and
  Brataas}(2015)}]{spin_motive_forces_current_induced_torques}
\bibinfo{author}{\bibfnamefont{K.~M.~D.} \bibnamefont{Hals}} \bibnamefont{and}
  \bibinfo{author}{\bibfnamefont{A.}~\bibnamefont{Brataas}},
  \bibinfo{journal}{Phys. Rev. B} \textbf{\bibinfo{volume}{91}},
  \bibinfo{pages}{214401} (\bibinfo{year}{2015}).

\bibitem[{\citenamefont{Freimuth
  et~al.}(2014{\natexlab{b}})\citenamefont{Freimuth, Bl\"ugel, and
  Mokrousov}}]{ibcsoit}
\bibinfo{author}{\bibfnamefont{F.}~\bibnamefont{Freimuth}},
  \bibinfo{author}{\bibfnamefont{S.}~\bibnamefont{Bl\"ugel}}, \bibnamefont{and}
  \bibinfo{author}{\bibfnamefont{Y.}~\bibnamefont{Mokrousov}},
  \bibinfo{journal}{Phys. Rev. B} \textbf{\bibinfo{volume}{90}},
  \bibinfo{pages}{174423} (\bibinfo{year}{2014}{\natexlab{b}}).

\bibitem[{\citenamefont{Freimuth et~al.}(2015)\citenamefont{Freimuth, Bl\"ugel,
  and Mokrousov}}]{invsot}
\bibinfo{author}{\bibfnamefont{F.}~\bibnamefont{Freimuth}},
  \bibinfo{author}{\bibfnamefont{S.}~\bibnamefont{Bl\"ugel}}, \bibnamefont{and}
  \bibinfo{author}{\bibfnamefont{Y.}~\bibnamefont{Mokrousov}},
  \bibinfo{journal}{Phys. Rev. B} \textbf{\bibinfo{volume}{92}},
  \bibinfo{pages}{064415} (\bibinfo{year}{2015}).

\bibitem[{\citenamefont{Ciccarelli et~al.}(2014)\citenamefont{Ciccarelli, Hals,
  Irvine, Novak, Tserkovnyak, Kurebayashi, Brataas, and
  Ferguson}}]{charge_pumping_Ciccarelli}
\bibinfo{author}{\bibfnamefont{C.}~\bibnamefont{Ciccarelli}},
  \bibinfo{author}{\bibfnamefont{K.~M.~D.} \bibnamefont{Hals}},
  \bibinfo{author}{\bibfnamefont{A.}~\bibnamefont{Irvine}},
  \bibinfo{author}{\bibfnamefont{V.}~\bibnamefont{Novak}},
  \bibinfo{author}{\bibfnamefont{Y.}~\bibnamefont{Tserkovnyak}},
  \bibinfo{author}{\bibfnamefont{H.}~\bibnamefont{Kurebayashi}},
  \bibinfo{author}{\bibfnamefont{A.}~\bibnamefont{Brataas}}, \bibnamefont{and}
  \bibinfo{author}{\bibfnamefont{A.}~\bibnamefont{Ferguson}},
  \bibinfo{journal}{Nature nanotechnology} \textbf{\bibinfo{volume}{10}},
  \bibinfo{pages}{50 } (\bibinfo{year}{2014}).

\bibitem[{\citenamefont{Mosendz et~al.}(2010)\citenamefont{Mosendz, Pearson,
  Fradin, Bauer, Bader, and Hoffmann}}]{prl_mosendz_spin_pumping}
\bibinfo{author}{\bibfnamefont{O.}~\bibnamefont{Mosendz}},
  \bibinfo{author}{\bibfnamefont{J.~E.} \bibnamefont{Pearson}},
  \bibinfo{author}{\bibfnamefont{F.~Y.} \bibnamefont{Fradin}},
  \bibinfo{author}{\bibfnamefont{G.~E.~W.} \bibnamefont{Bauer}},
  \bibinfo{author}{\bibfnamefont{S.~D.} \bibnamefont{Bader}}, \bibnamefont{and}
  \bibinfo{author}{\bibfnamefont{A.}~\bibnamefont{Hoffmann}},
  \bibinfo{journal}{Phys. Rev. Lett.} \textbf{\bibinfo{volume}{104}},
  \bibinfo{pages}{046601} (\bibinfo{year}{2010}).

\bibitem[{\citenamefont{Czeschka et~al.}(2011)\citenamefont{Czeschka, Dreher,
  Brandt, Weiler, Althammer, Imort, Reiss, Thomas, Schoch, Limmer
  et~al.}}]{prl_czeschka_spin_pumping}
\bibinfo{author}{\bibfnamefont{F.~D.} \bibnamefont{Czeschka}},
  \bibinfo{author}{\bibfnamefont{L.}~\bibnamefont{Dreher}},
  \bibinfo{author}{\bibfnamefont{M.~S.} \bibnamefont{Brandt}},
  \bibinfo{author}{\bibfnamefont{M.}~\bibnamefont{Weiler}},
  \bibinfo{author}{\bibfnamefont{M.}~\bibnamefont{Althammer}},
  \bibinfo{author}{\bibfnamefont{I.-M.} \bibnamefont{Imort}},
  \bibinfo{author}{\bibfnamefont{G.}~\bibnamefont{Reiss}},
  \bibinfo{author}{\bibfnamefont{A.}~\bibnamefont{Thomas}},
  \bibinfo{author}{\bibfnamefont{W.}~\bibnamefont{Schoch}},
  \bibinfo{author}{\bibfnamefont{W.}~\bibnamefont{Limmer}},
  \bibnamefont{et~al.}, \bibinfo{journal}{Phys. Rev. Lett.}
  \textbf{\bibinfo{volume}{107}}, \bibinfo{pages}{046601}
  (\bibinfo{year}{2011}).

\bibitem[{\citenamefont{Ralph and Stiles}(2008)}]{RALPH20081190}
\bibinfo{author}{\bibfnamefont{D.}~\bibnamefont{Ralph}} \bibnamefont{and}
  \bibinfo{author}{\bibfnamefont{M.}~\bibnamefont{Stiles}},
  \bibinfo{journal}{Journal of Magnetism and Magnetic Materials}
  \textbf{\bibinfo{volume}{320}}, \bibinfo{pages}{1190} (\bibinfo{year}{2008}).

\bibitem[{\citenamefont{Nakatsuji et~al.}(2015)\citenamefont{Nakatsuji,
  Kiyohara, and Higo}}]{ahe_Mn3Sn}
\bibinfo{author}{\bibfnamefont{S.}~\bibnamefont{Nakatsuji}},
  \bibinfo{author}{\bibfnamefont{N.}~\bibnamefont{Kiyohara}}, \bibnamefont{and}
  \bibinfo{author}{\bibfnamefont{T.}~\bibnamefont{Higo}},
  \bibinfo{journal}{Nature} \textbf{\bibinfo{volume}{527}},
  \bibinfo{pages}{212} (\bibinfo{year}{2015}).

\bibitem[{\citenamefont{Garate et~al.}(2009)\citenamefont{Garate, Gilmore,
  Stiles, and MacDonald}}]{nonadiabatic_stt_garate_macdonald}
\bibinfo{author}{\bibfnamefont{I.}~\bibnamefont{Garate}},
  \bibinfo{author}{\bibfnamefont{K.}~\bibnamefont{Gilmore}},
  \bibinfo{author}{\bibfnamefont{M.~D.} \bibnamefont{Stiles}},
  \bibnamefont{and} \bibinfo{author}{\bibfnamefont{A.~H.}
  \bibnamefont{MacDonald}}, \bibinfo{journal}{Phys. Rev. B}
  \textbf{\bibinfo{volume}{79}}, \bibinfo{pages}{104416}
  (\bibinfo{year}{2009}).

\bibitem[{\citenamefont{van~der Bijl and
  Duine}(2012)}]{current_induced_torques_rashba_ferromagnets}
\bibinfo{author}{\bibfnamefont{E.}~\bibnamefont{van~der Bijl}}
  \bibnamefont{and} \bibinfo{author}{\bibfnamefont{R.~A.} \bibnamefont{Duine}},
  \bibinfo{journal}{Phys. Rev. B} \textbf{\bibinfo{volume}{86}},
  \bibinfo{pages}{094406} (\bibinfo{year}{2012}).

\bibitem[{\citenamefont{Yang et~al.}(2009)\citenamefont{Yang, Beach, Knutson,
  Xiao, Niu, Tsoi, and Erskine}}]{universal_electromotive_force_induced_by_dwm}
\bibinfo{author}{\bibfnamefont{S.~A.} \bibnamefont{Yang}},
  \bibinfo{author}{\bibfnamefont{G.~S.~D.} \bibnamefont{Beach}},
  \bibinfo{author}{\bibfnamefont{C.}~\bibnamefont{Knutson}},
  \bibinfo{author}{\bibfnamefont{D.}~\bibnamefont{Xiao}},
  \bibinfo{author}{\bibfnamefont{Q.}~\bibnamefont{Niu}},
  \bibinfo{author}{\bibfnamefont{M.}~\bibnamefont{Tsoi}}, \bibnamefont{and}
  \bibinfo{author}{\bibfnamefont{J.~L.} \bibnamefont{Erskine}},
  \bibinfo{journal}{Phys. Rev. Lett.} \textbf{\bibinfo{volume}{102}},
  \bibinfo{pages}{067201} (\bibinfo{year}{2009}).

\bibitem[{\citenamefont{Barnes and
  Maekawa}(2007)}]{generalized_faraday_barnes_maekawa}
\bibinfo{author}{\bibfnamefont{S.~E.} \bibnamefont{Barnes}} \bibnamefont{and}
  \bibinfo{author}{\bibfnamefont{S.}~\bibnamefont{Maekawa}},
  \bibinfo{journal}{Phys. Rev. Lett.} \textbf{\bibinfo{volume}{98}},
  \bibinfo{pages}{246601} (\bibinfo{year}{2007}).

\bibitem[{\citenamefont{Freimuth
  et~al.}(2017{\natexlab{b}})\citenamefont{Freimuth, Bl\"ugel, and
  Mokrousov}}]{longitex}
\bibinfo{author}{\bibfnamefont{F.}~\bibnamefont{Freimuth}},
  \bibinfo{author}{\bibfnamefont{S.}~\bibnamefont{Bl\"ugel}}, \bibnamefont{and}
  \bibinfo{author}{\bibfnamefont{Y.}~\bibnamefont{Mokrousov}},
  \bibinfo{journal}{Phys. Rev. B} \textbf{\bibinfo{volume}{95}},
  \bibinfo{pages}{094434} (\bibinfo{year}{2017}{\natexlab{b}}).

\bibitem[{\citenamefont{Savrasov}(1998)}]{PhysRevLett.81.2570}
\bibinfo{author}{\bibfnamefont{S.~Y.} \bibnamefont{Savrasov}},
  \bibinfo{journal}{Phys. Rev. Lett.} \textbf{\bibinfo{volume}{81}},
  \bibinfo{pages}{2570} (\bibinfo{year}{1998}).

\bibitem[{\citenamefont{Gao et~al.}(2018)\citenamefont{Gao, Vanderbilt, and
  Xiao}}]{spin_toroidization}
\bibinfo{author}{\bibfnamefont{Y.}~\bibnamefont{Gao}},
  \bibinfo{author}{\bibfnamefont{D.}~\bibnamefont{Vanderbilt}},
  \bibnamefont{and} \bibinfo{author}{\bibfnamefont{D.}~\bibnamefont{Xiao}},
  \bibinfo{journal}{Phys. Rev. B} \textbf{\bibinfo{volume}{97}},
  \bibinfo{pages}{134423} (\bibinfo{year}{2018}).

\bibitem[{\citenamefont{Xiao and Niu}(2020)}]{PhysRevB.101.235430}
\bibinfo{author}{\bibfnamefont{C.}~\bibnamefont{Xiao}} \bibnamefont{and}
  \bibinfo{author}{\bibfnamefont{Q.}~\bibnamefont{Niu}},
  \bibinfo{journal}{Phys. Rev. B} \textbf{\bibinfo{volume}{101}},
  \bibinfo{pages}{235430} (\bibinfo{year}{2020}).

\bibitem[{\citenamefont{Freimuth
  et~al.}(2017{\natexlab{c}})\citenamefont{Freimuth, Bl\"ugel, and
  Mokrousov}}]{chigyromag}
\bibinfo{author}{\bibfnamefont{F.}~\bibnamefont{Freimuth}},
  \bibinfo{author}{\bibfnamefont{S.}~\bibnamefont{Bl\"ugel}}, \bibnamefont{and}
  \bibinfo{author}{\bibfnamefont{Y.}~\bibnamefont{Mokrousov}},
  \bibinfo{journal}{Phys. Rev. B} \textbf{\bibinfo{volume}{96}},
  \bibinfo{pages}{104418} (\bibinfo{year}{2017}{\natexlab{c}}).

\bibitem[{\citenamefont{Schulz et~al.}(2015)\citenamefont{Schulz, Alejos,
  Martinez, Hals, Garcia, Vila, Lee, Lo~Conte, Karnad, Moretti
  et~al.}}]{sots_perpendicular_to_wall}
\bibinfo{author}{\bibfnamefont{T.}~\bibnamefont{Schulz}},
  \bibinfo{author}{\bibfnamefont{O.}~\bibnamefont{Alejos}},
  \bibinfo{author}{\bibfnamefont{E.}~\bibnamefont{Martinez}},
  \bibinfo{author}{\bibfnamefont{K.~M.~D.} \bibnamefont{Hals}},
  \bibinfo{author}{\bibfnamefont{K.}~\bibnamefont{Garcia}},
  \bibinfo{author}{\bibfnamefont{L.}~\bibnamefont{Vila}},
  \bibinfo{author}{\bibfnamefont{K.}~\bibnamefont{Lee}},
  \bibinfo{author}{\bibfnamefont{R.}~\bibnamefont{Lo~Conte}},
  \bibinfo{author}{\bibfnamefont{G.~V.} \bibnamefont{Karnad}},
  \bibinfo{author}{\bibfnamefont{S.}~\bibnamefont{Moretti}},
  \bibnamefont{et~al.}, \bibinfo{journal}{Applied Physics Letters}
  \textbf{\bibinfo{volume}{107}}, \bibinfo{pages}{122405}
  (\bibinfo{year}{2015}).

\bibitem[{\citenamefont{Ju\'{e} et~al.}(2016)\citenamefont{Ju\'{e}, Safeer,
  Drouard, Lopez, Balint, Buda-Prejbeanu, Boulle, Auffret, Schuhl, Manchon
  et~al.}}]{chiral_damping_magnetic_domain_walls}
\bibinfo{author}{\bibfnamefont{E.}~\bibnamefont{Ju\'{e}}},
  \bibinfo{author}{\bibfnamefont{C.~K.} \bibnamefont{Safeer}},
  \bibinfo{author}{\bibfnamefont{M.}~\bibnamefont{Drouard}},
  \bibinfo{author}{\bibfnamefont{A.}~\bibnamefont{Lopez}},
  \bibinfo{author}{\bibfnamefont{P.}~\bibnamefont{Balint}},
  \bibinfo{author}{\bibfnamefont{L.}~\bibnamefont{Buda-Prejbeanu}},
  \bibinfo{author}{\bibfnamefont{O.}~\bibnamefont{Boulle}},
  \bibinfo{author}{\bibfnamefont{S.}~\bibnamefont{Auffret}},
  \bibinfo{author}{\bibfnamefont{A.}~\bibnamefont{Schuhl}},
  \bibinfo{author}{\bibfnamefont{A.}~\bibnamefont{Manchon}},
  \bibnamefont{et~al.}, \bibinfo{journal}{Nature materials}
  \textbf{\bibinfo{volume}{15}}, \bibinfo{pages}{272} (\bibinfo{year}{2016}).

\bibitem[{\citenamefont{Akosa et~al.}(2016)\citenamefont{Akosa, Miron, Gaudin,
  and Manchon}}]{phenomenology_chiral_damping}
\bibinfo{author}{\bibfnamefont{C.~A.} \bibnamefont{Akosa}},
  \bibinfo{author}{\bibfnamefont{I.~M.} \bibnamefont{Miron}},
  \bibinfo{author}{\bibfnamefont{G.}~\bibnamefont{Gaudin}}, \bibnamefont{and}
  \bibinfo{author}{\bibfnamefont{A.}~\bibnamefont{Manchon}},
  \bibinfo{journal}{Phys. Rev. B} \textbf{\bibinfo{volume}{93}},
  \bibinfo{pages}{214429} (\bibinfo{year}{2016}).

\end{thebibliography}

\end{document}